\documentclass[a4paper,11pt]{article}

\usepackage{jheppub} 

\usepackage{graphicx}

\usepackage{amsmath}
\usepackage[utf8]{inputenc}

\usepackage{amssymb}
\usepackage{amsfonts}
\usepackage{amsbsy}
\usepackage{url}
\usepackage{enumerate}
\usepackage{caption}
\usepackage{subcaption}
\usepackage{color}
\usepackage{ulem}
\usepackage{bm}
\usepackage{comment}
\usepackage{multirow,bigdelim}
\usepackage{braket}
\usepackage{tikz}
\usepackage{wrapfig}

\newcommand{\be}{\begin{eqnarray}}
\newcommand{\ee}{\end{eqnarray}}
\newcommand{\p}{\partial}

\makeatletter
\@addtoreset{equation}{section}
\makeatother

\newcommand{\bee}{\begin{equation}}
\newcommand{\eee}{(\end{equation})}

\newcommand\rsout{\bgroup\markoverwith{\textcolor{red}{\rule[0.5ex]{2pt}{0.4pt}}}\ULon}

\title{
Massless fermions and superconductivity of string-wall composites 
}

\author[a,b]{Minoru Eto}
\emailAdd{meto[at]sci.kj.yamagata-u.ac.jp}
\affiliation[a]{Department of Physics, Yamagata University, Kojirakawa-machi 1-4-12, Yamagata, Yamagata 990-8560, Japan}
\affiliation[b]{Research and Education Center for Natural Sciences, Keio University, 4-1-1 Hiyoshi, Yokohama, Kanagawa 223-8521, Japan}

\author[a]{and Yuito Suzuki}
\emailAdd{s231862d[at]st.yamagata-u.ac.jp}

\abstract{
An axion cosmic string is known to be a chiral superconductor when the axion couples to an electrically charged fermion.
After the QCD phase transition, a QCD axion string is attached by $N$ domain walls.
We would like to elucidate the fate of massless fermions on a global string after 
domain walls attached not only in the axion model but also in general models having string-wall composites.
We investigate the Dirac equation under various string-wall composite backgrounds both in the axion(-like) models and in the ${\cal N}=2$ supersymmetry inspired Abelian-Higgs models. We give an answer to
the elementary question of whether massless fermions exist, and if so, where they are localized.
The answer depends on fermion/boson masses in the models, and the massless fermion can be localized either on the string, on one of the domain walls, or in one of the vacua. We find analytic solutions for the fermion zero mode function by which we prove the existence of the massless fermion on the string-wall composites. We also show supercurrents flowing along the string-wall composites and anomalous electric currents flowing in from outside.
}
\preprint{YGHP-23-07}

\begin{document}
\maketitle


\section{Introduction}

Topological solitons are ubiquitous in vast area of modern science including not only physics but also chemistry, biology, optics, and so on. 
Even limiting ourself to physics, they play crucial roles in particle physics, nuclear physics, condensed matter physics, and cosmology.
Such great utility is due to simple but universal phenomenon called spontaneous symmetry breaking (SSB), which does not depend on the scale of the system's size, energy, or dimension etc. 
If SSB in the system under consideration is driven in sufficiently simple situation, i.e. by a single condensate at single stage, then topological solitons of a single species will appear. However, it sometimes happens that multiple types of solitons appear simultaneously as composite solitons. In fact, it is not uncommon for several different SSBs to occur at once or in succession in systems with multiple fields developing non-zero vacuum expectation values (VEVs), so the appearance of composite solitons might be quite common. There are any number of examples of composite solitons in the literature, and it would be impossible to keep track of them all, so we will only mention two here that are relevant to our work. One is the axion string-wall composites, and the other is Bogomol'nyi-Prasad-Sommerfield (BPS) domain wall junctions in ${\cal N}=2$ supersymmetric Abelian-Higgs models.

Topological defects would be generated at phase transitions in the early Universe \cite{Vilenkin:2000jqa}. The cosmic strings are one of the most intensively studied topological solitons. Perhaps, the axion, a pseudo Nambu-Goldstone (NG) particle produced by the broken $U(1)_{\rm PQ}$ and a dark matter candidate, is one of the most important fields, since it can solve
the strong CP problem by the Peccei-Quinn (PQ) mechanism where the anomalous global $U(1)_{\rm PQ}$ symmetry breaking sets the QCD $\theta$-angle to zero \cite{Peccei:1977hh,Weinberg:1977ma,Wilczek:1977pj}.
When $U(1)_{\rm PQ}$ is spontaneously broken in the early universe, it gives rise to cosmic strings by the Kibble-Zurek mechanism \cite{Kibble:1976sj,Zurek:1985qw}. 
Lifetime of the cosmic strings is important because it affects axion relic abundance through evolution of the universe.
When two strings come across, they reconnect and form loops. The loops shrink and disappear due to self-tension. The number of strings reduces  through this process as the Universe cools down.
There is an important possibility that the cosmic strings can be superconducting \cite{WITTEN1985557}.  When the axions couple to fermions, the fermion zero modes appear on the axion strings \cite{Jackiw:1981ee,CALLAN1985427}, and if the fermions carry a non-zero electric charge,  an electric supercurrent can be built up.
Once persistent supercurrents are built up on the string loops (they are called vortons), the supercurrents prevent loop shrinkage and can make the string lifetime longer \cite{Lazarides:1984zq,Lazarides:1987rq,Carter:1993wu,Brandenberger:1996zp,Martins:1998gb,Martins:1998th,Carter:1999an}. If the vortons survive for sufficiently long period, their contribution to the density of axion dark mater could be quite large and non-negligible in cosmology.
Recently, the superconducting axion string has been revived \cite{Fukuda:2020kym,Abe:2020ure,Agrawal:2020euj,Ibe:2021ctf}.
They are the chiral superconductor, and when they move in a primordial magnetic field a drag force makes them slowdown, which leads to an orders of magnitude enhancement in the number density of the axion strings \cite{Fukuda:2020kym}.

The universe further expands and cools down, and at the time of QCD phase transition the $U(1)_{\rm PQ}$ symmetry explicitly breaks down to $Z_N$ and the axion gets massive. The $Z_N$ symmetry is also spontaneously broken, so that the $N$ domain walls emerge and attach to the axion string \cite{Sikivie:1982qv,Chang:1998tb,Vilenkin:1981zs}, resulting in appearance of composite solitons of the string and domain walls which are the main targets of this paper. We want to give an answer to the elementary question of whether massless fermions exist on the string-wall composites and, if so, where they are localized. Are they localized on the string, the domain wall, or somewhere else?
The answer depends on the model, but all of these are correct.

In seeking the answer, we will be naturally invited to ${\cal N}=2$ supersymmetric (SUSY) Abelian-Higgs models.
Indeed, it is an ideal platform for studying more general string-wall composites \cite{Eto:2005cp,Eto:2005fm}. There, the string-wall composites are often called domain wall junctions. They are special in the sense that they preserve a part of supersymmetries and satisfy first order differential equations called the BPS equations \cite{Gibbons:1999np,Carroll:1999wr}. Static inter-soliton forces between constituent solitons are exactly cancelled, so that the BPS solutions are infinitely degenerate in energy. They are parametrized by so-called moduli parameters. The moduli parameters typically relate to positions of multiple solitons. Hence, varying them, the solitonic configuration changes accordingly. The BPS equations are well controlled by mathematically sophisticated structures, making it easy to construct even the very complex networks. In a certain limit, even analytic solutions can be available. All these nice properties are due to the supersymmetry from which we will get a lot of benefits in this paper. 

In order to make a relation between the axion models and the SUSY gauge theories, we will abandon all the SUSY fermions and only take the bosons of the ${\cal N}=2$ supersymmetric Abelian-Higgs models. Then we will rebuild up our theory by including a new fermion which interacts with the $U(1)$ vector multiplet scalar via a normal Yukawa term. The resulting theory has no SUSYs. However, since the bosonic part is unchanged, we can fully make use of all the virtues of the BPS string-wall composites. Then, we will succeed in obtaining analytic solutions to the Dirac equation and find analytic fermion zero mode functions for all the BPS string-wall composites. We will also clarify the normalization condition for the zero modes. By giving the analytic solutions together with the normalization condition, we complete proving the existence of fermion zero modes on the string-wall composites.  
In addition, we will study effects of a bulk fermion mass in the axion-like models and show that localization of the fermion zero mode function depends sensitively on it. We also show the supercurrents flowing on the string-wall composites and the anomalous electric currents which flows in from outside and compensates violation of the electric charge conservation law on the string-wall composites.

The organization of this paper is as follows. In Sec.~\ref{sec:review}, we will review the string-wall composites both in the axion models and in the ${\cal N}=2$ supersymmetric Abelian-Higgs models. In Sec.~\ref{sec:DWSF}, first we will convert the Dirac equation with the string-wall composite backgrounds into a tractable form, and second we will discuss generic features about chirality of the massless fermions. Third we will clarify solvable conditions for the Dirac equation. Fourth, we will obtain numerical solutions of the fermion zero mode functions for the axion string-wall composites, and fifth we will explain the analytic solutions of the massless fermion for all the BPS string-wall composite backgrounds. In Sec.~\ref{sec:bulkmass}, we will study effect of the bulk fermion mass, and in Sec.~\ref{sec:supercurrrent}, we will discuss the superconducting currents flowing along the string-wall composites and anomalous electric currents flowing in. Sec.~\ref{sec:summary} is devoted to summarize the results and discuss possible future directions.

\section{Composite solitons of domain walls and vortex strings}
\label{sec:review}

In this section, we will review two models which admit composite solitons consisting of two kinds of solitons, domain walls and vortex strings.
One is the axion models where an axion string is attached by $N$ domain walls. The other is a bosonic part of an ${\cal N}=2$ supersymmetric Abelian-Higgs model with $N_{\rm F}$ flavors (hypermultiplets) which is one of the ideal theoretical laboratories for studying string-wall composites which are sometimes called the domain wall junctions.

\subsection{Axion string-wall composites}
\label{sec:axion_striing_wall}

\subsubsection{The axion models}
\label{sec:axion_model}

Let $\varphi$ be an complex scalar field. 
We consider the Lagrangian in \(3+1\) dimensions 
\begin{align}
\mathcal{L}_\mathrm{axion}&=|\partial_\mu\varphi|^2-\frac{\lambda}{4}(|\varphi|^2-v^2)^2-\eta(\varphi^N+\bar \varphi^N) \,,
\label{eq:Lag_axion}
\end{align}
where \(\lambda,\eta\) are coupling constants. It turns out that the integer \(N\) corresponds to the number of domain walls attaching a vortex string.
When $\eta = 0$, the Lagrangian is invariant under the global $U(1)$ transformation $\varphi \to e^{i\zeta} \varphi$ which we identify to the Peccei-Quinn $U(1)_{\rm PQ}$ symmetry.
The $U(1)$ symmetry is spontaneously broken in the vacuum $|\varphi| = v$, and the corresponding Nambu-Goldstone boson (axion) appears $\zeta = \arg \varphi$. Furthermore, SSB of the $U(1)$ symmetry gives rise to topological vortex strings, so-called the axion strings.
In sufficiently low energy scale where we cannot ignore $\eta$, the second term in the potential explicitly breaks the $U(1)$ symmetry. 
We can approximate $\varphi$ by $\varphi \simeq v e^{i\zeta}$, and the scalar potential reduces to
\be
V \simeq 2 \eta v^N \cos N \zeta.
\ee
The axion become pseudo NG boson with a non-zero mass $m_{\rm axion}^2 = \eta  v^N$. Furthermore, when we once go around the vortex string, $\zeta$ traverses the potential barrier $N$ times. This results in the $N$ domain walls attaches the vortex string.

For later convenience, we will use the dimensionless notation
\begin{gather}
\phi \equiv \varphi/v,\quad \tilde x^\mu \equiv \sqrt{\lambda}v x^\mu.
\end{gather}
The Lagrangian and the  equation of motion are expressed as
\begin{gather}
\frac{\mathcal{L}}{\lambda v^4}=|\tilde{\partial}_\mu\phi|^2-\frac{1}{4}(|\phi|^2-1)^2-\alpha(\phi^N+\bar\phi^N),  \\
\tilde{\partial}^2\phi+\frac{1}{2}(|\phi|^2-1)\phi+\alpha N \bar\phi^{N-1}=0, 
\label{eq:axion_eom}
\end{gather}
where \(\tilde{\partial}_\mu=\partial/\partial \tilde x^\mu\) and we define
\be
\alpha \equiv \frac{\eta v^{N-4}}{\lambda}.
\ee
We will decompose the complex scalar field $\phi$ into the real and imaginary parts as $\phi = \phi_1 + i \phi_2$.

\subsubsection{The axion strings and domain walls}

Let us go tour of the composite solitons in the axion models with various $N$'s.

\begin{wrapfigure}{r}[0pt]{0.3\textwidth}
  \centering
 \includegraphics[width=0.3\textwidth]{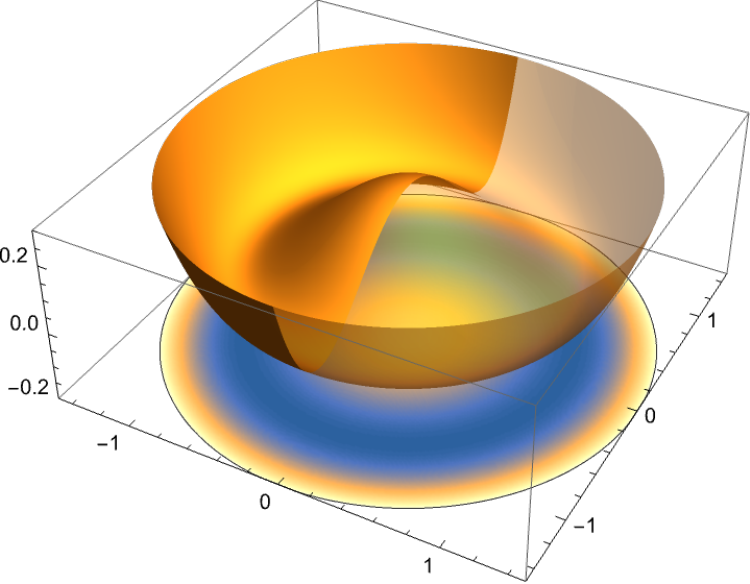}
 \caption{The dimensionless potential for \(N=0\).}
 \label{fig:axion_N=0_vac}
\end{wrapfigure}
\begin{figure}[h]
\centering
\begin{minipage}[t]{0.32\linewidth}
\centering
\includegraphics[keepaspectratio,scale=0.4]{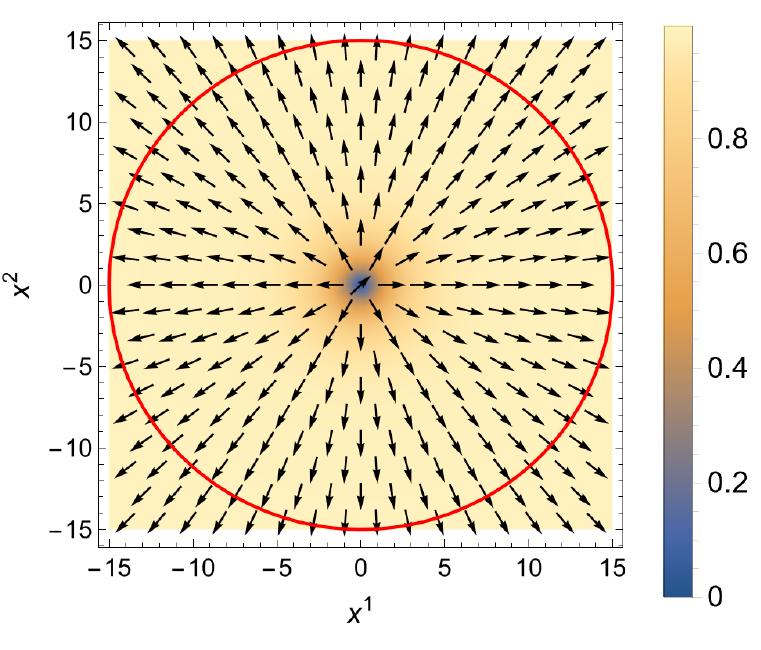}
\subcaption{}
\end{minipage}
\begin{minipage}[t]{0.32\linewidth}
\centering
\includegraphics[keepaspectratio,scale=0.4]{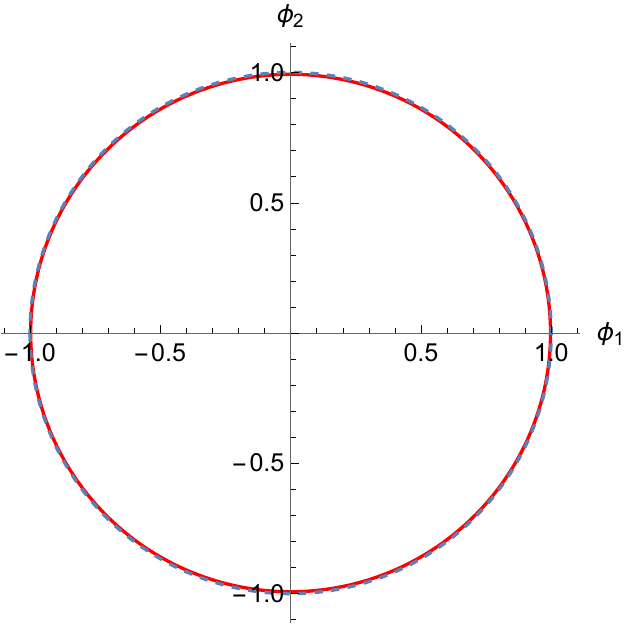}
\subcaption{}
\end{minipage}
\begin{minipage}[t]{0.32\linewidth}
\centering
\includegraphics[keepaspectratio,scale=0.4]{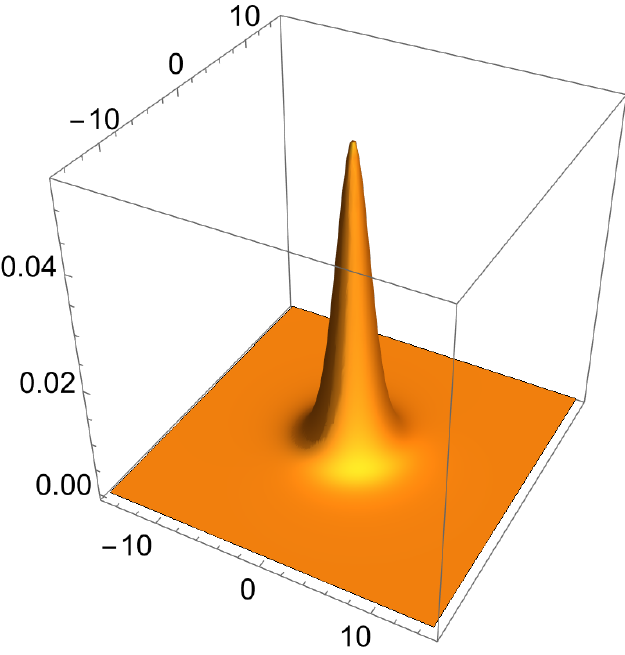}
\subcaption{}
\end{minipage}
\caption{A string in the axion model for \(N=0\).
(a) The amplitude and phase of \(\phi\) are shown by density and vector plots, respectively.
(b) The red curve is image of the red circle in (a). The dashed circle is the vacuum manifold $S^1$.
(c) The charge density $q$.
}
\label{fig:axion_N=0_BG}
\end{figure}

\paragraph{The model with $N=0$}
The first site is $N=0$ where the global strings exist alone without domain walls.
The potential is a wine bottle type and its vacuum manifold is \(S^1\) as shown in Fig.~\ref{fig:axion_N=0_vac}.
For the fundamental group of the vacuum manifold is $\pi_1(S^1) = \mathbb{Z}$, there exist topologically stable strings. 
A nontrivial string solution to (\ref{eq:axion_eom}) is available only numerically. 
The axially symmetric string can be obtained by making an Ansatz
\be
\phi(\tilde x) = F(\tilde \rho) e^{i\theta}\,,\quad \tilde x^1 + i \tilde x^2 = \tilde \rho e^{i\theta}\,,
\label{eq:axion_string}
\ee
with the appropriate boundary condition $f(0) = 0$ and $f(\infty) = 1$.
We show a numerical solution in Fig.~\ref{fig:axion_N=0_BG} where the minimal winding string extending along the $z$-axis is located at the origin on the $xy$ plane.
From the color density plot in the panel (a) one can see the amplitude $F(\tilde\rho)$, and the radially spreading arrows which correspond to the two component vector field $(\phi_1,\phi_2)$ indicate the winding number is one. 
The field $\phi$ gives a map from $\mathbb{R}^2$ (the $xy$ plane) to $\mathbb{C}$ (the complex field $\phi$).
We are interested in only maps belonging to a class that the boundary condition is $|\phi| \to 1$ as $\tilde \rho = |\tilde x+i\tilde y| \to \infty$.
Therefore, the whole $xy$ plane is mapped on to the unit disk in the $\phi$ plane.
The panel (b) shows the image of the large red circle in the panel (a). 
A quantity that directly shows topological property of $\phi$ is the following topological charge density
\be
q(\tilde x^i) = \frac{1}{2\pi} \epsilon^{ab}\epsilon^{ij} \tilde\p_i \phi_a \tilde\p_j\phi_b\,,\quad(a,b,i,j=1,2)\,.
\label{eq:topological_charge}
\ee
The integration of $q$ on the $x^1x^2$ plane is quantized and takes value in the integer as $Q = \int d^2\tilde x\, q \in \mathbb{Z}$.
Fig.~\ref{fig:axion_N=0_BG}(c) shows the topological charge density $q$ which is well-localized about the string center.

\paragraph{The model with $N=1$}

\begin{wrapfigure}{r}[0pt]{0.3\textwidth}
  \centering
 \includegraphics[width=0.3\textwidth]{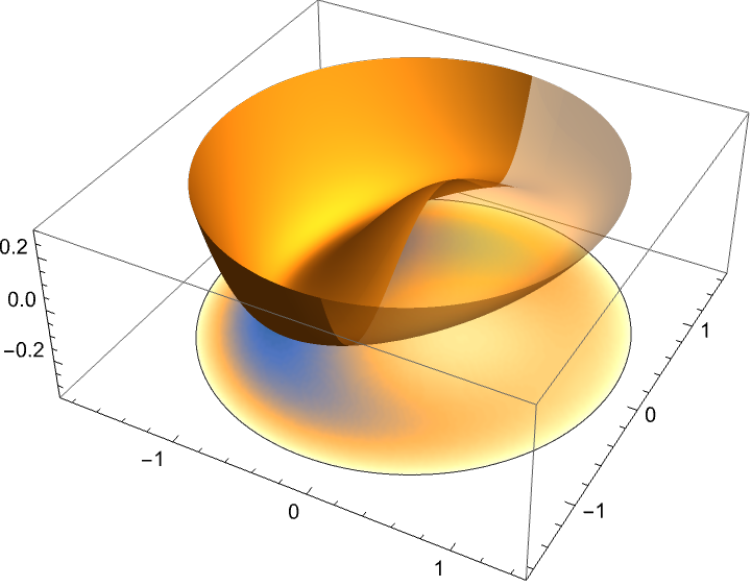}
 \caption{The dimensionless potential for \(N=1\) for $\alpha = 1/10$.}
 \label{fig:axion_N=1_pot}
\end{wrapfigure}
\begin{figure}[htbp]
\centering
\begin{minipage}[t]{0.32\linewidth}
\centering
\includegraphics[keepaspectratio,scale=0.4]{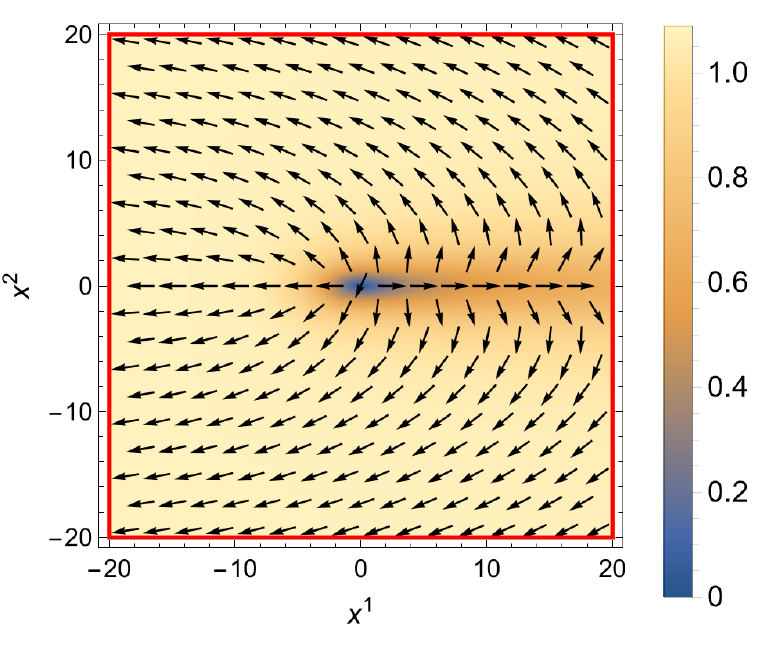}
\subcaption{}
\end{minipage}
\begin{minipage}[t]{0.32\linewidth}
\centering
\includegraphics[keepaspectratio,scale=0.4]{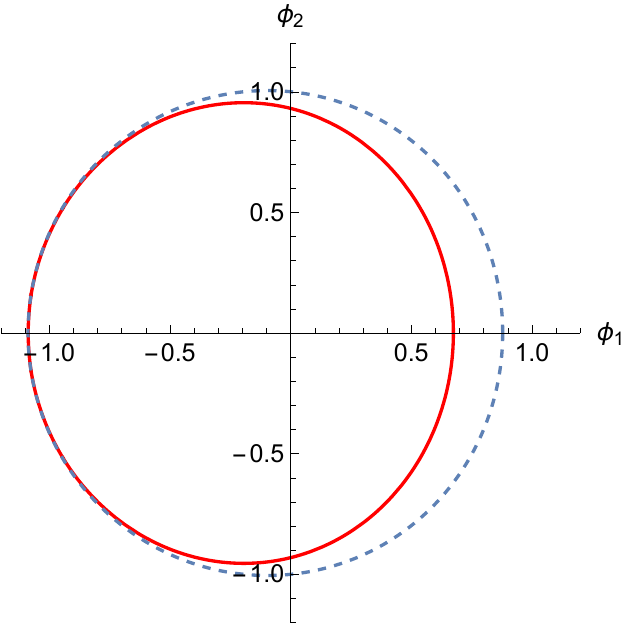}
\subcaption{}
\end{minipage}
\begin{minipage}[t]{0.32\linewidth}
\centering
\includegraphics[keepaspectratio,scale=0.4]{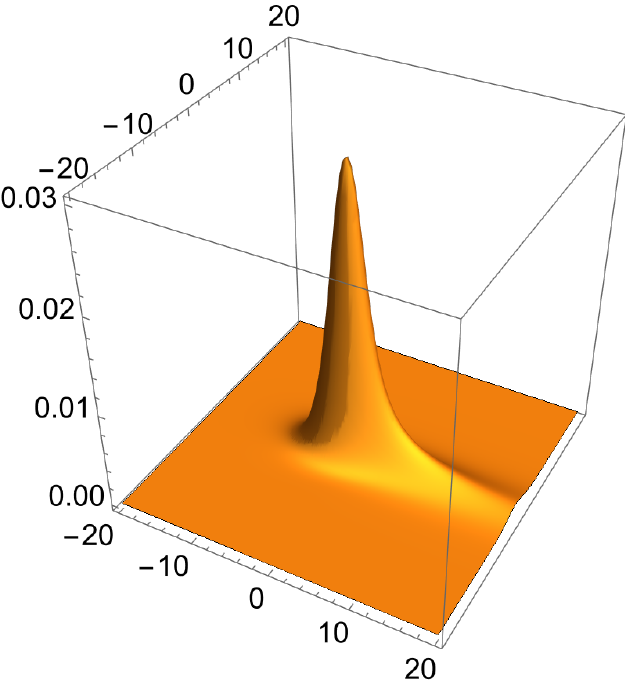}
\subcaption{}
\end{minipage}
\caption{The string-wall composite in the axion model with \(N=1\). $\alpha = 1/10$. The string is located at the origin and the domain wall extends along the positive side of the $x^1$ axis. (a) shows $\vec\phi$. (b) shows the image of the red square in (a) onto the $\phi_1\phi_2$ plane. The dashed curve shows the quasi vacuum manifold. (c) shows $q$ with outflow into the wall.}
\label{fig:axion_N=1_BG}
\end{figure}

The next tour site is $N=1$ where the string is attached by a single domain wall.
The second term in the potential is
\be
V \supset \alpha (\phi + \bar \phi) = 2 \alpha \phi_1,
\ee
so that the bottom of the wine bottle potential is linearly lifted along the direction of $\phi_1$ as
shown in Fig.~\ref{fig:axion_N=1_pot}.
The vacuum manifold consists of one point, and if $0<\alpha \ll 1$, the vacuum expectation value (VEV)
is approximately
\be
\phi_1 \simeq -1 - \alpha\,,\quad
\phi_2 = 0\,.
\ee
There are two hierarchical mass$^2$ scales $\lambda v^2$ and $\eta/v$.
The former is related to curvature of the potential along the radial direction of $\varphi$ while the second one is that along the NG boson $\zeta$.  Their ratio $\alpha = (\eta/v) /(\lambda v^2)$ is a unique dimensionless parameter of the model.
Therefore, the condition $\alpha \ll 1$ implies the potential along the NG boson $\zeta$ is only slightly lifted.  
Hence, despite the true vacuum is the one point, a slightly deformed $S^1$-like structure remains 
at the bottom of the scalar potential. We will call it the quasi vacuum manifold $S^1$.
More precisely, it can be defined as the set of $\phi$ satisfying the condition
\be
\frac{\delta V}{\delta |\phi|} = 0\,,\quad |\phi| = \sqrt{\phi_1^2+\phi_2^2}\,.
\ee
If the quasi $S^1$ is at the bottom of the deep enough potential, 
there could exist strings wrapping around the quasi vacuum manifold $S^1$. 
Strictly speaking, they are neither topological nor static because each string is inevitably attached by the single domain wall which pulls the string toward the attached domain wall. We show a numerical configuration of the single string attached by the single domain wall in Fig.~\ref{fig:axion_N=1_BG} for $\alpha = 1/10$.

\begin{figure}[hbtp]
\centering
\begin{minipage}[b]{0.45\linewidth}
\centering
\includegraphics[keepaspectratio,scale=0.4]{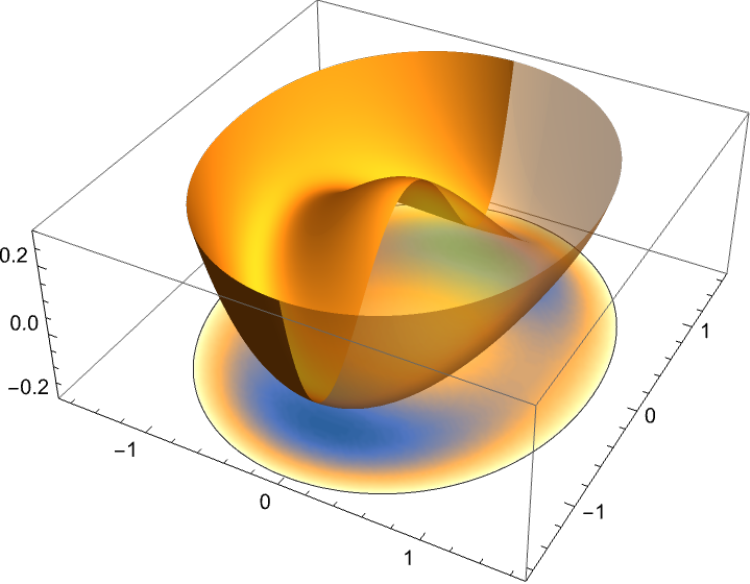}
\subcaption{$\alpha=1/20$}
\end{minipage}
\begin{minipage}[b]{0.45\linewidth}
\centering
\includegraphics[keepaspectratio,scale=0.4]{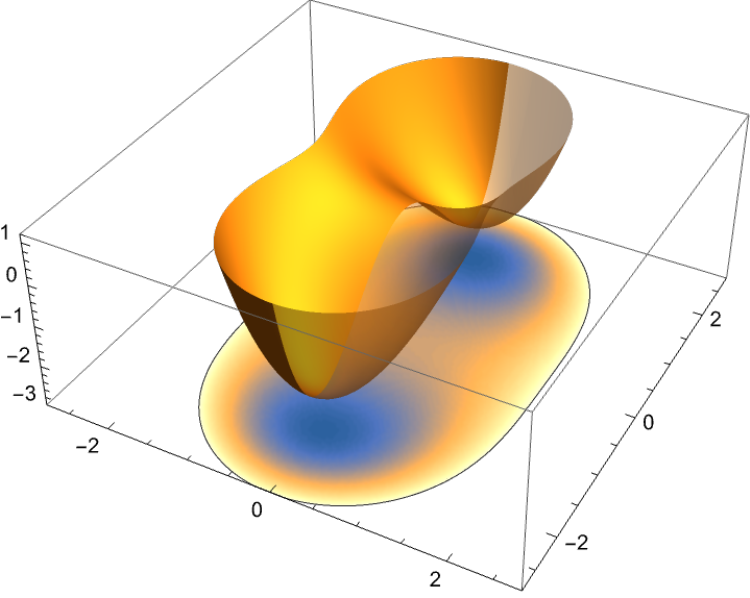}
\subcaption{$\alpha=1/2$}
\end{minipage}
\caption{The scalar potentials for \(N=2\)}
\label{fig:axion_N=2_pot}
\end{figure}

\paragraph{The model with $N= 2$}
The third site of our tour is the case of $N=2$. 
The potential can be rewritten as
\be
V = \frac{1}{4}\left(\phi_1^2+\phi_2^2 - (1+4\alpha)\right)^2 + 4\alpha \phi_1^2 - 2\alpha(1+2\alpha)\,,
\ee
and one can see that there are two discrete vacua $(\phi_1,\phi_2) = (0,\pm \sqrt{1+4\alpha})$.
The Lagrangian has \(\mathbb{Z}_2^{(1)}\times\mathbb{Z}_2^{(2)}\) symmetry
\begin{align}
\mathbb{Z}_2^{(1)}&: (\phi_1,\phi_2)\rightarrow(-\phi_1,\phi_2) \\
\mathbb{Z}_2^{(2)}&: (\phi_1,\phi_2)\rightarrow(\phi_1,-\phi_2).
\end{align}
Thus, only $\mathbb{Z}_2^{(2)}$ is spontaneously broken in the vacua.
As in the case of $N=1$ the quasi vacuum manifold $S^1$ remains for $\alpha \ll1$.
We can take $\alpha = 1/4$ as a threshold. 
The potential has two saddle points \((\phi_1,\phi_2)=(\pm\sqrt{1-4\alpha},0)\) for  \(\alpha<1/4\), while there is only one saddle point at \((\phi_1,\phi_2)=(0,0)\) for \(1/4\leq\alpha\). 
Hence the potential has the quasi $S^1$ structure for $\alpha < 1/4$ as shown in Fig.~\ref{fig:axion_N=2_pot} (a) whereas no $S^1$-like structure is found for $\alpha \ge 1/4$ as in Fig.~\ref{fig:axion_N=2_pot} (b).

The presence of the discrete and degenerate vacua together with SSB of $\mathbb{Z}_2^{(2)}$ immediately leads to topologically stable domain walls. There are two possibilities for $\alpha < 1/4$.
Suppose the domain wall is perpendicular to the $x$-axis. Then the coordinate $x$ can be thought of as the parameter for the curve $\phi(x)$ connecting two points $\phi = \pm i \sqrt{1+4\alpha}$ in the field space.
We should clarify which is energetically smaller: a straight path or a curved path along about the quasi vacuum manifold.
From the view point of the potential energy cost, it is obvious that the latter has smaller energy. However, we should also take a kinetic energy into account. Fortunately, the analytic solutions of domain walls have been known in this model \cite{Montonen:1976yk,Sarkar:1976vr}.
The true threshold can be analytically obtained: It is not $\alpha = 1/4$ but $\alpha = 1/12$. 
For $0 < \alpha < 1/12$ the domain wall solution is given by
\be
\phi_1 = \epsilon_1 \sqrt{1-12\alpha}\, {\rm sech}\, 2 \sqrt\alpha \tilde x\,,\quad
\phi_2 = \epsilon_2 \sqrt{1+4\alpha} \tanh 2 \sqrt\alpha \tilde x\,,
\label{eq:wall_n2}
\ee
with $\epsilon_{1,2} = \pm 1$.
While the choice of $\epsilon_2$ is related to the spontaneously broken $\mathbb{Z}_2^{(2)}$ in the vacua,
we can further chose $\epsilon_1 = \pm 1$ due to the fact that $\phi_1$ condensation inside the domain wall spontaneously breaks
$\mathbb{Z}_2^{(1)}$ as well \cite{Eto:2023gfn}. 
For $\alpha \ge 1/12$ the domain wall solution is given by
\be
\phi_1 = 0\,,\quad
\phi_2 = \epsilon_2 \sqrt{1+4\alpha} \tanh \frac{\sqrt{1+4\alpha}}{2} \tilde x\,.
\ee
Although this solves the equation of motion for any $\alpha$ but is not energetically lowest state for $\alpha < 1/12$.
Since $\phi_1 = 0$, the $\mathbb{Z}_2^{(1)}$ is unbroken everywhere in contrast to the domain wall for $\alpha < 1/12$.

The domain wall solution in the region $0 < \alpha < 1/12$ can be regarded to have a half winding number around the quasi $S^1$. To be concrete, let us fix $\epsilon_2 = +1$. Then the phase $\zeta = \arg \phi$ is $-\pi/2$ at $\tilde x=-\infty$ and $\pi/2$ at $\tilde x = + \infty$.  Hence, one sees that the winding number is $\pm 1/2$ for $\epsilon_1 = \pm1$.
\begin{figure}[htbp]
\centering
\begin{minipage}[t]{0.32\linewidth}
\centering
\includegraphics[keepaspectratio,scale=0.4]{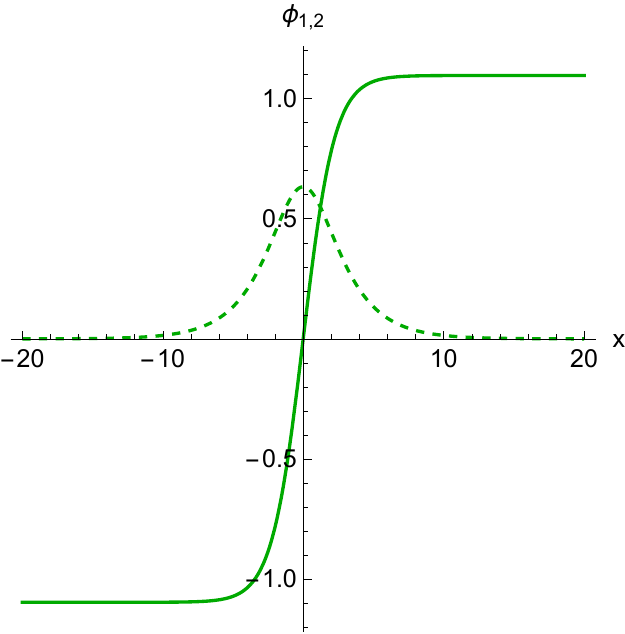}
\subcaption{}
\end{minipage}
\begin{minipage}[t]{0.32\linewidth}
\centering
\includegraphics[keepaspectratio,scale=0.4]{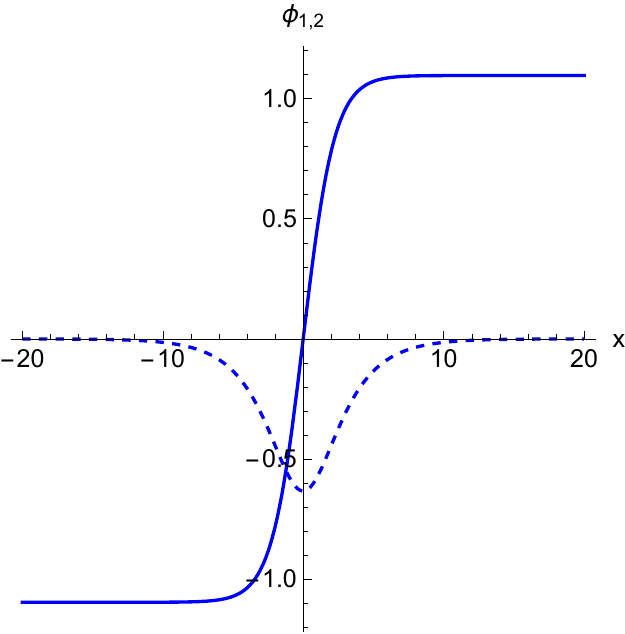}
\subcaption{}
\end{minipage}
\begin{minipage}[t]{0.32\linewidth}
\centering
\includegraphics[keepaspectratio,scale=0.4]{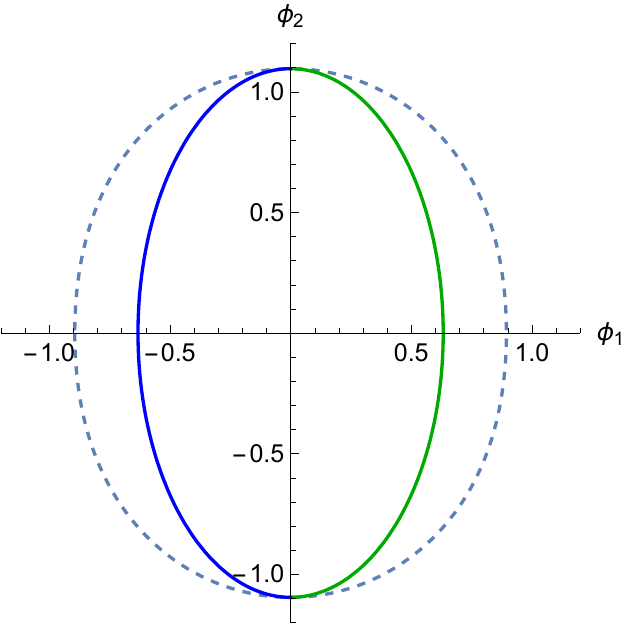}
\subcaption{}
\end{minipage}
\caption{The domain walls in the axion model with \(N=2\) and $\alpha = 1/20$. 
(a) shows $\phi_1$ (green-dashed) and $\phi_2$ (green-solid) for $\epsilon_1=1$, and (b) shows $\phi_1$ (blue-dashed) and $\phi_2$ (blue-solid) for $\epsilon_1=-1$. (c) The green (blue) half-oval corresponds to the solution in (a) [(b)]. The dashed-oval shows the quasi vacuum manifold.}
\label{fig:wall_N=2}
\end{figure}

\begin{figure}[h]
\centering
\begin{minipage}[t]{0.32\linewidth}
\centering
\includegraphics[keepaspectratio,scale=0.4]{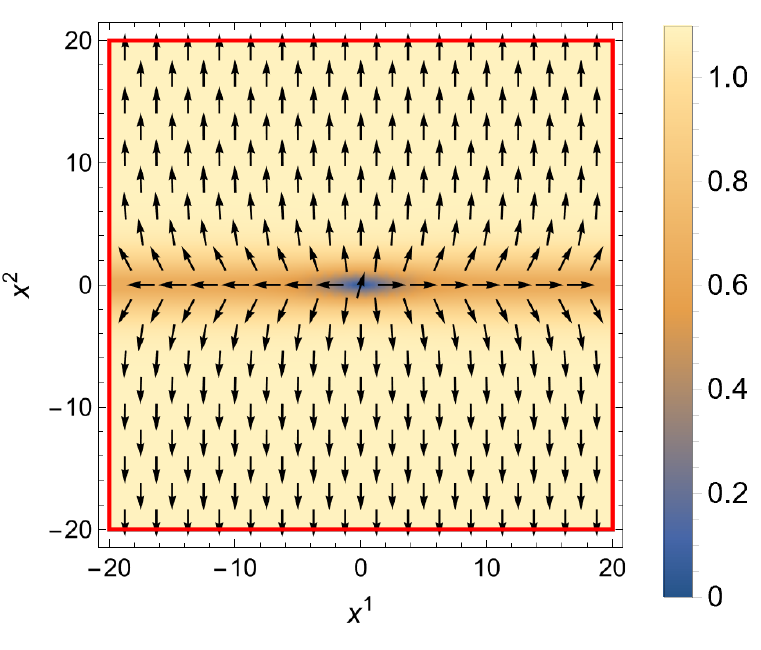}
\subcaption{}
\end{minipage}
\begin{minipage}[t]{0.32\linewidth}
\centering
\includegraphics[keepaspectratio,scale=0.4]{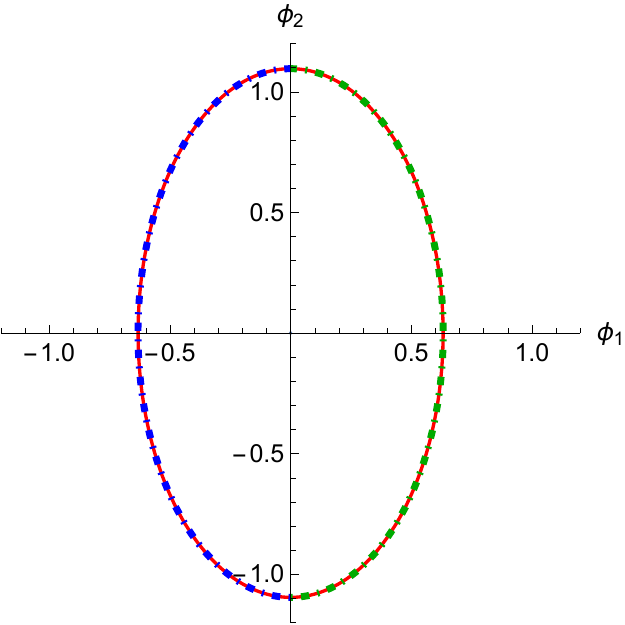}
\subcaption{}
\end{minipage}
\begin{minipage}[t]{0.32\linewidth}
\centering
\includegraphics[keepaspectratio,scale=0.4]{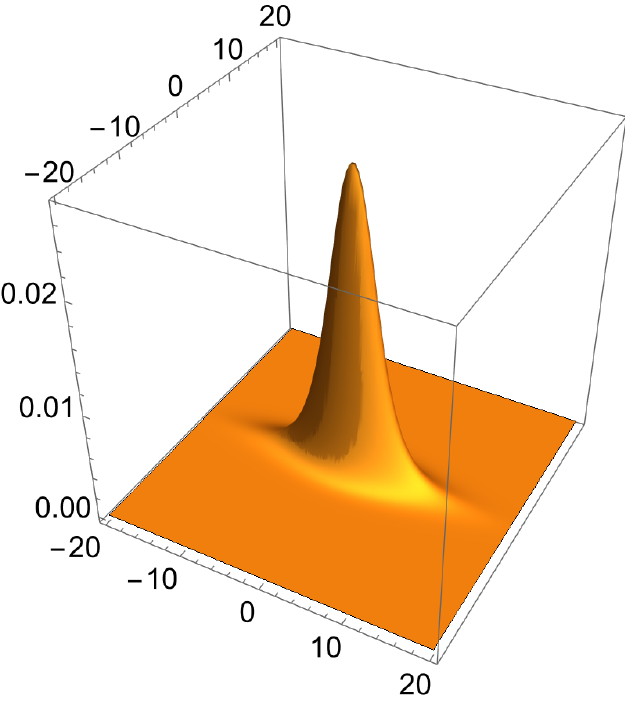}
\subcaption{}
\end{minipage}
\caption{
The numerical solution of the single string attached  by the two domain walls in the axion model with \(N=2\) and $\alpha = 1/20$. 
(a) shows the scalar field $\vec\phi$, and (b) shows the image of the red square in (a) on the $\phi_1\phi_2$ plane. The green and blue dashed-dotted half ovals correspond to the isolated domain walls given in the panel (c) of Fig.~\ref{fig:wall_N=2}. (c) The charge density $q$ is plotted.}
\label{fig:axion_N=2_a=0.1_BG}
\end{figure}

Moreover, we can combine these two domain walls of the winding number $1/2$ to make a configuration with the winding number $+1$. Namely, the resulting soliton is the axion string with the $N=2$ domain walls. The axion string can be regarded as the junction connecting two different domain walls \cite{Eto:2023gfn}.
A numerical solution for \(\alpha=1/20\) is shown in Fig.~\ref{fig:axion_N=2_a=0.1_BG}.
The string is located at $(x^1,x^2) = (0,0)$. The domain wall on the right hand has $(\epsilon_1,\epsilon_2) = (1,1)$,
and that on the left has $(1,-1)$. The composite soliton is not only static but also topologically stable due to the SSB of $\mathbb{Z}_2^{(2)}$.

\begin{figure}[h]
\centering
\begin{minipage}[b]{0.45\linewidth}
\centering
\includegraphics[keepaspectratio,scale=0.4]{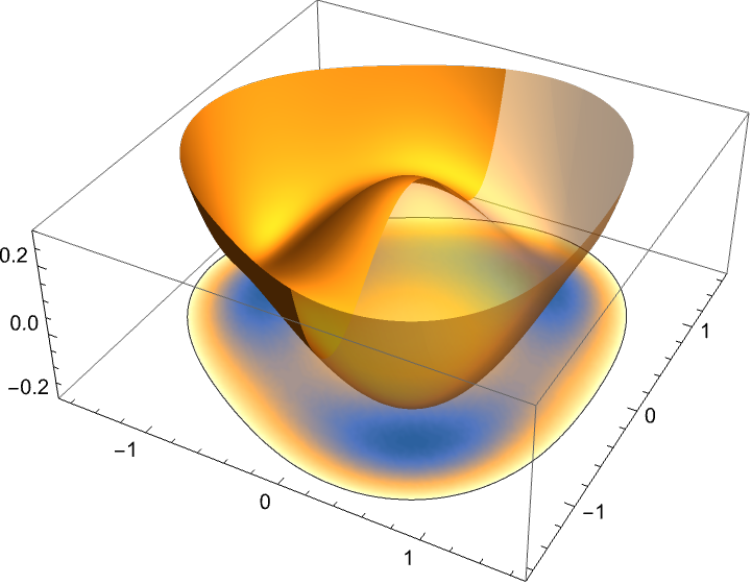}
\subcaption{$N=3,~\alpha=1/30$}
\end{minipage}
\begin{minipage}[b]{0.45\linewidth}
\centering
\includegraphics[keepaspectratio,scale=0.4]{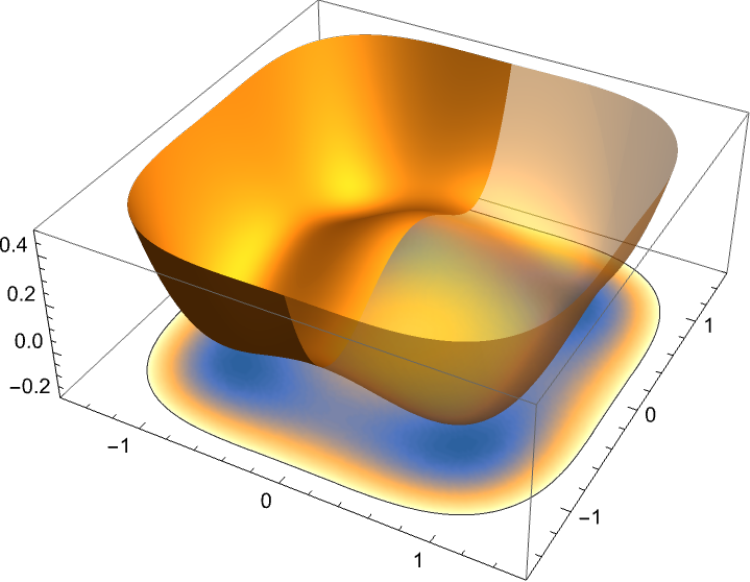}
\subcaption{$N=4,~\alpha=1/40$}
\end{minipage}
\caption{The scalar potentials for \(N=3,4\)}
\label{fig:axion_N=34_pot}
\end{figure}

\paragraph{The model with $N\ge 3$}
Finally, we explain the composites of strings and domain walls in the cases with $N=3$ and $4$ for the examples of higher $N>2$.
The number of discrete vacua is $N$ as shown in Fig.~\ref{fig:axion_N=34_pot}.

\begin{figure}[h]
\centering
\begin{minipage}[t]{0.32\linewidth}
\centering
\includegraphics[keepaspectratio,scale=0.4]{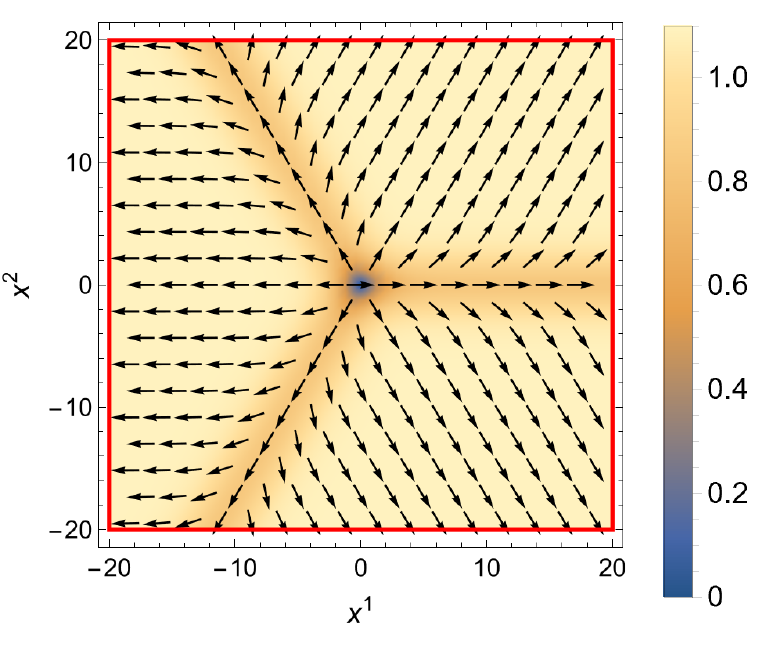}
\subcaption{}
\vspace*{.5cm}
\end{minipage}
\begin{minipage}[t]{0.32\linewidth}
\centering
\includegraphics[keepaspectratio,scale=0.4]{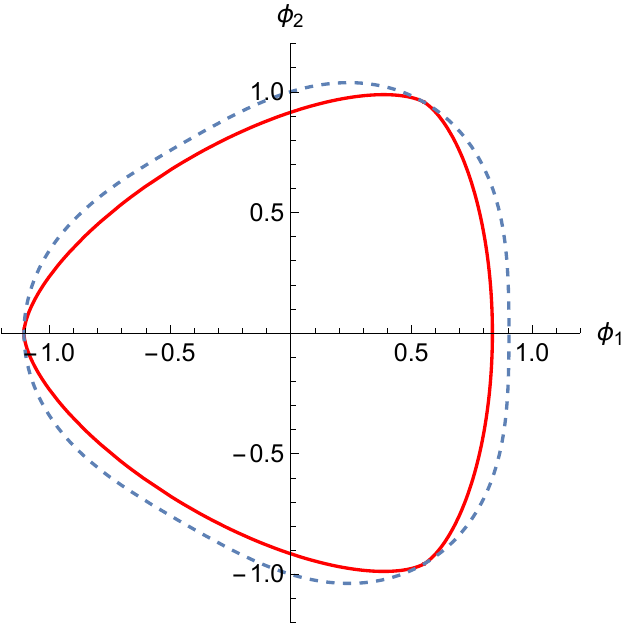}
\subcaption{}
\end{minipage}
\begin{minipage}[t]{0.32\linewidth}
\centering
\includegraphics[keepaspectratio,scale=0.4]{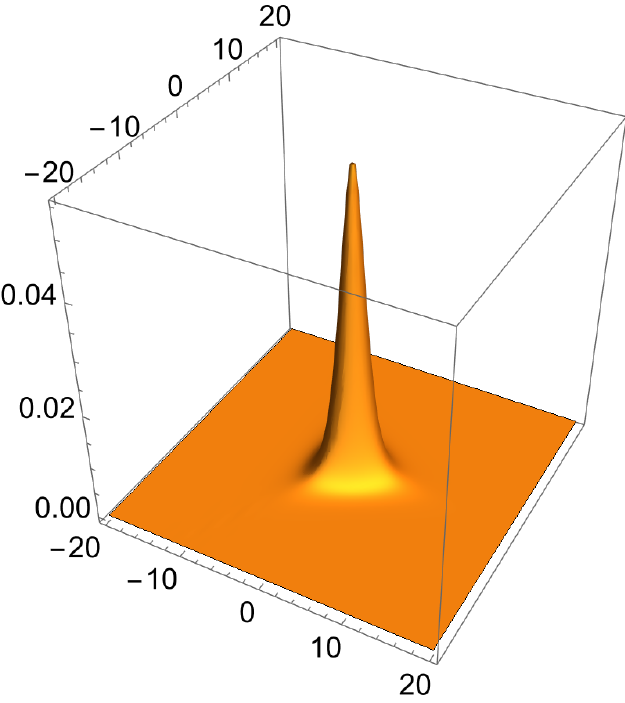}
\subcaption{}
\end{minipage}
\begin{minipage}[t]{0.32\linewidth}
\centering
\includegraphics[keepaspectratio,scale=0.4]{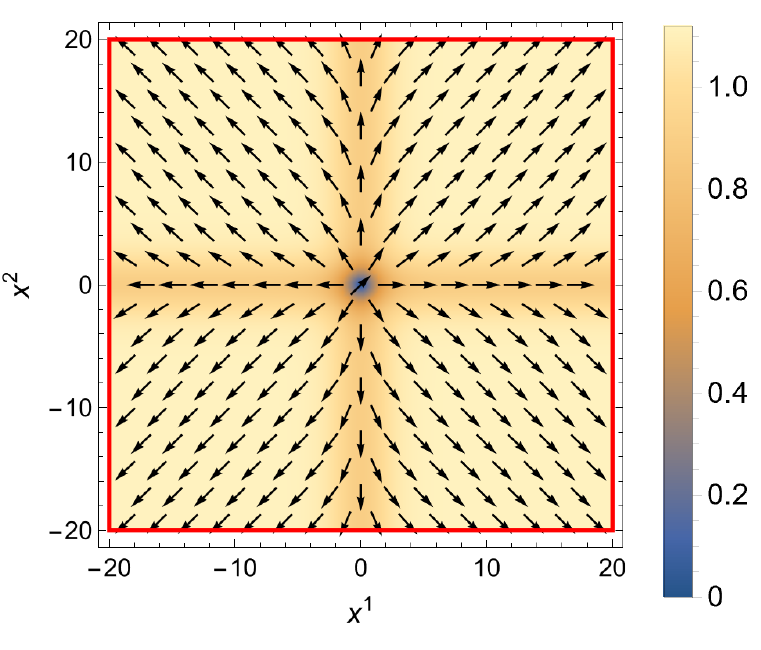}
\subcaption{}
\end{minipage}
\begin{minipage}[t]{0.32\linewidth}
\centering
\includegraphics[keepaspectratio,scale=0.4]{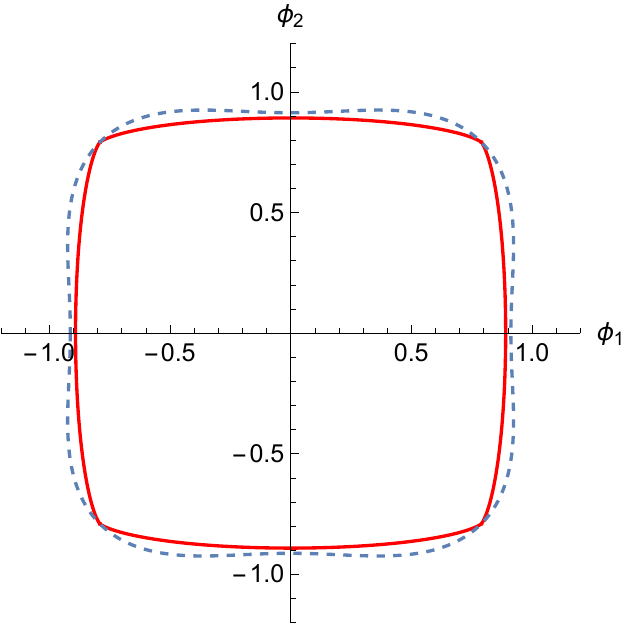}
\subcaption{}
\end{minipage}
\begin{minipage}[t]{0.32\linewidth}
\centering
\includegraphics[keepaspectratio,scale=0.4]{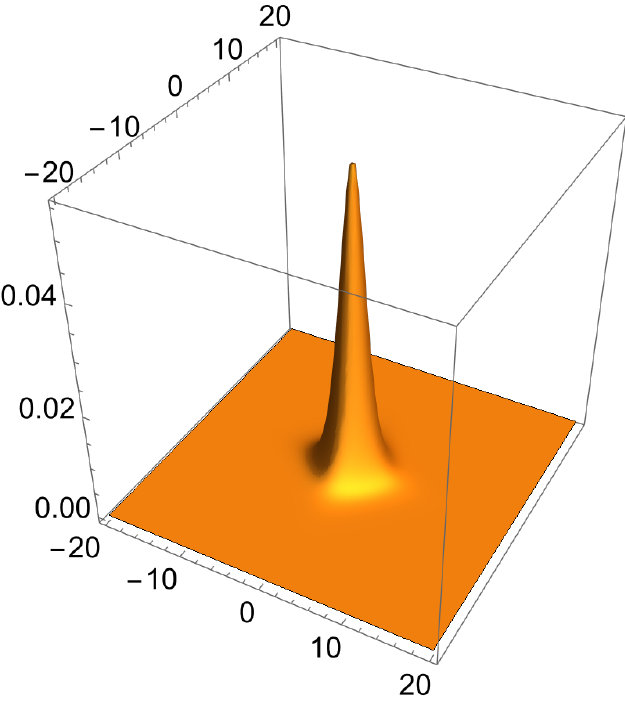}
\subcaption{}
\end{minipage}
\caption{The top (bottom) row shows the numerical solution of the $N=3$ and $\alpha=1/30$ ($N=4$ and $\alpha =1/40$).
What is shown is the same as that in Figs.~\ref{fig:axion_N=1_BG} and \ref{fig:axion_N=2_a=0.1_BG}.}
\label{fig:axion_N=3and4}
\end{figure}

An elementary domain wall is the one connecting adjacent vacua. 
Hence, the number of different domain walls is also $N$. Each of them has $1/N$ winding number. 
The axion string with the unit winding number appears at the junction point of the $N$ domain walls.
Note that, since  the $N(\ge 3)$ vacua are located at the $N$ vertices of an $N$-gon, the stable string-wall composite always exists
independent of the value of $\alpha$. This is contrast to the case of $N=2$ that the string appears only when $\alpha$ is sufficiently small $(\alpha < 1/12)$. We show the numerical solutions for $N=3$ and $4$ in Fig.~\ref{fig:axion_N=3and4}.

\subsection{BPS domain wall junctions in SUSY Abelian-Higgs models}
\label{sec:susy}

In this section we will review the BPS domain wall junctions in ${\cal N}=2$ supersymmetric Abelian-Higgs models.
Although there seems to be no relations between the axion models and the SUSY gauge theories, they are actually similar to each other in the composite solitons as we will clarify below. The SUSY gauge theories have the following advantages.
First, the ${\cal N}=2$ supersymmetric Abelian-Higgs models are one of the ideal theoretical platforms for studying the composite solitons. In addition to the axion string-wall composite that is a string with $N$ domain walls attached radially, more complicated composite solitons can be easily constructed as the BPS states. This is advantageous in clarifying the question of where massless fermions are localized. Second, and this is more important, we can solve the Dirac equation analytically and find zero mode functions of the fermions analytically. This is contrast to the axion models where we can only find numerical solutions for the fermion zero modes. These will be explained in Sec.~\ref{sec:DWSF}.

\subsubsection{BPS equations and the moduli matrix formalism}
We here consider the \(3+1\) dimensional \(\mathcal{N}=2\) supersymmetric Abelian-Higgs model.
We are interested in topological solitons, so we concentrate on the bosonic Lagrangian which is given by
\begin{align}
\mathcal{L}&=-\frac{1}{4g^2}F_{\mu\nu}F^{\mu\nu}+\frac{1}{2g^2}\sum_{\alpha=1}^{2}\partial_\mu \varphi_\alpha \partial^\mu\varphi_\alpha+|D_\mu H|^2-V, 
\label{eq:susy_Lag}\\
V&=\frac{g^2}{2}(|H|^2 - v^2)^2+\sum_{\alpha=1}^{2}|HM_\alpha-\varphi_\alpha H|^2\,.
\label{eq:susy_pot}
\end{align}
The bosonic fields in the vector multiplet is the $U(1)$ gauge field $A_\mu$ and the real scalar fields $\varphi_\alpha$ ($\alpha=1,2$).
$H^A$ ($A=1,2,\cdots,N_{\rm F}$) is the $N_{\rm F}$ complex scalar fields which we deal with as the $N_{\rm F}$ component row vector.
$H$ belongs to the hypermulitplet and form the $SU(2)_{\rm R}$ doublet together with the other complex scalar field $\tilde H^A$.
Since $\tilde H^A$ do not play any role for the solitons, we will omit $\tilde H$ throughout this paper.
The $U(1)$ gauge coupling constant is $g$, the Fayet-Iliopoulos $D$ parameter is $v^2$, and $M_\alpha$ is the mass matrices for the hypermultiplet. We consider the diagonal mass matrices \(M_1=\mathrm{diag}(m_1,m_2,\cdots,m_{N_\mathrm{F}})\) and \(M_2=\mathrm{diag}(n_1,n_2,\cdots,n_{N_\mathrm{F}})\).
For later convenience, we introduce the $N_{\rm F}$ mass vectors defined by
\be
\vec m_A = \left(
\begin{array}{c}
m_A\\
n_A
\end{array}
\right)\,.
\ee

We will consider generic cases where the mass vectors are not fully degenerate, namely $\vec m_A \neq \vec m_B$ if $A\neq B$.
Then, there are $N_{\rm F}$ discrete and degenerate SUSY vacua satisfying $V=0$. The $A$-th vacuum is given by
\be
H^B = v \delta^B_A\,,\quad
\vec \varphi = \vec m_A\,,
\ee
where we defined $\vec \varphi = \left(\begin{array}{c}\varphi_1\\ \varphi_2\end{array}\right)$.
Hence the vacua correspond to the $N_{\rm F}$ points ($\vec\varphi=\vec m_A$) in the two dimensional $\vec \varphi$ plane.

The parallel BPS domain walls in this theory have been extensively studied in \cite{Lambert:1999ix,Tong:2002hi,Shifman:2002jm,Shifman:2003uh,Lee:2002gv,Isozumi:2003rp,Isozumi:2003uh,Isozumi:2004jc,Isozumi:2004va,Eto:2004vy}. The BPS domain wall junctions have been also studied in \cite{Kakimoto:2003zu,Eto:2005cp,Eto:2005fm}.
In the following we consider static configurations which are independent on \(x^3\) (\(\partial_0=\partial_3=0\)) and set \(A_0=A_3=0\).
The energy density can be cast into the well-known Bogomol'nyi completion as \cite{Kakimoto:2003zu,Eto:2005cp,Eto:2005fm}
\begin{align}
\begin{aligned}
\mathcal{E}=&~\frac{1}{2g^2}B_3^2+\frac{1}{2g^2}(\partial_1\varphi_2-\partial_2\varphi_1)^2+\frac{1}{2g^2}\left\{\vec\partial \cdot \vec\varphi-g^2\left(v^2-|H^A|^2\right)\right\}^2 \\
&+ \sum_A|(\vec D - \vec m_A + \vec \varphi)H^A|^2+ v^2 \mathcal{Z}_1+ v^2\mathcal{Z}_2+\frac{1}{g^2} \mathcal{Y}+\vec\partial \cdot \vec{\mathcal{J}}\nonumber \\
\ge& ~
v^2 \mathcal{Z}_1+ v^2 \mathcal{Z}_2 - \frac{1}{g^2} \mathcal{Y}+\vec\partial \cdot \vec{\mathcal{J}}
 \label{eq:2.3}
\end{aligned}
\end{align}
where the topological charge densities are defined by
\begin{gather}
\mathcal{Y}\equiv \varepsilon^{\alpha\beta} \partial_\alpha\varphi_1\partial_\beta\varphi_2,\quad \mathcal{Z}_\alpha\equiv \partial_\alpha\varphi_\alpha\,.
\end{gather}
Here summation is not taken over $\alpha$ for ${\cal Z}_\alpha$.
Note that ${\cal Y}$ is essentially the same quantity as $q$ defined in Eq.~(\ref{eq:topological_charge}).
Hence, the domain wall junction is essentially same as the global string, though this fact does not seem to be well recognized in the literature.
We also introduced \(\vec{\mathcal{J}} = \sum_A(\vec m_A- \vec\varphi)|H^A|^2\) which vanishes in the vacua.

The energy bound is saturated when the BPS equations are satisfied
\be
B_3 =0\,,\quad
\partial_1\varphi_2-\partial_2\varphi_1=0\,, \quad
\vec D H^A = (\vec m_A - \vec \varphi)H^A\,,
\label{eq:BPS1}
\\
\vec \partial \cdot \vec \varphi = g^2\left(v^2-\sum_{A=1}^{N_F}|H^A|^2\right)\,. \label{eq:BPS2}
\ee
The three equations in Eq.~(\ref{eq:BPS1}) can be solved by the moduli matrix method \cite{Isozumi:2004jc,Isozumi:2004va,Isozumi:2004vg,Eto:2005cp,Eto:2005fm,Eto:2006pg}
\be
H^A= v S^{-1}H_0^A\mathrm{e}^{M_\alpha^A x^\alpha},\quad A_\alpha-i\varphi_\alpha=-iS^{-1}\partial_\alpha S \label{eq:2.9}
\ee
where $\{H_0^A\}$ are  a set of $N_{\rm F}$  arbitrary complex constants (except for $H_0^A = 0$ for all $A$), and $S$ is a complex scalar field. However, there is redundancy that $H^A$, $A_\alpha$, and $\varphi_\alpha$ are unchanged under multiplication
$(S, H_0^A) \sim c (S,H_0^A)$ with arbitrary non-zero complex constant $c$.
Thus, $H_0$ takes the value from $\mathbb{C}^{N_{\rm F}}/\mathbb{C}^* \simeq \mathbb{C}P^{N_{\rm F}-1}$.
Plugging Eq.~(\ref{eq:2.9}) into Eq.~(\ref{eq:BPS2}), we find the master equation which determines $S$ for a given $H_0$
\be
\frac{1}{2g^2v^2}\partial^2 \log \Omega
=1-\Omega_0\Omega^{-1}\,,\quad \Omega = |S|^2\,,
\label{eq:master}
\ee
with
\be
\Omega_0 =  \sum^{N_\mathrm{F}}_{A=1}|H_0^A|^2e^{2\vec m_A \cdot \vec x}.
\label{eq:Omega0}
\ee
This should be solved with the boundary condition $\Omega \to \Omega_0$ at $|\vec x| \to \infty$.
We can fix $S$ to be real by using the $U(1)$ gauge symmetry, so that $A_\alpha = 0$ and $\Omega = S^2$.
Then the real scalar field $\varphi_\alpha$ is given by
\begin{gather}
\vec\varphi= \frac{1}{2} \vec \partial \log \Omega\,.
\label{eq:phi_omega}
\end{gather}

\subsubsection{Several examples}

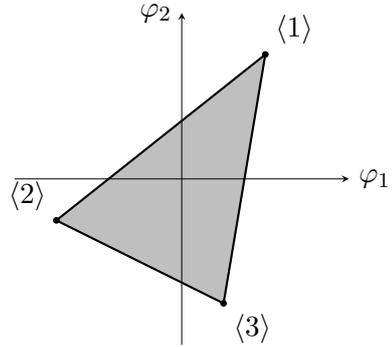
\begin{wrapfigure}{r}[0pt]{0.4\textwidth}
  \centering
 \begin{tikzpicture}[scale=0.55]
\coordinate(A)at(2,3);
\coordinate(B)at(-3,-1);
\coordinate(C)at(1,-3);
\filldraw[fill=lightgray,thick](A)--(B)--(C)--cycle;
\fill[black](A)circle(0.08)node[above right]{\(\braket{1}\)};
\fill[black](B)circle(0.08)node[above left]{\(\braket{2}\)};
\fill[black](C)circle(0.08)node[below right]{\(\braket{3}\)};
\draw[->,>=stealth,very thin](-4,0)--(4,0)node[right]{\(\varphi_1\)};
\draw[->,>=stealth,very thin](0,-4)--(0,4)node[left]{\(\varphi_2\)};
\end{tikzpicture}
 \caption{The three vacua (mass tri-gon) shown in the $\vec\varphi$ plane.}
 \label{fig:vac_NF=3}
\end{wrapfigure}
The BPS domain wall junctions (string-wall composites) are controlled by the mass vectors $\{\vec m_A\}$ and
the moduli matrix $H_0$. Instead of giving detail explanations \cite{Eto:2005cp,Eto:2005fm}, we give only several examples here.

Let us consider the case with $N_{\rm F} = 3$ as the simplest example. One can randomly chose the three mass vectors $\{\vec m_1,\,\vec m_2,\,\vec m_3\}$. We assume they are linearly independent. There are three vacua which correspond to the points $\vec \varphi = \vec m_A$ $(A=1,2,3)$ on the two dimensional $\vec \varphi$ plane as shown in Fig.~\ref{fig:vac_NF=3}.

The $\left<A\right>$-th vacuum  can be expressed by the moduli matrix: 
\be
H_0^B = \delta^B_A\,.
\ee
Plugging this into Eq.~(\ref{eq:Omega0}), we have $\Omega_0 = e^{2\vec m_A\cdot \vec x}$, and the master equation (\ref{eq:master}) is solved by $\Omega = \Omega_0$. Now, one can correctly reproduce 
$\vec\varphi = \vec m_A$ from Eq.~(\ref{eq:phi_omega}).

\begin{wrapfigure}{R}[0pt]{0.4\textwidth}
\centering
 \includegraphics[width=0.35\textwidth]{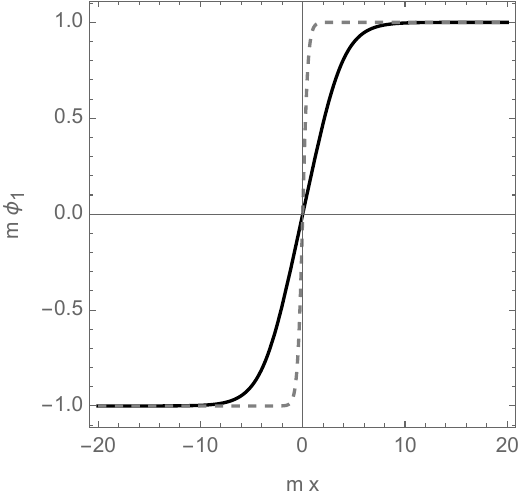}
 \caption{The domain wall configuration in $\varphi_1$. The solid curve is the numerical solution for $m/gv = 2$,
 and the dashed one is the approximation with $\Omega = \Omega_0$.}
 \label{fig:susy_dw}
\end{wrapfigure}
The single domain wall connecting the two vacua, say $\left<1\right>$ and $\left<2\right>$, can be obtained by
the following moduli matrix
\be
H_0 =  (1,\,1,\,0)\,.
\ee
Similarly, domain walls connecting the other pair of vacua are generated by $H_0 =  (1,\,0,\,1)$ or $H_0 =  (0,\,1,\,1)$.
To be concrete, let us take $\vec m_1 = (m,0)$ and $\vec m_2 = (0,-m)$. Then the source $\Omega_0$ for the domain wall connecting $\left<1\right>$ and $\left<2\right>$ is given by
\be
\Omega_0 = e^{2mx}+e^{-2mx}\,.
\ee

In general, the master equation (\ref{eq:master}) cannot analytically be solved except for several special cases \cite{Isozumi:2003rp,Blaschke:2016lyj}. However, the solution $\Omega = \Omega_0$ at $gv\to\infty$ limit \cite{Gauntlett:2000ib} captures almost all the important properties. Hence, $\Omega_0$ can be used as an instantaneous checking tool for figuring out the shape of solitons roughly. We compare the numerically obtained solution for $m/gv = 2$ and the configuration obtained by the approximation $\Omega = \Omega_0$ in Fig.~\ref{fig:susy_dw}.

\begin{figure}[h]
\centering
\begin{minipage}[t]{0.24\linewidth}
\centering
\includegraphics[keepaspectratio,scale=0.3]{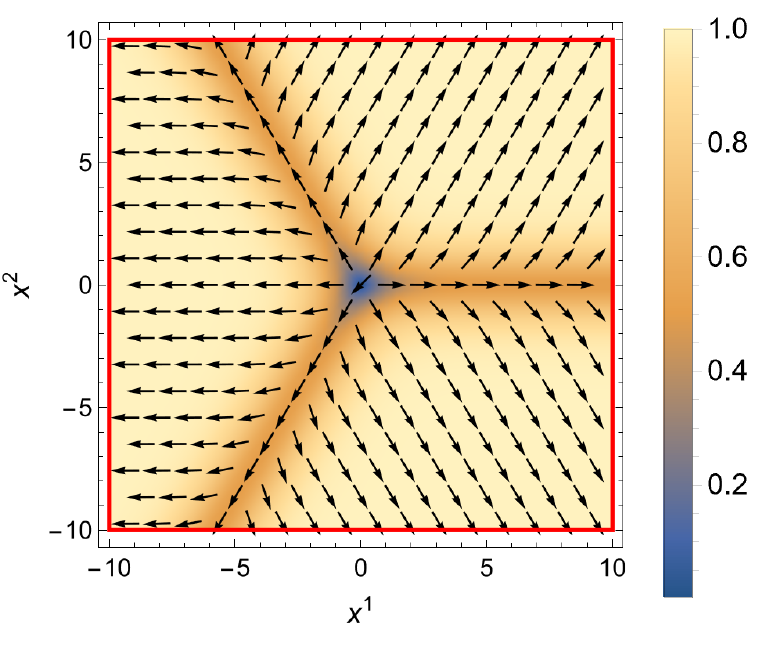}
\subcaption{$x^1x^2$ plane}
\end{minipage}
\begin{minipage}[t]{0.24\linewidth}
\centering
\includegraphics[keepaspectratio,scale=0.3]{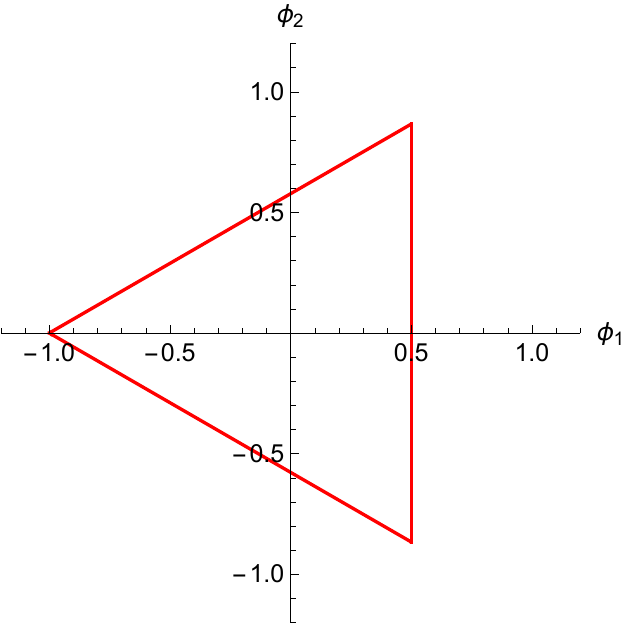}
\subcaption{$\varphi^1\varphi^2$ plane}
\end{minipage}
\begin{minipage}[t]{0.24\linewidth}
\centering
\includegraphics[keepaspectratio,scale=0.3]{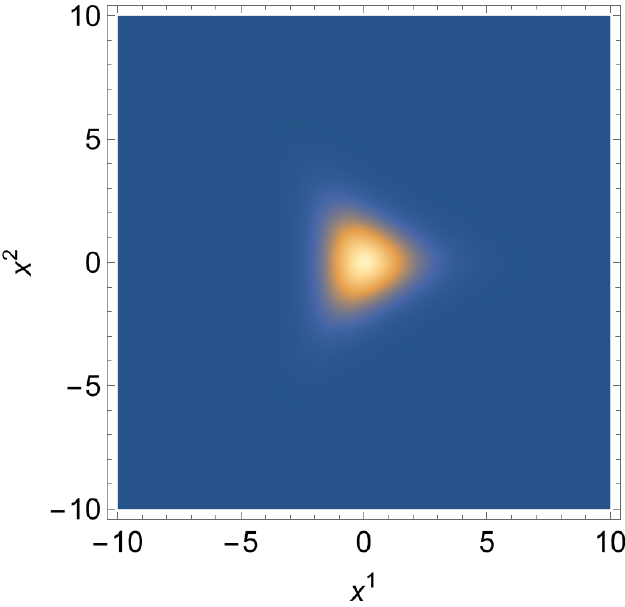}
\subcaption{${\cal Y}$}
\end{minipage}
\begin{minipage}[t]{0.24\linewidth}
\centering
\includegraphics[keepaspectratio,scale=0.3]{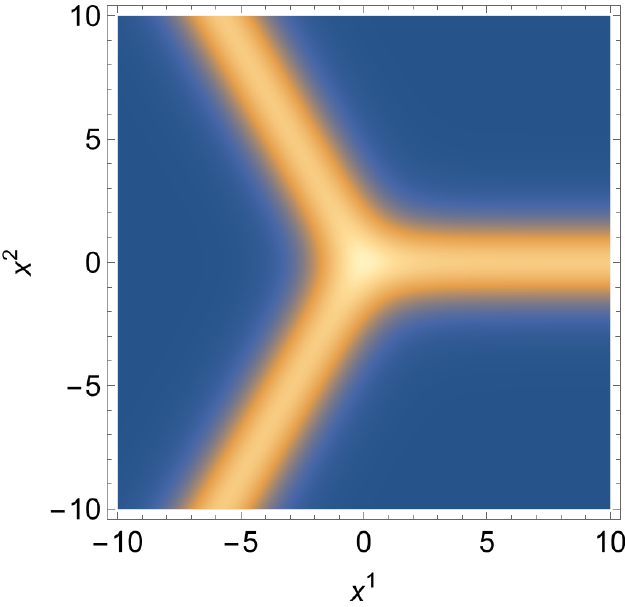}
\subcaption{${\cal Z}$}
\end{minipage}
\caption{The $S_3$ symmetric BPS string-wall composite in the model with $N_{\rm F}=3$ with the mass vectors  $m (-1,0)$, $m(-\cos\frac{2\pi}{3},\sin\frac{2\pi}{3})$, and $m(-\cos\frac{2\pi}{3},-\sin\frac{2\pi}{3})$. (a) shows $\varphi(x^1,x^2)$, and (b) shows $\varphi$ on the spacial boundary [the red square in (a)]. (c) and (d) show the topological charge densities ${\cal Y}$ and ${\cal Z}_1 + {\cal Z}_2$, respectively. We set $m/gv = 1$.}
\label{fig:susy_dwj_n3_1}
\end{figure}

\begin{figure}[hbtp]
\centering
\begin{minipage}[t]{0.24\linewidth}
\centering
\includegraphics[keepaspectratio,scale=0.3]{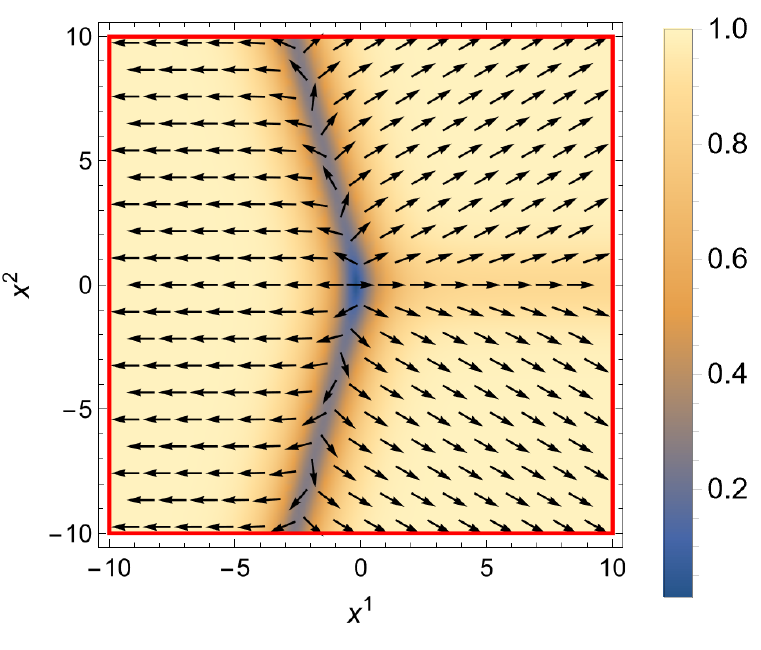}
\subcaption{}
\end{minipage}
\begin{minipage}[t]{0.24\linewidth}
\centering
\includegraphics[keepaspectratio,scale=0.3]{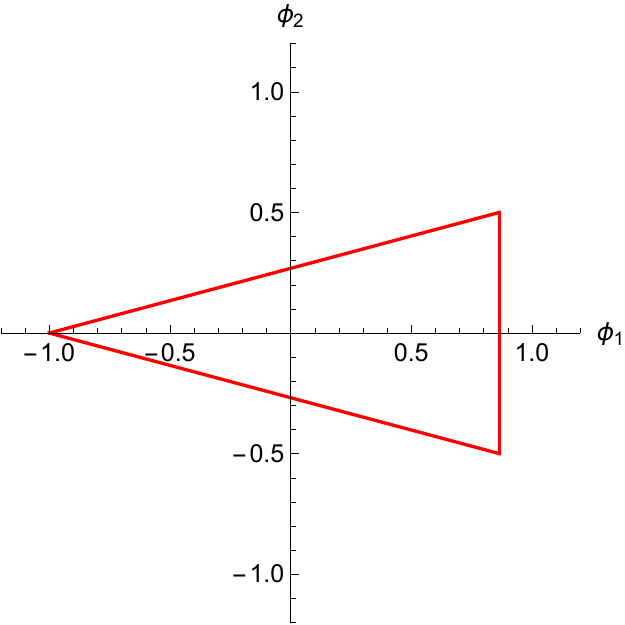}
\subcaption{}
\end{minipage}
\begin{minipage}[t]{0.24\linewidth}
\centering
\includegraphics[keepaspectratio,scale=0.3]{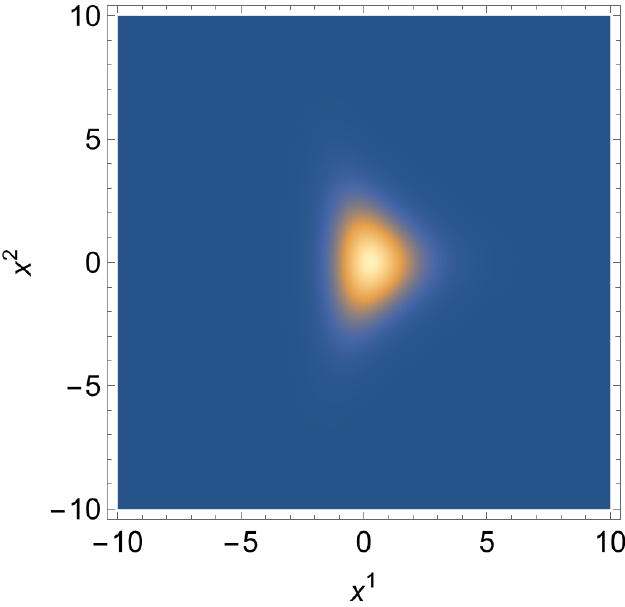}
\subcaption{}
\end{minipage}
\begin{minipage}[t]{0.24\linewidth}
\centering
\includegraphics[keepaspectratio,scale=0.3]{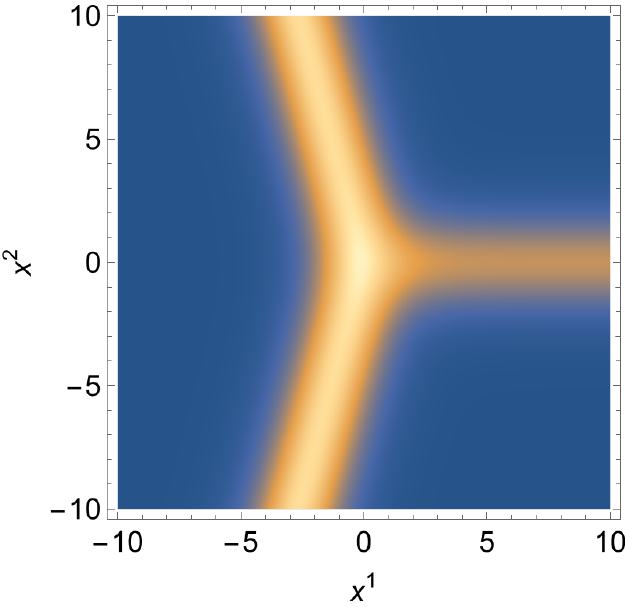}
\subcaption{}
\end{minipage}
\caption{The $S_3$ asymmetric BPS string-wall composite in the model with $N_{\rm F}=3$ with the mass vectors  $m (-1,0)$, $m(-\cos\frac{5\pi}{6},\sin\frac{5\pi}{6})$, and $m(-\cos\frac{5\pi}{6},-\sin\frac{5\pi}{6})$. We set $m/gv = 1$.}
\label{fig:susy_dwj_n3_2}
\end{figure}

The domain wall junction composed of all the three vacua can also easily be generated by the moduli matrix
\be
H_0 = (1,1,1).
\ee
The master equation can be solved numerically (The analytic solution was obtained for a special case \cite{Kakimoto:2003zu}).  
The first solution given in Fig.~\ref{fig:susy_dwj_n3_1} is $S_3$ symmetric by arranging the mass vectors on an equilateral triangle as $\vec m = m (-1,0)$, $m(-\cos\frac{2\pi}{3},\sin\frac{2\pi}{3})$, and $m(-\cos\frac{2\pi}{3},-\sin\frac{2\pi}{3})$. 
This supersymmemtric composite soliton should be compared with those in the axion model with $N=3$ given in Fig.~\ref{fig:axion_N=3and4}.
Indeed, they are very similar. However, the supersymmetric model admits more generic solutions: The three domain walls have to have the common tension in the axion model whereas the three walls can have different tensions in the SUSY model. As an example, we show an asymmetric solution with $\vec m = m (-1,0)$, $m(-\cos\frac{5\pi}{6},\sin\frac{5\pi}{6})$, and $m(-\cos\frac{5\pi}{6},-\sin\frac{5\pi}{6})$ in Fig.~\ref{fig:susy_dwj_n3_2}.

All the important informations about the string-wall composite can readily be obtained from the mass polygon \cite{Eto:2005cp,Eto:2005fm} whose $N_{\rm F}$ vertices are the mass vector $\vec m_A$ drawn in the $\varphi_1\varphi_2$ plane as Fig.~\ref{fig:vac_NF=3}.
Each vertex corresponds to the discrete vacuum, each edge to the domain wall, and each triangle to the string (three-pronged domain wall junction). This correspondence works at a more qualitative level: length of the edge is tension of the domain wall, and area of the triangle is tension of the string.
Furthermore, the shape of string-wall composite in the $x^1x^2$ plane is dual to the polygon in the $\varphi^1\varphi^2$ plane.
These properties can be clearly seen in Figs.~\ref{fig:susy_dwj_n3_1} and \ref{fig:susy_dwj_n3_2}.
Indeed, the red square at the boundary of the $x^1x^2$ plane in Fig.~\ref{fig:susy_dwj_n3_1}(a) is mapped onto the triangle connecting three vacua in the $\varphi^1\varphi^2$ plane in Fig.~\ref{fig:susy_dwj_n3_1}(b). When we gradually shrink the red square toward the origin, its image is almost unchanged until the red square gets close to the string at the origin. Hence, when we sweep the whole $x^1x^2$ plane once, the whole polygon in the $\varphi^1\varphi^2$ plane is swept once.

\begin{figure}[h]
\centering
\begin{minipage}[t]{0.24\linewidth}
\centering
\includegraphics[keepaspectratio,scale=0.3]{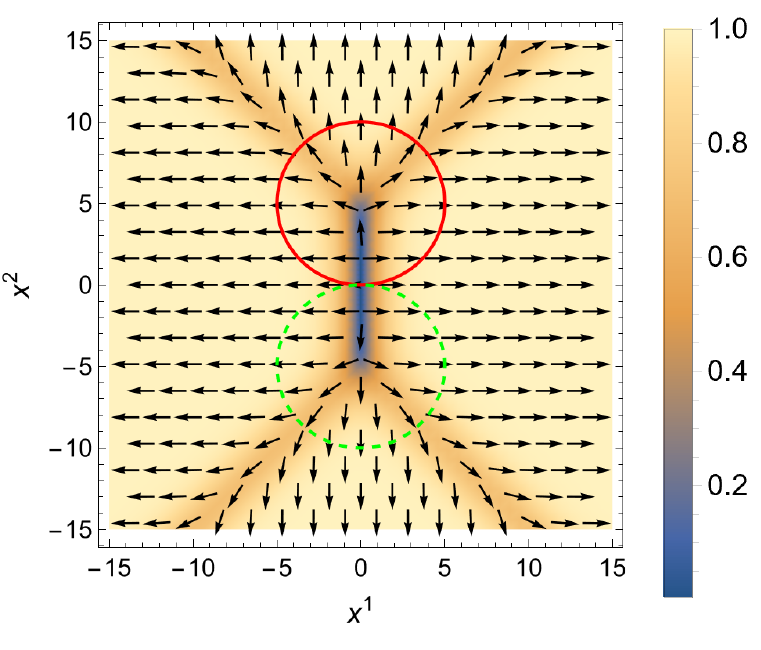}
\subcaption{$\vec\varphi$}
\end{minipage}
\begin{minipage}[t]{0.24\linewidth}
\centering
\includegraphics[keepaspectratio,scale=0.3]{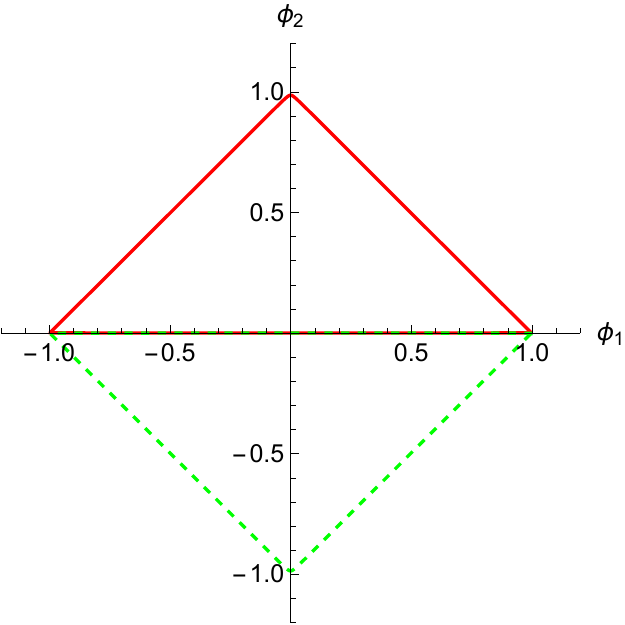}
\subcaption{$\vec\varphi$}
\end{minipage}
\begin{minipage}[t]{0.24\linewidth}
\centering
\includegraphics[keepaspectratio,scale=0.3]{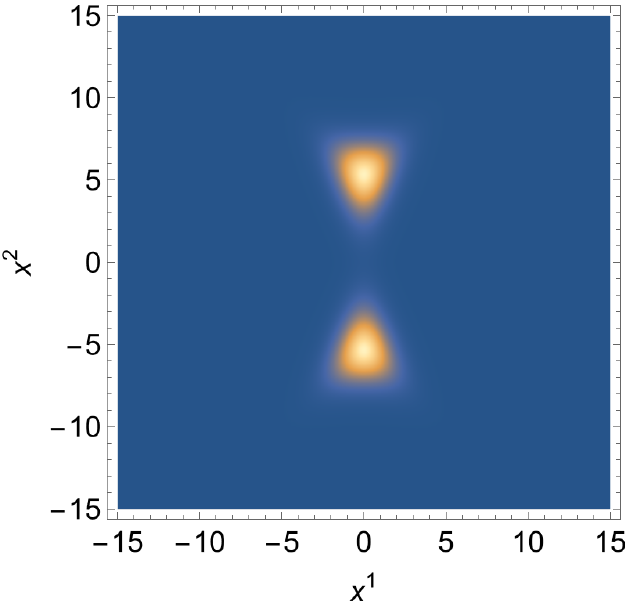}
\subcaption{${\cal Y}$}
\end{minipage}
\begin{minipage}[t]{0.24\linewidth}
\centering
\includegraphics[keepaspectratio,scale=0.3]{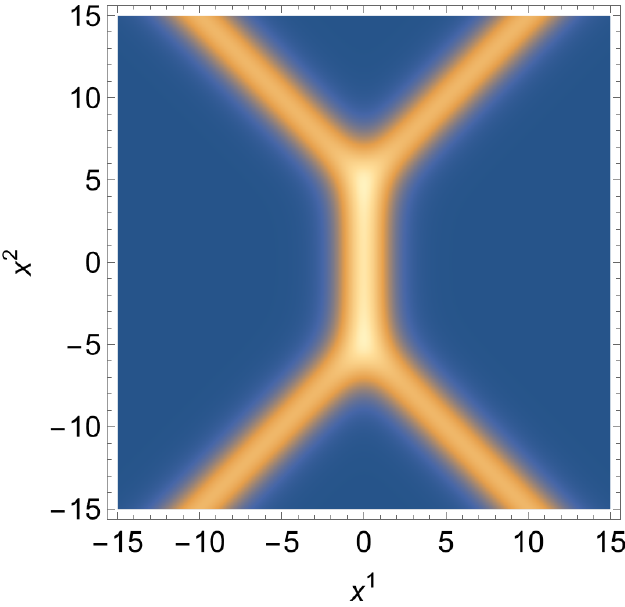}
\subcaption{${\cal Z}_1 + {\cal Z}_2$}
\vspace*{.5cm}
\end{minipage}
\begin{minipage}[t]{0.24\linewidth}
\centering
\includegraphics[keepaspectratio,scale=0.3]{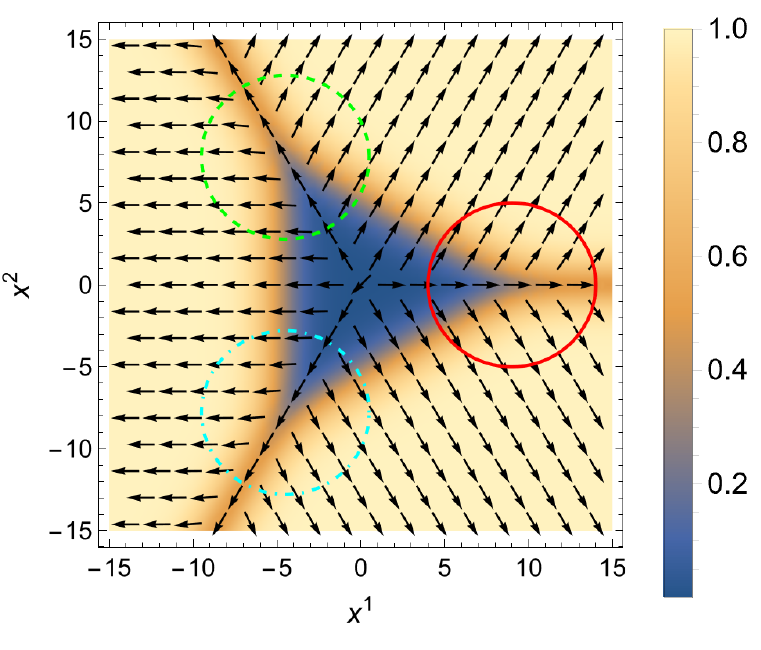}
\subcaption{$\vec\varphi$}
\end{minipage}
\begin{minipage}[t]{0.24\linewidth}
\centering
\includegraphics[keepaspectratio,scale=0.3]{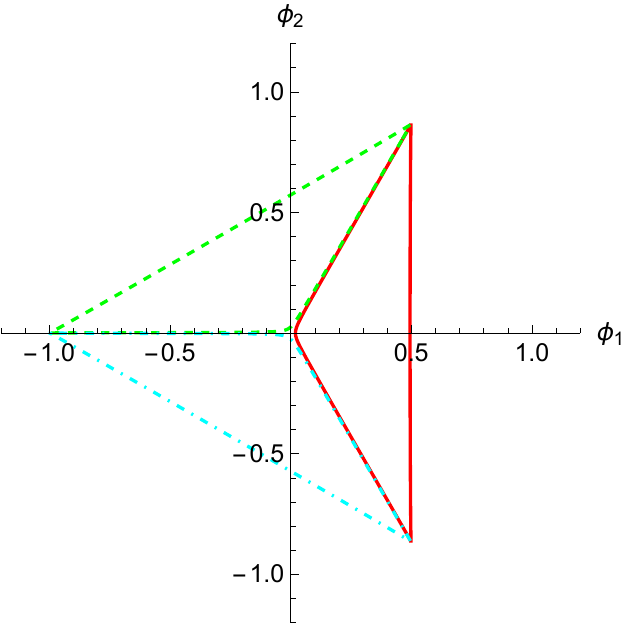}
\subcaption{$\vec\varphi$}
\end{minipage}
\begin{minipage}[t]{0.24\linewidth}
\centering
\includegraphics[keepaspectratio,scale=0.3]{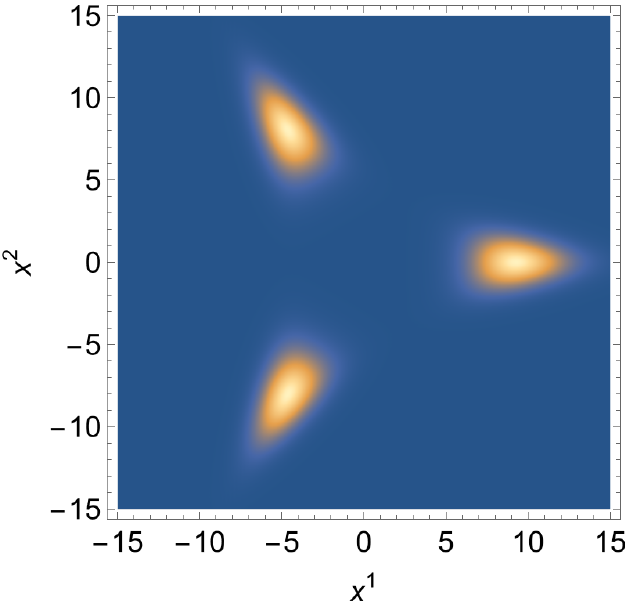}
\subcaption{${\cal Y}$}
\end{minipage}
\begin{minipage}[t]{0.24\linewidth}
\centering
\includegraphics[keepaspectratio,scale=0.3]{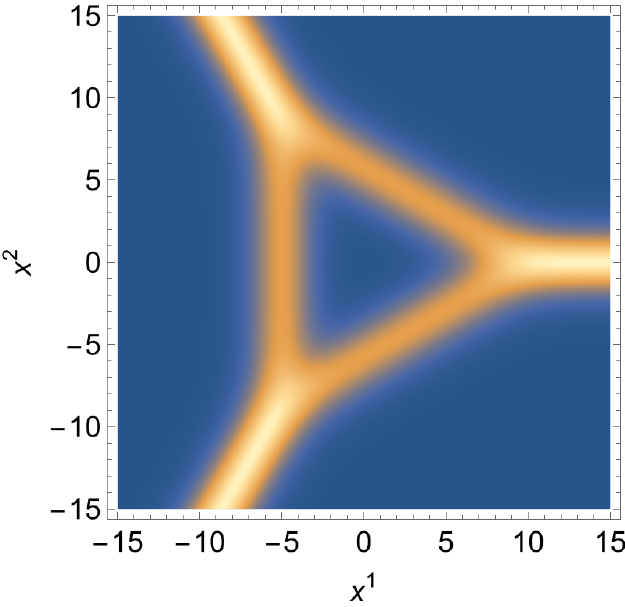}
\subcaption{${\cal Z}_1 + {\cal Z}_2$}
\end{minipage}
\caption{The BPS composite solitons of the domain walls and strings in the model of $N_{\rm F}=4$. We set $m/gv = 1$.
The panels on the top row are for the mass vectors arranged on the vertices of a square, and those on the bottom row are for those
placed on the vertices of an equilateral triangle in addition to the origin.}
\label{fig:susy_dwj_Nf4}
\end{figure}

As we increase $N_{\rm F}$, the number of the vacua increases. Accordingly, the solitons show complicated network structures.
The smallest networks appear in $N_{\rm F}=4$. We show two different solutions ($m/gv = 1$) in Fig.~\ref{fig:susy_dwj_Nf4}. One is obtained by arranging  the mass vectors on the four vertices of a square as $\vec m_A = (m,0)$, $(0,m)$, $(-m,0)$, and $(0,-m)$. The moduli matrix is taken to be $H_0 = (e^{5},1,e^{5},1)$. The resulting configuration has four semi-infinite domain walls and a finite domain wall with two strings at the junctions of domain walls as shown in Fig.~\ref{fig:susy_dwj_Nf4} (a--d).
The other solution is obtained by arranging  the mass vectors on the three vertices of an equilateral triangle as before, and also at the origin $\vec m_4 = (0,0)$. The moduli matrix is taken to be $H_0 = (1,1,1,e^5)$. As shown in Fig.~\ref{fig:susy_dwj_Nf4} (e--h), the network consists of the three semi-infinite domain walls and three finite domain walls with three strings at the junction points. 

\subsubsection{A constant shift of the mass vectors}
\label{sec:shift}

We here would like to make a remark about a seemingly redundant degree of freedom, an overall shift in the mass vectors. Consider two sets of the mass vectors, one is $\{\vec m_A\}$ and the other is $\{\vec m_A + \vec m_0\}$ for $A=1,2,\cdots,N_F$. Two models with different combinations of the model parameters $\{\vec m_A\}$ and $\{\vec m_A + \vec m_0\}$  are different models. However, the overall shift $\vec m_0$ can be absorbed by redefining the field as $\vec \varphi \to \vec \varphi - \vec m_0$ in the Lagrangian (\ref{eq:susy_Lag}). Therefore, the string-wall composites in the two different theories are indistinguishable by physical observables such as energy density and topological charge densities. Indeed, the overall constant shift has been ignored in the previous works \cite{Eto:2005cp,Eto:2005fm,Eto:2020vjm,Eto:2020cys}.
Nevertheless, we will see in the next sections that in the presence of fermions coupled with $\vec\varphi$, the overall mass shift cannot be ignored.

\subsubsection{The strong gauge coupling limit}
\label{sec:strong}

The strong gauge coupling limit $g^2 \to \infty$ where the Lagrangian (\ref{eq:susy_Lag}) corresponds to the massive $\mathbb{C}P^{N_{\rm F}-1}$ nonlinear sigma model is very special because the master equation (\ref{eq:master}) reduces to an algebraic equation.
It is analytically solvable for all the moduli matrix $H_0$. The analytic solution is given by
\be
\Omega = \Omega_0 = \sum_{A=1}^{N_{\rm F}} |H_0^A|^2 e^{2\vec m_A \cdot \vec x}\,,\quad
\varphi_a = \frac{1}{2}\p_a \log \Omega_0\,.
\label{eq:Omega_infty}
\ee
This is quite powerful since we have the analytic solution no matter how complicated the string-wall composite is.

\section{String-wall fermions}
\label{sec:DWSF}

It has been shown in Ref.~\cite{Jackiw:1975fn} that massless fermions are localized on domain walls, and also 
in Ref.~\cite{Jackiw:1981ee} that they are localized in vortex strings. In this section, 
we will answer to the elementary question of whether a massless fermion is localized when strings and walls coexist to form composite solitons, and if so, where it is localized.

\subsection{The Dirac equations}

Hereafter, we deal with $\varphi$ as a background field of the string-wall composites parallel to the $z$ axis. $\varphi$ can be the field of either the axion models in Sec.~\ref{sec:axion_striing_wall} or the bosonic part of the supersymmetric Abelian-Higgs models in Sec.~\ref{sec:susy}.
We consider the simplest fermionic Lagrangian
\be
\mathcal{L}_\mathrm{F} = i \bar \chi \bar \sigma^\mu D_\mu \chi + i \xi \sigma^\mu D_\mu \bar \xi - h \varphi \chi \xi - h \varphi^*  \bar\chi \bar\xi\,,
\label{eq:L_fermion}
\ee
where $\xi$ and $\chi$ are two component Weyl spinors, and  \(h\) is a Yukawa coupling constant which we can set real and positive without loss of generality.\footnote{\label{foot:susy}
Note that ${\cal L}_{\rm F}$ is not the supersymmetric partner of the bosonic Lagrangian (\ref{eq:susy_Lag}) of the ${\cal N}=2$ supersymmetric Abelian-Higgs model. So we abandon SUSY at all in this paper and indeed we do not need SUSY. What we need is only the string-wall composites in the boson field $\varphi$, and the bosonic Lagrangian  (\ref{eq:susy_Lag}) is suitable for that purpose. 
The fermion zero mode for the domain wall junction in the genuine ${\cal N}=2$ supersymmetric Abelian-Higgs model which has more fermions from the $N_F$ hypermultiplets and the vector multiplet was studied in \cite{Kakimoto:2003zu}. The massless fermion in the SUSY model \cite{Kakimoto:2003zu} is quite different from those we obtain in this paper. The difference comes from SUSY. In the SUSY model the massless fermions are NG fermions associated with spontaneously broken supersymmetry, and at the same time they are supersymmetric partner of the translational NG bosons, so that the massless fermions are localized over the all domain walls. In contrast, the massless fermions are nothing  to do with the translational NG bosons and they appear as genuine topological states in our models.
}
Here, we included the electromagnetic $U(1)_{\rm EM}$ symmetry with $D_\mu \chi = (\p_\mu + i e A_\mu)\chi$ and $D_\mu \bar\xi = (\p_\mu + i e A_\mu)\bar\xi$.
The four-dimensional Pauli matrices are given by $\sigma^\mu = ({\bf 1}_2,\vec \sigma)$, and $\bar\sigma^\mu = ({\bf 1}_2,-\vec \sigma)$.
In the standard four component notation
\be
\psi = \left(\begin{array}{c}
\chi\\
\bar\xi
\end{array}
\right)\,,\quad
\gamma^\mu = \begin{pmatrix}
0 & \sigma^\mu \\
\bar{\sigma}^\mu & 0
\end{pmatrix}\,,\quad
\gamma_5 = \begin{pmatrix}
1 & 0 \\
0 & -1
\end{pmatrix}\,,
\ee
the above fermionic Lagrangian can be expressed as
\be
{\cal L}_{\rm F} = i \bar\psi \gamma^\mu D_\mu \psi - h \bar\psi\left(\varphi_1 + i \gamma_5 \varphi_2\right)\psi\,,
\ee
with $\varphi = \varphi_1 + i \varphi_2$ and $D_\mu \psi = (\p_\mu + i e A_\mu)\psi$.
The fermionic Lagrangian is invariant under \(U(1)_{\rm A} \times U(1)_{\rm EM}\) 
\begin{align}
U(1)_\mathrm{A}:&\quad\psi\rightarrow\mathrm{e}^{-i\frac{\zeta}{2}\gamma_5}\psi\quad\varphi\rightarrow \mathrm{e}^{i\zeta}\varphi\,,
\label{eq:U(1)A}\\
U(1)_{\mathrm{EM}}:&\quad \psi \rightarrow\mathrm{e}^{i\beta(x)} \psi,\quad\varphi\rightarrow\mathrm \varphi\,.
\label{eq:U(1)EM}
\end{align}
When this Lagrangian is coupled with the axion models, $U(1)_{\rm A}$ is identical to $U(1)_{\rm PQ}$.
The Dirac equation reads
\begin{gather}
i\gamma^\mu
\partial_\mu
\begin{pmatrix}
\chi \\
\bar{\xi}
\end{pmatrix}
= h
\begin{pmatrix}
\varphi(x,y) & 0 \\
0 & \varphi^*(x,y)
\end{pmatrix}
\begin{pmatrix}
\chi \\
\bar{\xi}
\end{pmatrix}\,.
\label{eq:3.6}
\end{gather}
Hereafter, we assume the electromagnetic field $A_\mu = 0$.

Dealing with $\varphi(x,y)$ as the background field, we are going to find out the mass spectra of the fermions.
To this end, let us assume the following $x^\mu$ dependence
\begin{gather}
\begin{pmatrix}
\chi(x^\mu) \\
\bar{\xi}(x^\mu)
\end{pmatrix}
=\mathrm{e}^{-i\omega t+ikz}
\begin{pmatrix}
\chi_{\omega,k}(x,y) \\
\bar{\xi}_{\omega,k}(x,y)
\end{pmatrix}\,.
\end{gather}
Plugging this into (\ref{eq:3.6}), we have the eigenvalue equation
\be
H_k
\begin{pmatrix}
\chi_{\omega,k}(x,y) \\
\bar{\xi}_{\omega,k}(x,y)
\end{pmatrix}
= \omega
\begin{pmatrix}
\chi_{\omega,k}(x,y) \\
\bar{\xi}_{\omega,k}(x,y)
\end{pmatrix}\,,
\label{eq:eigeneq1}
\ee
where we have defined
\begin{gather}
H_k =  kZ+M\,,
\end{gather}
and
\begin{gather}
Z=
\begin{pmatrix}
-\sigma^3 & 0\\
0 & \sigma^3
\end{pmatrix}
,\quad M=
\begin{pmatrix}
-i\bar{\sigma}^a\partial_a & h\varphi^* \\
h\varphi & -i\sigma^a \partial_a
\end{pmatrix}\,.
\label{eq:Z,M}
\end{gather}
$M$ and $Z$ are Hermite, so is \(H_k\).
Note that \(Z\) and \(M\) anti-commute as $\{Z,M\}=0$. 
The square of \(H_k\) reads
\begin{gather}
H_k^2=k^2Z^2+k\{Z,M\}+M^2=k^2+M^2\,.
\label{eq:Hk2}
\end{gather}
Thus, $H_k^2$ and $M$ can be simultaneously diagonalized.
Let $m_n$ be a real eigenvalue of the Hermite operator $M$ as
\begin{gather}
M
\begin{pmatrix}
\chi_{\omega,k}^{(n)} \\
\bar{\xi}_{\omega,k}^{(n)}
\end{pmatrix}
=m_n
\begin{pmatrix}
\chi_{\omega,k}^{(n)} \\
\bar{\xi}_{\omega,k}^{(n)}
\end{pmatrix}
\,.
\label{eq:M2}
\end{gather}
Then, Eq.~(\ref{eq:Hk2}) gives a dispersion relation
\begin{gather}
\omega_n^2=k^2+m_n^2\,.
\label{eq:dispersion}
\end{gather}
Obviously, the eigenvalue of $M^2$ is positive semidefinite, so the eigenvalues satisfy $m_n^2 \ge 0$.

Let us write down the explicit form of the operator $M$
\begin{gather}
M=
\begin{pmatrix}
0 & i(\partial_1-i\partial_2) & h\varphi^* & 0 \\
i(\partial_1+i\partial_2) & 0 & 0 & h\varphi^* \\
h\varphi & 0 & 0 & -i(\partial_1-i\partial_2) \\
0 & h\varphi & -i(\partial_1+i\partial_2) & 0 \\
\end{pmatrix}
\,.
\end{gather}
Therefore, Eq.~(\ref{eq:M2}) can be decomposed into two independent equations as
\begin{gather}
\begin{pmatrix}
\chi_{\omega,k}^{(n)} \\
\bar{\xi}_{\omega,k}^{(n)}
\end{pmatrix}=
\begin{pmatrix}
0 \\
f^{(n)} \\
\tilde f^{(n)} \\
0
\end{pmatrix}
\,,~
\begin{pmatrix}
g^{(n)} \\
0 \\
0 \\
\tilde g^{(n)}
\end{pmatrix}
\,,
\label{eq:two_zeros}
\end{gather}
where the components $\{f^{(n)},\tilde f^{(n)},g^{(n)},\tilde g^{(n)}\}$ are complex valued functions of $x$ and $y$.
Note that the eigenstates of $M$ given in Eq.~(\ref{eq:two_zeros}) are simultaneously eigenstates of $Z$ given in Eq.~(\ref{eq:Z,M}) with the eigenvalues are $+1$ and $-1$.
Action of the mass square operator $M^2$  on the eigenstates can be compactly expressed as
\be
Q^\dag Q 
\left(\begin{array}{c}
f^{(n)}\\
\tilde f^{(n)}
\end{array}
\right)
= m_n^2 
\left(\begin{array}{c}
f^{(n)}\\
\tilde f^{(n)}
\end{array}
\right)\,,\quad
Q Q^\dag 
\left(\begin{array}{c}
g^{(n)}\\
\tilde g^{(n)}
\end{array}
\right)
= m_n^2 
\left(\begin{array}{c}
g^{(n)}\\
\tilde g^{(n)}
\end{array}
\right)\,,
\label{eq:QQ}
\ee
where we have defined
\be
Q = \left(
\begin{array}{cc}
i\left(\p_1 - i \p_2\right) & h\varphi^*\\
h\varphi & - i\left(\p_1 + i \p_2\right)
\end{array}
\right)\,,\quad
Q^\dag = \left(
\begin{array}{cc}
i\left(\p_1 + i \p_2\right) & h\varphi^*\\
h\varphi & - i\left(\p_1 - i \p_2\right)
\end{array}
\right)\,.
\ee
The explicit forms are given by
\be
Q^\dag Q &=& \left(
\begin{array}{cc}
-\p_a^2 + h^2|\varphi|^2 & i h \left(\p_1 + i \p_2\right)\varphi^* \\
- ih \left(\p_1 - i \p_2\right)\varphi & -\p_a^2 + h^2|\varphi|^2
\end{array}
\right)\,,\\
Q Q^\dag &=& \left(
\begin{array}{cc}
-\p_a^2 + h^2|\varphi|^2 & i h \left(\p_1 - i \p_2\right)\varphi^* \\
- ih \left(\p_1 + i \p_2\right)\varphi & -\p_a^2 + h^2|\varphi|^2
\end{array}
\right)\,.
\ee

As is well-known in  supersymmetric quantum mechanics, the eigenvalues of $Q^\dag Q$ and $QQ^\dag$ coincide except for the zero mode $m_0^2 = 0$. Thus, it is sufficient for us to study mainly the mass spectrum of $Q^\dag Q$ for $h > 0$ (or $QQ^\dag$ for $h<0$).
Making the following ansatz
\be
\left(
\begin{array}{c}
f^{(n)}\\
\tilde f^{(n)}
\end{array}
\right)
=
\left(
\begin{array}{c}
f^{(n)}\\
if^{(n)*}
\end{array}
\right)\,,
\label{eq:af+-}
\ee
the first equation in Eq.~(\ref{eq:QQ}) reduces to
\be
\left(- \p_a^2 + h^2 |\varphi|^2\right)f^{(n)} - h \left[\left(\p_1 + i \p_2\right)\varphi^*\right] f^{(n)*} = m_n^2 f^{(n)}\,.
\label{eq:f}
\ee
Let us further rewrite this by decomposing $f$ and $\varphi$ into their real and imaginary parts as
$f^{(n)} = f^{(n)}_1 + i f^{(n)}_2$ and $\varphi = \varphi_1 + i \varphi_2$. Then Eq.~(\ref{eq:f}) is expressed as
\be
D^\dag D 
\left(
\begin{array}{c}
f_1^{(n)}\\
f_2^{(n)}
\end{array}
\right)
= m_n^2
\left(
\begin{array}{c}
f_1^{(n)}\\
f_2^{(n)}
\end{array}
\right)\,,
\label{eq:DDm}
\ee
with
\be
D
= \left(
\begin{array}{cc}
\p_1 + h \varphi_1 & \p_2 - h \varphi_2\\
- \p_2 - h \varphi_2 & \p_1 - h \varphi_1
\end{array}
\right)\,,\quad
D^\dag
= \left(
\begin{array}{cc}
-\p_1 + h \varphi_1 & \p_2 - h \varphi_2\\
- \p_2 - h \varphi_2 & -\p_1 - h \varphi_1
\end{array}
\right)\,,
\ee
and
\be
D^\dag D =
\left(
\begin{array}{cc}
- \p_a^2 - h \p_a\varphi_a + h^2 \varphi_a^2 & h \epsilon_{ab}\p_a\varphi_b \\
h \epsilon_{ab}\p_a\varphi_b & - \p_a^2 + h \p_a\varphi_a + h^2 \varphi_a^2
\end{array}
\right)\,.
\label{eq:DdD}
\ee
Eq.~(\ref{eq:DDm}) consists of real variables only, so that it is numerically tractable.

\subsection{Chiral fermion zero modes}
\label{sec:chiral_fermion}

We are primally interested in a zero mode ($m_0 = 0$). Let us assume that it is normalizable for a while (we will give the normalizability condition later).
Combining Eqs.~(\ref{eq:two_zeros}) and (\ref{eq:af+-}), its explicit expression in the four component notation is
\begin{gather}
\begin{pmatrix}
\chi_{\omega_0,k}^{(0)} \\
\bar{\xi}_{\omega_0,k}^{(0)}
\end{pmatrix} = 
\begin{pmatrix}
0 \\
f^{(0)} \\
if^{(0)*} \\
0
\end{pmatrix}\,.
\end{gather}
So far, we have not mentioned the frequency and the momentum along $z$-axis. Let us take them into account from now on.
From Eq.~(\ref{eq:dispersion}) the frequency is either $\omega_0 = k$ or $\omega_0 = -k$.
Since the zero mode is the eigenstate of $Z$ with the eigenvalue $+1$, this is also the eigenstate of $H_k = kZ + M$ with the eigenvalue $+k$.
Then Eq.~(\ref{eq:eigeneq1}) allows only $\omega_0 = + k$. Hence, the fermion zero mode including $t$ and $z$ is given by
\be
\begin{pmatrix}
\chi^{(0)} \\
\bar{\xi}^{(0)}
\end{pmatrix} = 
e^{-i k(t - z)}
\begin{pmatrix}
0 \\
f^{(0)}(x,y) \\
if^{(0)*}(x,y) \\
0
\end{pmatrix}\,.
\label{eq:zeromode}
\ee
This corresponds to a massless chiral fermion moving at the speed of light in the positive $z$-direction (right-moving by convention). 
The two dimensional ($tz$) chirality can be also understood from the chirality matrix $\gamma_0\gamma_3$ in the two dimensions. 
Eq.~(\ref{eq:zeromode}) is the eigenstate of $\gamma_0\gamma_3$ with the eigenvalue $+1$.
These facts are well known \cite{WITTEN1985557,CALLAN1985427} for axisymmetric strings whose cross section through the $xy$ plane is point-like, and the low-energy effective theory of the world volume of the string is a $1+1$-dimensional theory with the right-moving chiral fermion.

In our case, however, the background solitons are composites of strings and domain walls which have nontrivial and infinite expanse on the $xy$ plane. As we will show below, consequence of this is that the zero mode function can have non-trivial shapes such as triangles, squares, etc. In some cases, it may have a finite structure, such as a finite one-dimensional segment or a finite two-dimensional polygonal shape. 
Hence, the massless fermion in general is neither axisymmetric nor point-like. Nevertheless, the solution (\ref{eq:zeromode}) tells that 
the massless fermion can be regarded as the right-moving chiral fermion from the view point of an observer in $1+1$ dimensions.

Furthermore, we will encounter cases where the zero mode function even spreads out in an infinite region such as a half-line or a half-plane.
In such cases that the zero mode function does not diverge but approaches to a constant at spatial infinity, so it is a physical mode (like a plane wave)  though it is non-normalizable with respect to integration over the $xy$ plane. In these cases, the low energy effective theories are not simple $1+1$ dimensional theories but become $1+2$ or $1+3$ dimensional theories with several boundaries. Hence the fermion chirality should be discussed with some care. Despite of this point, Eq.~(\ref{eq:zeromode}) is always valid which has the eigenvalue $+1$ of $\gamma_0\gamma_3$, and is right-moving state along the $z$-axis.

\subsection{Analytic solutions of fermion zero mode for $\epsilon_{ab}\p_a\varphi_b=0$}
\label{sec:fermion_zero_diagonal}

What we have to do next is to find solutions to Eq.~(\ref{eq:DDm}) for $m_0=0$. 
In this section we will accomplish this task for the background configurations satisfying the condition $\epsilon_{ab}\p_a\varphi_b=0$.
There are two cases satisfying the condition. The one is the normal axisymmetric axion string in the $N=0$ axion model reviewed in Sec.~\ref{sec:axion_model}. We have $\varphi= v F(\rho) e^{i \theta}$, and it is easy to verify the condition for this. The other is all the solutions to the BPS equations (\ref{eq:BPS1}) and (\ref{eq:BPS2}) in Sec.~\ref{sec:susy}.

When the condition is satisfied, the off-diagonal component  of $D^\dag D$ vanishes, and Eq.~(\ref{eq:DDm}) is particularly simplified. 
Let us introduce the following differential operators 
\be
d_a = \p_a + h \varphi_a\,,\quad d_a^\dag = - \p_a + h \varphi_a\,,
\ee
the diagonal elements of $D^\dag D$ can be expressed as
\be
d_a^\dag d_a f_1^{(n)} = m_n^2 f_1^{(n)}\,,\quad
d_a d_a^\dag f_2^{(n)} = m_n^2 f_2^{(n)}\,.
\label{eq:ddm}
\ee
Therefore, the zero modes are the kernels of $d_a$ and $d_a^\dag$ as
\be
d_a f_1^{(0)} = 0\,,\quad
d_a^\dag f_2^{(0)} = 0\,,\qquad(a=1,2)\,.
\label{eq:fermion_zero_mode_simple}
\ee
In general, the kernels of $d_a$ and $d_a^\dag$ cannot simultaneously be normalizable. 
As we will see below, when we assume $h>0$ and the background solution $\varphi_a$ has positive winding number,
the imaginary part must be $f^{(0)}_2 = 0$ because otherwise it diverges.

The condition $\epsilon_{ab}\p_a\varphi_b = 0$ is rotation-free condition, so it implies that there exists a scalar function $U$ satisfying
\be
\varphi_a(x) = \p_a U(x)\,,
\label{eq:phi_U}
\ee
and this ensures that the following line integral on the $xy$ plane
\be
\int^{x}_{x_0} d\vec y\cdot \vec \varphi(y) = U(x) - U(x_0)\,,
\ee
is independent of the path between $\vec x_0$ and $\vec x$.
Now, the analytic solution of Eq.~(\ref{eq:fermion_zero_mode_simple}) can be obtained as
\be
f_1^{(0)}(x) = A \exp\left(-h \int^x d\vec y\cdot \vec \varphi(y)\right) = A' e^{-hU(x)}\,,
\label{eq:f10}
\ee
with $A$ and $A' = A e^{hU(x_0)}$ being normalization constants.
This is a generalization of the analytic solution for the axially symmetric string obtained in Ref.~\cite{CALLAN1985427} to the non-axially symmetric string-wall composites satisfying the rotation-free condition.
We will entirely make use of this formula in Sec.~\ref{sec:axisymmetric_axion_string_fermion}, 
\ref{sec:fermion_susy}, and \ref{sec:delocalization}.

Normalizability of the fermion zero mode function fully owes to $U(x)$. 
Indeed, we will see that $U$ linearly diverges as $U \sim \rho$ ($\rho = \sqrt{x^2+y^2}$) at spatial infinity, so that $e^{-hU}$ converges to zero exponentially fast. In order to figure out a criterion for normalizability, let us take the gradient of $f_1^{(0)}$,
\be
\p_a f_1^{(0)} = -h(\p_aU) A'e^{-hU}= - h A' e^{-hU} \varphi_a \,.
\ee 
Due to the positive definiteness $e^{-h U} > 0$, $\p_a f_1^{(0)} = 0$ is realized if and only if there exists a point at which $\varphi_a = 0$ is satisfied.
Hence, if the zero mode function has a peak, the peak is located at the point where $\varphi_a$ vanishes.
Conversely, when $\varphi_a = 0$ does not occur at any point, the zero mode function is not normalizable.
This can be also stated as follows. From Eq.~(\ref{eq:L_fermion}), the position-dependent fermion mass reads
\be
m_{\rm f}(x) = h \varphi(x) = h (\p_1 + i \p_2) U(x) \,,\quad (\varphi = \varphi_1 + i \varphi_2)\,.
\label{eq:mf}
\ee
Thus, the normalizability condition can be rephrased as follows: The fermion zero mode on the string-wall composite is normalizable if and only if
there exists a point in the $xy$ plane at which $m_{\rm f}$ vanishes. The zero mode function peaks at that point. 
This is a natural extension of the fermion zero mode on the domain walls \cite{Jackiw:1975fn} and the strings \cite{Jackiw:1981ee,CALLAN1985427}.

Giving the analytic solution is important because it is nothing but the mathematical proof of the existence of massless fermion. In this respect the existence of fermion zero mode is proved for all the BPS string-wall solutions to the BPS equations (\ref{eq:BPS1}) and (\ref{eq:BPS2}), in addition to the $N=0$ axion strings as a reconfirmation of \cite{Jackiw:1981ee,CALLAN1985427}.

\subsection{Fermion zero modes on axion string-wall composites}

\subsubsection{$N=0$: An axisymmetric string}
\label{sec:axisymmetric_axion_string_fermion}

\begin{wrapfigure}{r}[0pt]{0.3\textwidth}
\vspace*{-.2cm}
 \centering
 \includegraphics[width=0.3\textwidth]{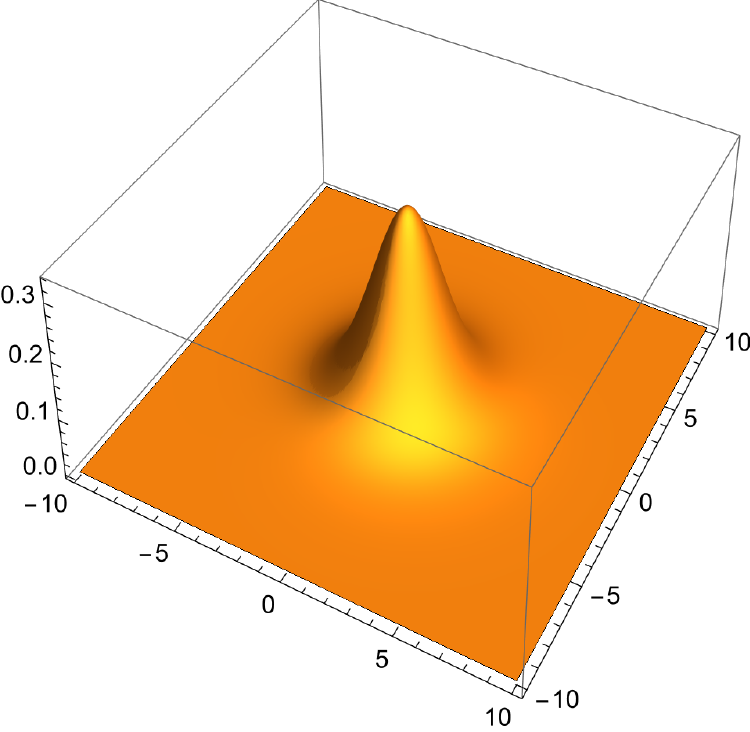}
 \caption{The fermionic zero mode function $f_1^{(0)}$ on the axion string given in Fig.~\ref{fig:axion_N=0_BG}.}
 \label{fig:fermion_axion_N=0}
\end{wrapfigure}
Let us first verify our formula given in Sec.~\ref{sec:fermion_zero_diagonal} to the axion model (\ref{eq:Lag_axion}) 
with $N=0$. The background solution is the axially symmetric string given in Eq.~(\ref{eq:axion_string})
\begin{gather}
\varphi_1 = v  F(\tilde \rho) \cos\theta\,,\quad \varphi_2 = v  F(\tilde \rho) \sin\theta\,,
\label{eq:axisym_axion_striing}
\end{gather}
with $\tilde \rho = \sqrt{\lambda}v\rho$ and $x^1 + i x^2 = \rho e^{i\theta}$.
Since this satisfies $\epsilon_{ab}\p_a\varphi_b = 0$, we can adopt the formula (\ref{eq:f10}) by introducing
\be
U(x) = \int^\rho v F(\sqrt{\lambda}v\rho')\, d\rho'\,.
\label{eq:U_axion}
\ee
Then $\varphi_a$ in Eq.~(\ref{eq:axisym_axion_striing}) can be given by $\varphi_a = \p_a U$.
The asymptotic behavior of $U$ reads as $U \to v\rho$ as $\rho \to \infty$. Hence,
under the assumption $h>0$, we find the normalizable zero mode function \cite{CALLAN1985427}
\be
f_1^{(0)}(x) = A \exp\left(- hv \int^\rho F(\sqrt{\lambda}v\rho')\, d\rho'\right)\,,
\ee
with $f_2^{(0)} = 0$. 
The numerical solution is shown in Fig.~\ref{fig:fermion_axion_N=0}.
The zero mode function peaks at the vanishing point of $\varphi_a$, namely the center of the axion string.

\subsubsection{$N\ge1$: The non-axisymmetric cases}
\label{sec:axion_N>=1}

\begin{figure}[htbp]
\centering
\begin{minipage}[b]{0.32\linewidth}
\centering
\includegraphics[keepaspectratio,scale=0.35]{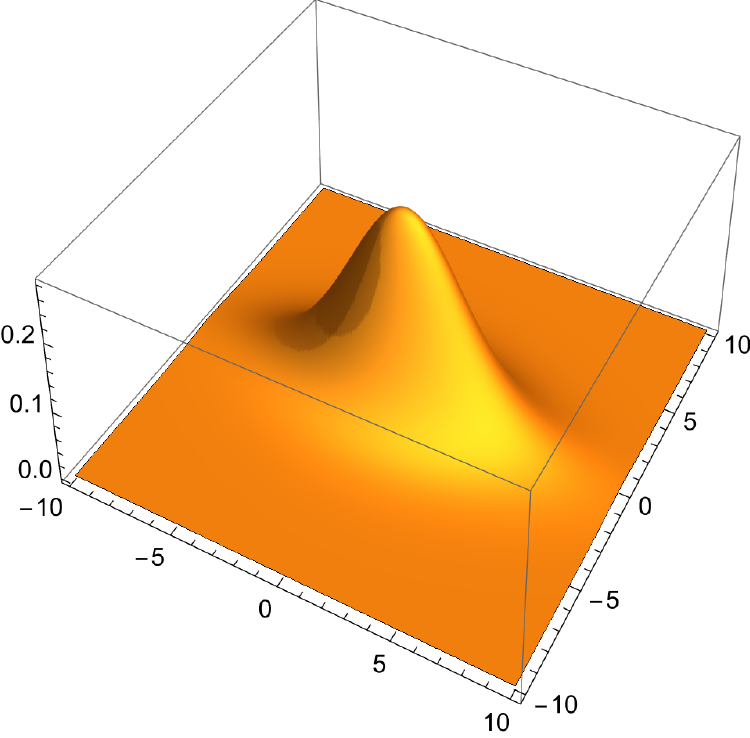}
\subcaption{$|f_1^{(0)}+if_2^{(0)}|$}
\end{minipage}
\begin{minipage}[b]{0.32\linewidth}
\centering
\includegraphics[keepaspectratio,scale=0.35]{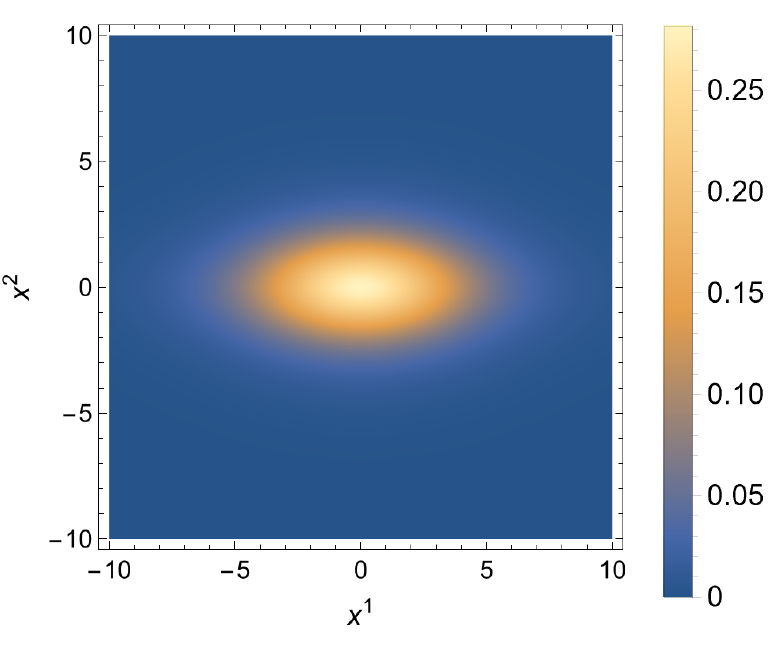}
\subcaption{$f_1^{(0)}$}
\end{minipage}
\begin{minipage}[b]{0.32\linewidth}
\centering
\includegraphics[keepaspectratio,scale=0.35]{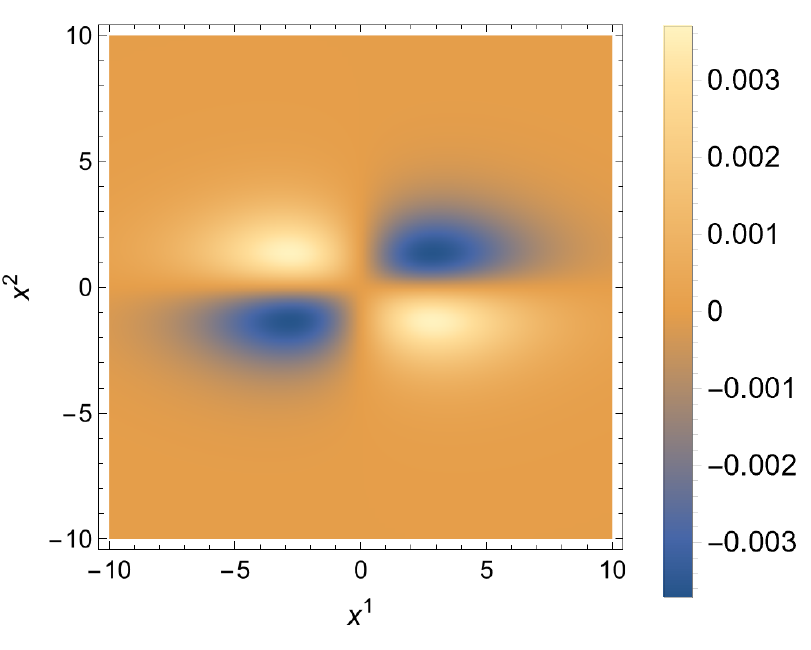}
\subcaption{$f_2^{(0)}$}
\end{minipage}
\caption{The fermionic zero mode on the axion string-wall composite given in Fig.~\ref{fig:axion_N=2_a=0.1_BG} for the axion model with $N=2$. We set $h = \sqrt\lambda$.}
\label{fig:axion_zeromode_N=2}
\end{figure}

Next, we consider the non-axisymmetric background solutions. 
The first example is the fermion zero mode for the string with two domain walls in the case of $N=2$.
The corresponding background solution is shown in Fig.~\ref{fig:axion_N=2_a=0.1_BG} for $\alpha = 1/20$.
Since this does not satisfy the rotation-free condition $\epsilon_{ab}\p_a \varphi_b = 0$, we cannot make use of the formula (\ref{eq:f10}). Therefore, we have to numerically solve the mode equation (\ref{eq:DDm}) for the background solution $\varphi_a$ which is also obtained by the previous numerical calculation.
The resulting zero mode function is shown in Fig.~\ref{fig:axion_zeromode_N=2}. The eigenvalue is $\omega_0 = 2.4 \times 10^{-6}$ in the unit of $\sqrt{\lambda}v$. It is localized not on the domain wall but on the string with small outflows toward the domain walls on the both sides of the string. Reflecting the fact that the string is attached by the $N=2$ domain walls, the zero mode function has the oval shape.
Note that both real and imaginary parts, $f_{1}^{(0)}$ and $f_{2}^{(0)}$, are nontrivial in contrast to the axisymmetric background solution for which $f_2^{(0)}$ vanishes everywhere.

\begin{figure}[htbp]
\centering
\begin{minipage}[b]{0.32\linewidth}
\centering
\includegraphics[keepaspectratio,scale=0.3]{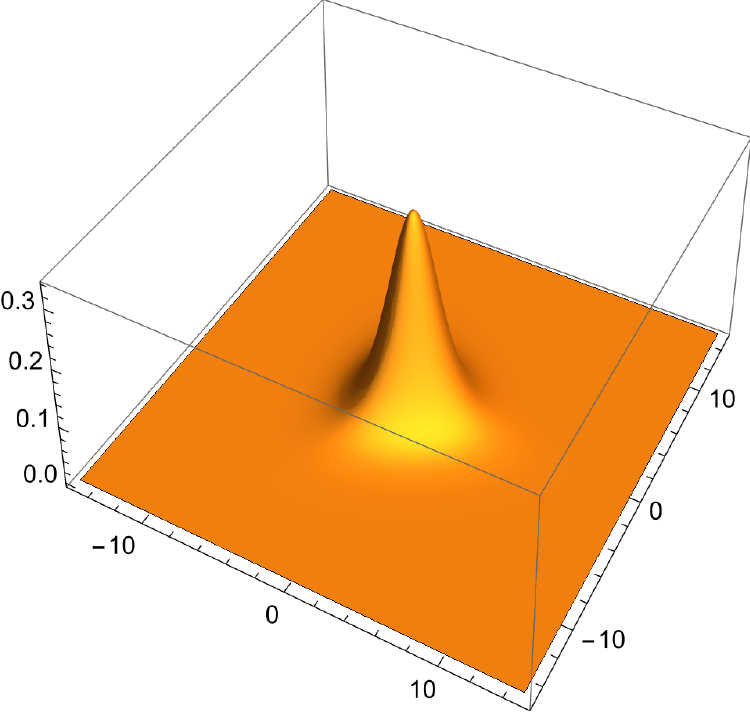}
\subcaption{$|f_1^{(0)}+if_2^{(0)}|$}
\vspace*{.5cm}
\end{minipage}
\begin{minipage}[b]{0.32\linewidth}
\centering
\includegraphics[keepaspectratio,scale=0.3]{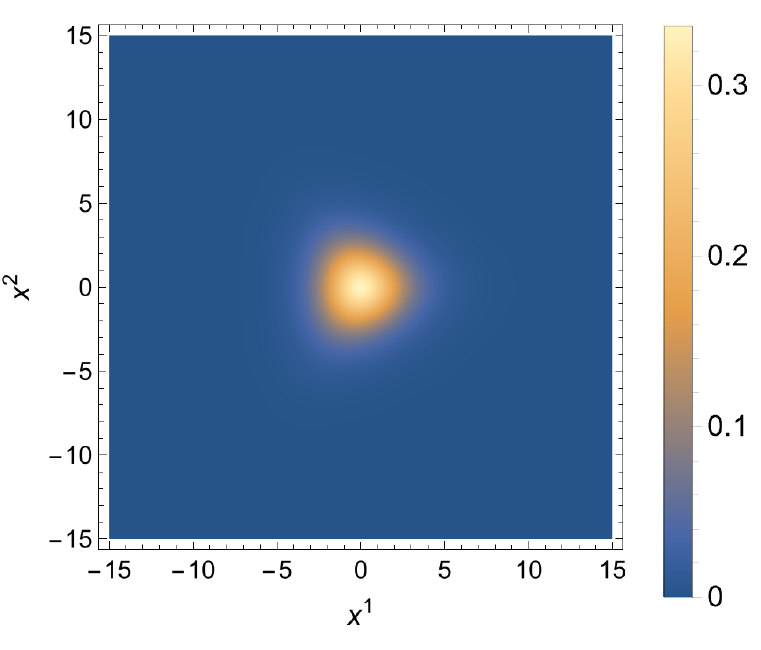}
\subcaption{$f_1^{(0)}$}
\vspace*{.5cm}
\end{minipage}
\begin{minipage}[b]{0.32\linewidth}
\centering
\includegraphics[keepaspectratio,scale=0.3]{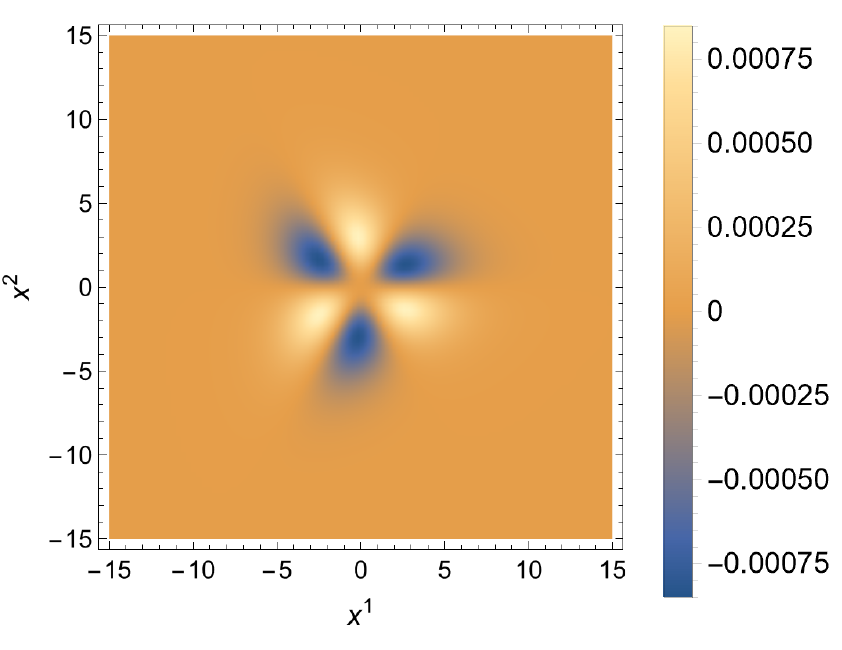}
\subcaption{$f_2^{(0)}$}
\vspace*{.5cm}
\end{minipage}
\begin{minipage}[b]{0.32\linewidth}
\centering
\includegraphics[keepaspectratio,scale=0.3]{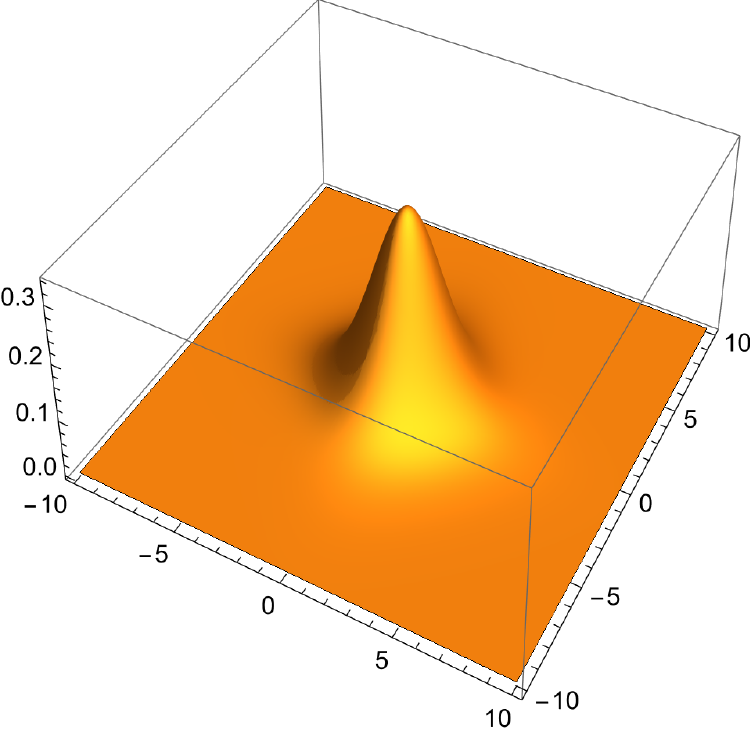}
\subcaption{$|f_1^{(0)}+if_2^{(0)}|$}
\end{minipage}
\begin{minipage}[b]{0.32\linewidth}
\centering
\includegraphics[keepaspectratio,scale=0.3]{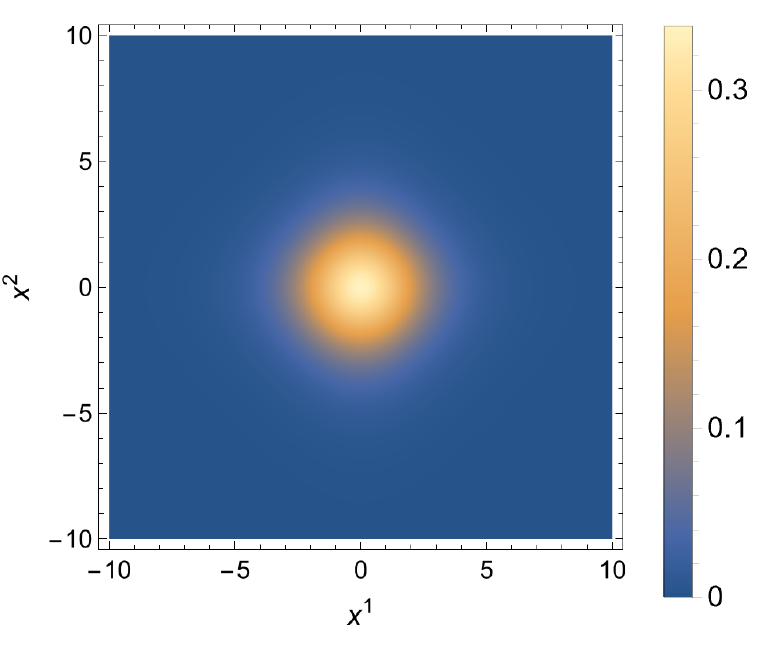}
\subcaption{$f_1^{(0)}$}
\end{minipage}
\begin{minipage}[b]{0.32\linewidth}
\centering
\includegraphics[keepaspectratio,scale=0.3]{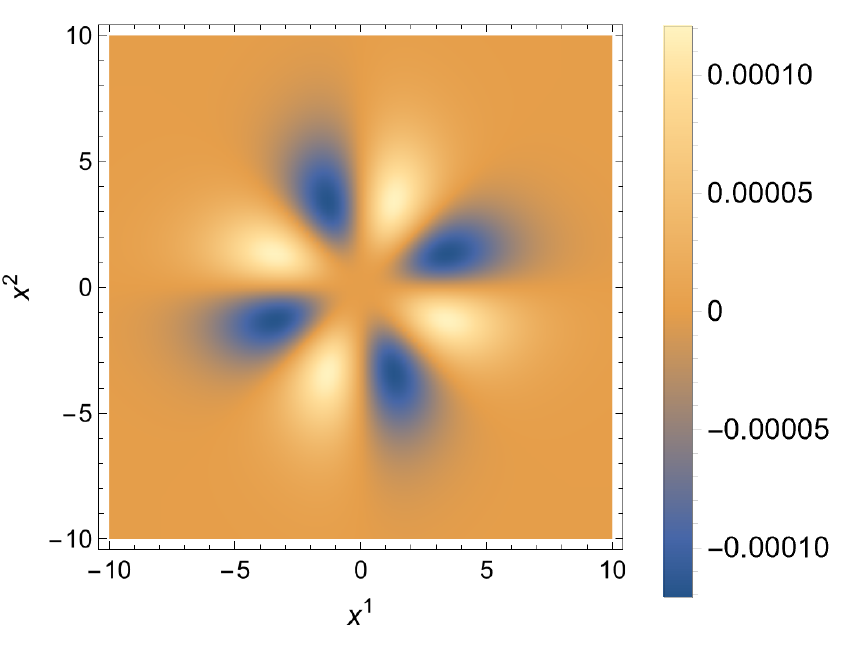}
\subcaption{$f_2^{(0)}$}
\end{minipage}
\caption{The panels in the top (bottom) row show the fermion zero mode on the axion string-wall composites given in the top (bottom) row of Fig.~\ref{fig:axion_N=3and4} for the axion model with $N=3$ and $\alpha = 1/30$ ($N=4$ with $\alpha = 1/40$).}
\label{fig:axion_zeromode_N=3and4}
\end{figure}

\begin{figure}[htbp]
\centering
\begin{minipage}[b]{0.32\linewidth}
\centering
\includegraphics[keepaspectratio,scale=0.3]{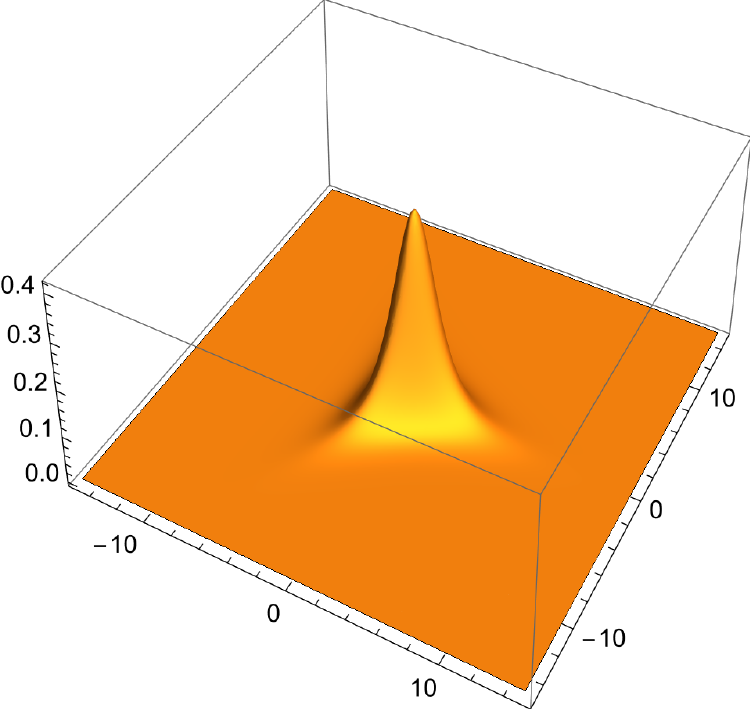}
\subcaption{$|f_1^{(0)}+if_2^{(0)}|$}
\vspace*{.5cm}
\end{minipage}
\begin{minipage}[b]{0.32\linewidth}
\centering
\includegraphics[keepaspectratio,scale=0.3]{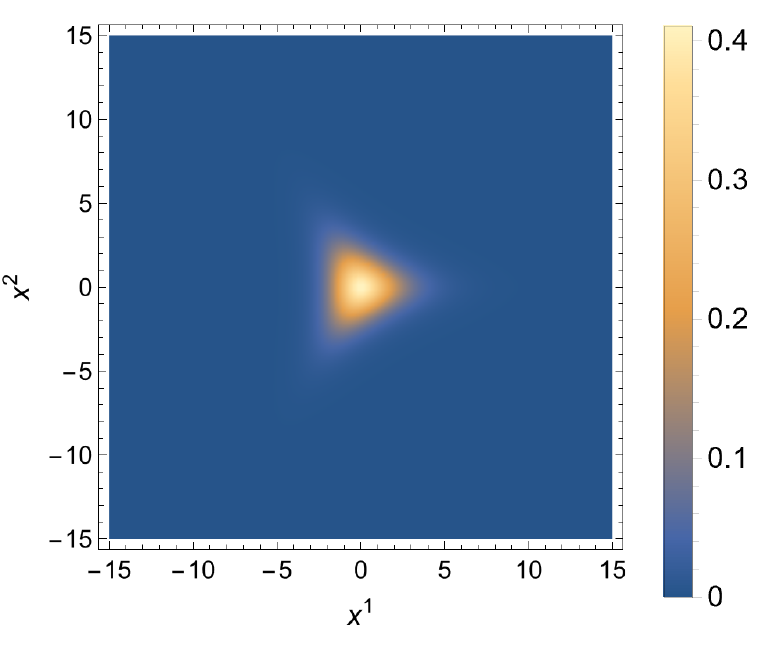}
\subcaption{$f_1^{(0)}$}
\vspace*{.5cm}
\end{minipage}
\begin{minipage}[b]{0.32\linewidth}
\centering
\includegraphics[keepaspectratio,scale=0.3]{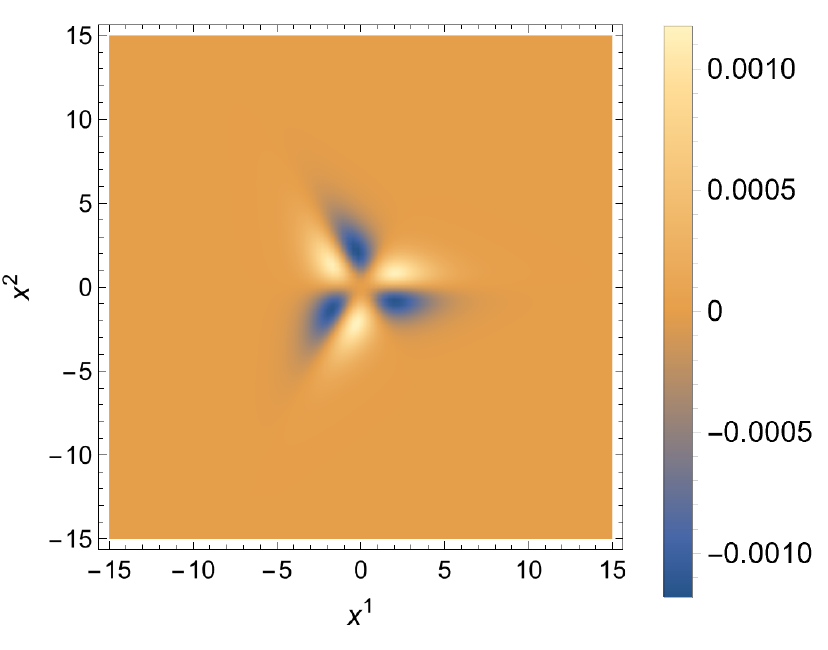}
\subcaption{$f_2^{(0)}$}
\vspace*{.5cm}
\end{minipage}
\begin{minipage}[b]{0.32\linewidth}
\centering
\includegraphics[keepaspectratio,scale=0.3]{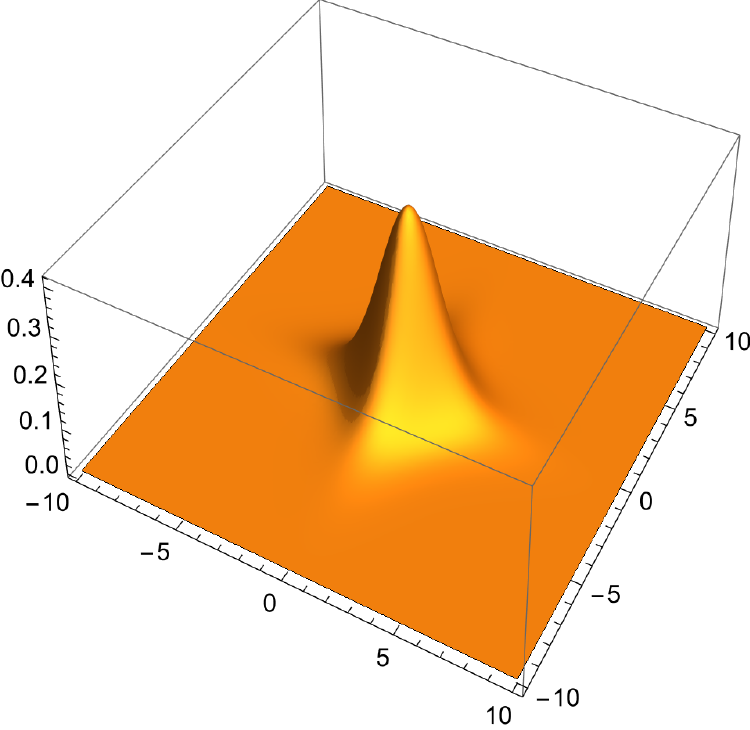}
\subcaption{$|f_1^{(0)}+if_2^{(0)}|$}
\vspace*{.5cm}
\end{minipage}
\begin{minipage}[b]{0.32\linewidth}
\centering
\includegraphics[keepaspectratio,scale=0.3]{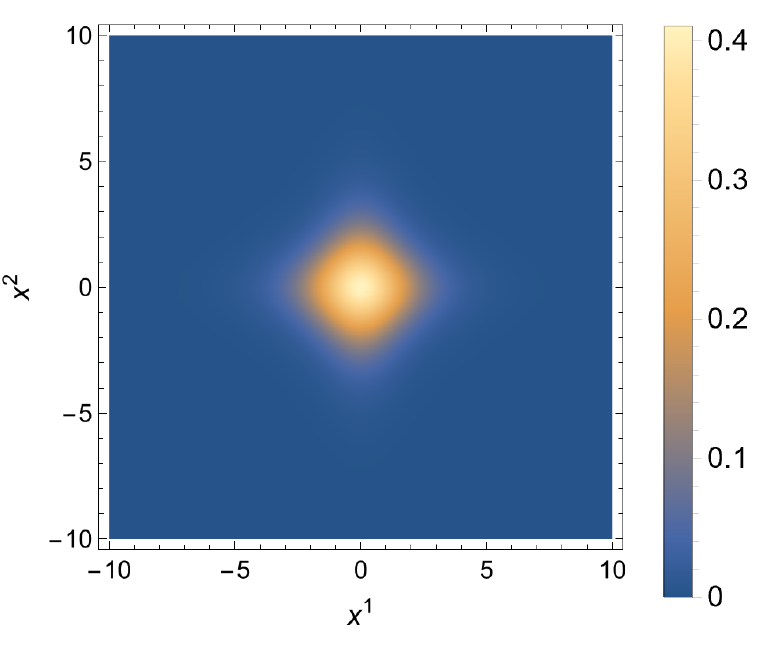}
\subcaption{$f_1^{(0)}$}
\vspace*{.5cm}
\end{minipage}
\begin{minipage}[b]{0.32\linewidth}
\centering
\includegraphics[keepaspectratio,scale=0.3]{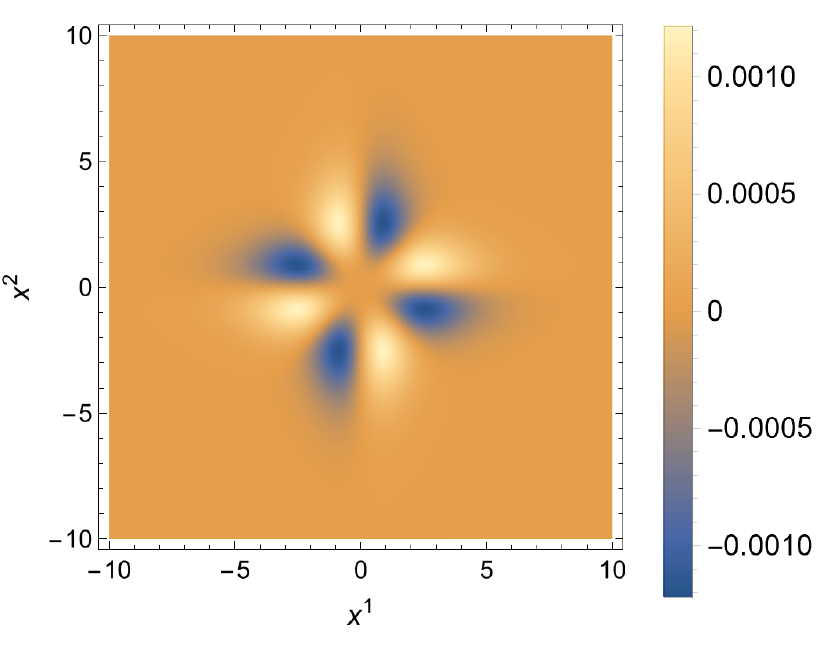}
\subcaption{$f_2^{(0)}$}
\vspace*{.5cm}
\end{minipage}
\caption{The panels in the top (bottom) row show the fermion zero mode on the string-wall composites given in the top (bottom) row of Fig.~\ref{fig:axion_N=3and4} for the axion model with $N=3$ and $\alpha = 1/5$ ($N=4$ with $\alpha = 1/10$).}
\label{fig:axion_zeromode_N=3_2}
\end{figure}

The next examples are the string-wall composites in the axion model with $N=3$ and $4$. The string is attached by three and four domain walls as given in Fig.~\ref{fig:axion_N=3and4}.
The corresponding fermion zero mode functions are shown in Fig.~\ref{fig:axion_zeromode_N=3and4}.
Reflecting geometric structures of the background solutions, the zero mode functions are triangle and square shapes.
But the polygon-like shape in Fig.~\ref{fig:axion_zeromode_N=3and4} is quite fuzzy for small $\alpha$. In contrast, it gets sharpen by
increasing $\alpha$ as shown in Fig.~\ref{fig:axion_zeromode_N=3_2}.

All in all, the fermion zero modes commonly localize at the center of the strings. This can be readily understood from
the image of the function $\varphi_a(x)$ given in Figs.~\ref{fig:axion_N=2_a=0.1_BG}(b) and \ref{fig:axion_N=3and4}(b), and (e).
These figures shows the position-dependent mass $m_{\rm f} = h\varphi$ (up to the Yukawa coupling constant $h$) for each configurations. 
The borders and interiors of the oval, rounded triangle and rounded square are the image of the $xy$ plane by the string-wall composite solutions. Since all these images include $m_{\rm f} = 0$ point ($\varphi = 0$) in their insides, the fermion zero mode is normalizable, and the peak appears at the point where $m_{\rm f}$ vanishes. Namely the zero mode appears at the string center.

\begin{figure}[htbp]
\centering
\begin{minipage}[b]{0.32\linewidth}
\centering
\includegraphics[keepaspectratio,scale=0.35]{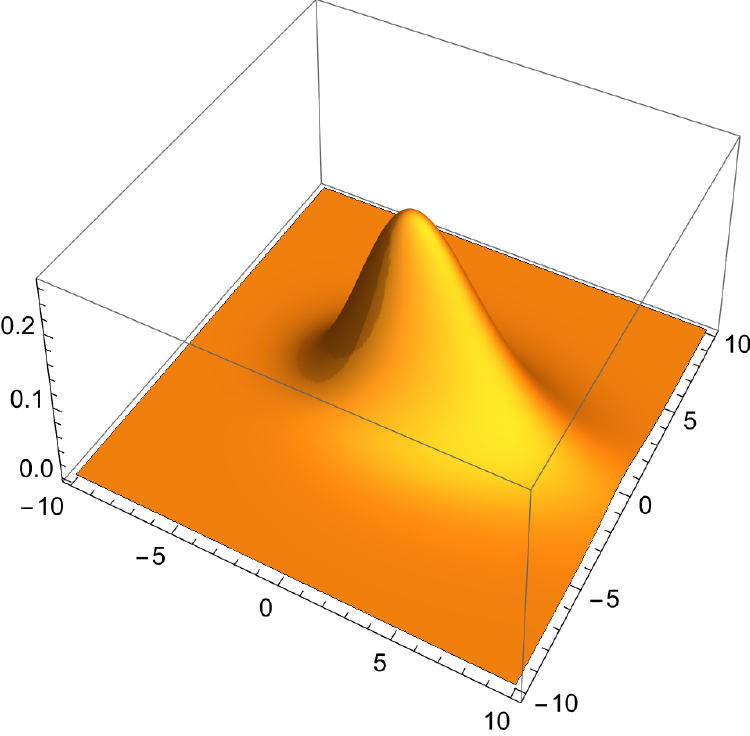}
\subcaption{$|f_1^{(0)}+if_2^{(0)}|$}
\end{minipage}
\begin{minipage}[b]{0.32\linewidth}
\centering
\includegraphics[keepaspectratio,scale=0.35]{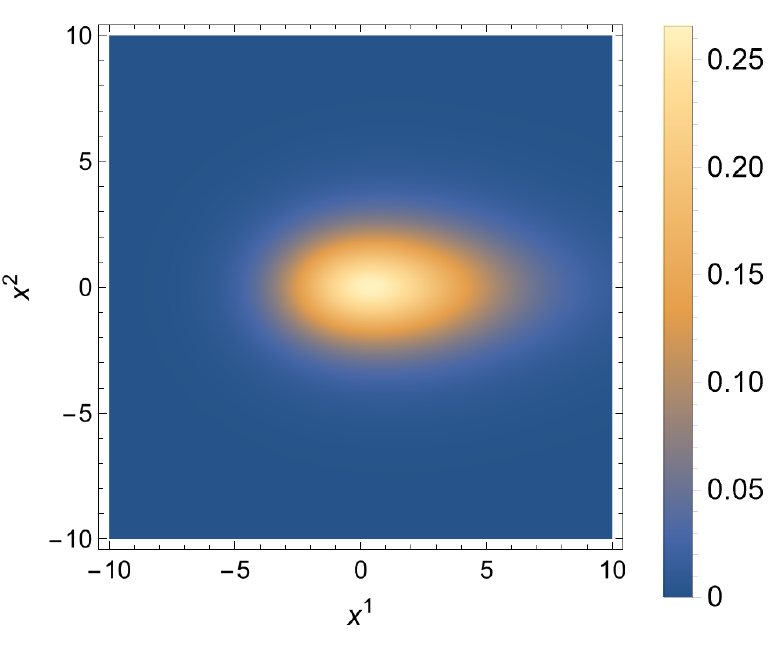}
\subcaption{$f_1^{(0)}$}
\end{minipage}
\begin{minipage}[b]{0.32\linewidth}
\centering
\includegraphics[keepaspectratio,scale=0.35]{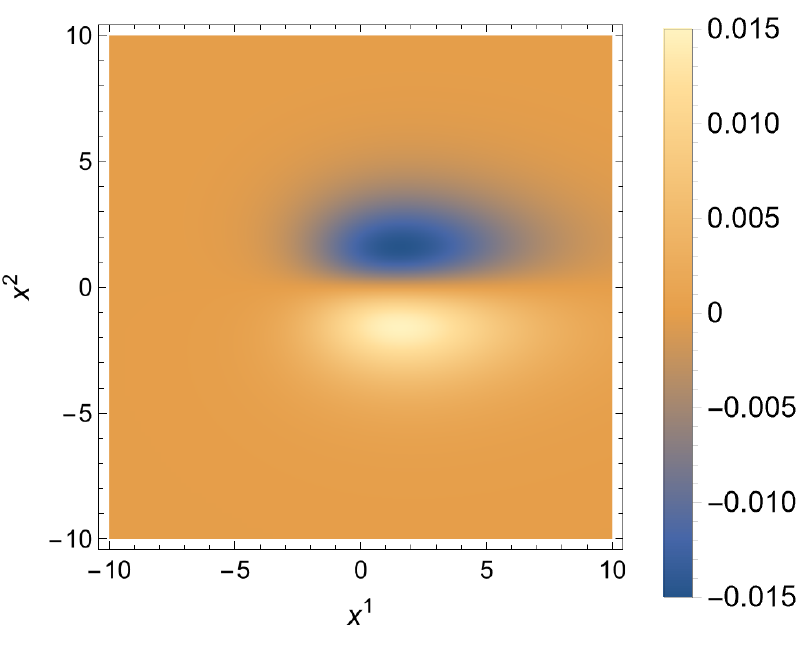}
\subcaption{$f_2^{(0)}$}
\end{minipage}
\caption{The fermionic zero mode on the axion string-wall composite given in Fig.~\ref{fig:axion_N=1_BG} for the axion model with $N=1$.}
\label{fig:axion_zeromode_N=1}
\end{figure}

The last example is the string with the single domain wall in the axion model with $N=1$.
Let us take the numerical background solution shown in Fig.~\ref{fig:axion_N=1_BG} for $\alpha = 1/10$.
The corresponding numerical solution to Eq.~(\ref{eq:DDm}) is shown in Fig.~\ref{fig:axion_zeromode_N=1}.
However, this solution of $N=1$ should be received with a caution that the background solution cannot be static and we have ignored time dependence at both levels of obtaining the bosonic background and the fermionic zero mode. 

In this section we gave the numerical solutions of the fermion zero modes for $N \ge 1$ axion string-wall composites. 
This is certainly good progress but it is far from giving a mathematically rigorous proof because any numerical solutions are approximations. Indeed, numerically obtained eigenvalues in our computations are of order $10^{-6}$ in the unit of $\sqrt{\lambda} v$. We can improve the numerical calculation to have the eigenvalues closer to 0 but they are never equal to exactly zero.
Hence, our achievement here is modest in the sense that we obtained numerical solutions for the fermion zero modes for $N\ge 1$ axion string-wall composites.

\subsection{Fermion zero modes on the BPS string-wall composites}
\label{sec:fermion_susy}

Next we will investigate the fermion zero modes for the BPS composites of strings and domain walls in the bosonic part of \(\mathcal{N}=2\) supersymmetric Abelian-Higgs model (\ref{eq:susy_Lag}) explained in Sec.~\ref{sec:susy}. 
Owing to the BPS nature, the background solution is rotation-free, namely $\epsilon_{ab}\p_a\varphi_b = 0$.
Hence, the zero mode function is analytically given in the form of Eq.~(\ref{eq:f10}). It is mathematical proof of the existence of massless fermion but is also useful in a practical sense, we do not need any additional numerical analysis other than those for solving the master equation (\ref{eq:master}). What we need to do is determining the scalar function $U(x)$ in Eq.~(\ref{eq:phi_U}). But it has been already done: The scalar function $U$ can be found by Eq.~(\ref{eq:phi_omega}) as 
\be
U(x) = \frac{1}{2}\log \Omega\,.
\ee
Then, we have 
\be
f_1^{(0)}(x) = \exp \left(- \frac{h}{2} \log \Omega \right) = \Omega^{-\frac{h}{2}}\,,\quad f_2^{(0)} = 0\,.
\label{eq:susy_fermion_zero_mode}
\ee
Thus, by solving the master equation (\ref{eq:master}), we have both the bosonic background configuration and the fermionic zero mode function simultaneously. This is a virtue of the BPS composite of the strings and domain walls.

\begin{figure}[hbtp]
\centering
\begin{minipage}[b]{0.45\linewidth}
\centering
\includegraphics[keepaspectratio,scale=0.4]{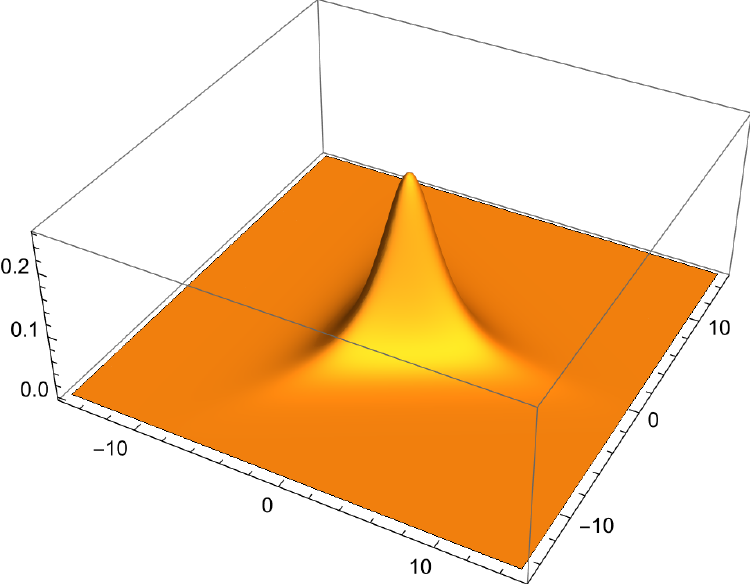}
\subcaption{}
\end{minipage}
\begin{minipage}[b]{0.45\linewidth}
\centering
\includegraphics[keepaspectratio,scale=0.4]{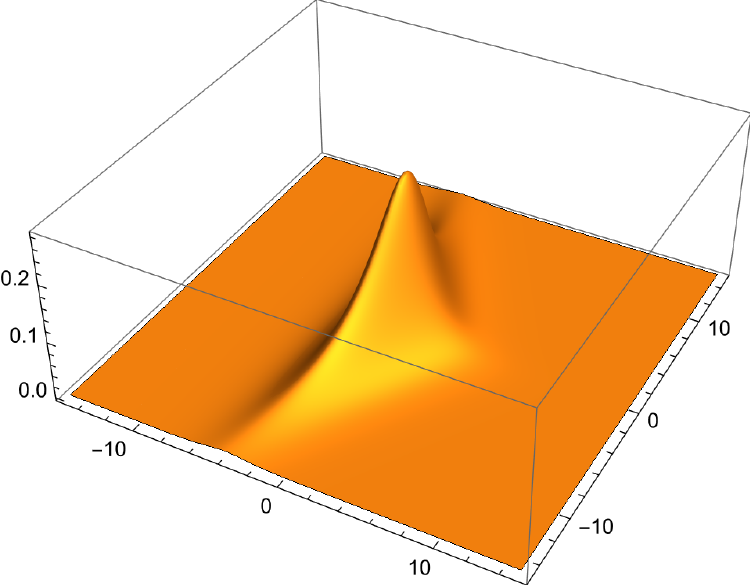}
\subcaption{}
\end{minipage}
\caption{The fermion zero mode functions on the BPS string-wall composites: (a) The equiangular three-pronged junction in Fig.~\ref{fig:susy_dwj_n3_1}. (b) The non-equiangular three-pronged junction in Fig.~\ref{fig:susy_dwj_n3_2}.}
\label{fig:fermion_susy_nf3}
\end{figure}

Fig.~\ref{fig:fermion_susy_nf3} shows the fermion zero modes for the background solutions given in Figs.~\ref{fig:susy_dwj_n3_1}  and \ref{fig:susy_dwj_n3_2}. The background solution in Fig.~\ref{fig:susy_dwj_n3_1} has the equiangular three-pronged junction which is quite similar to the string-wall composite given in Fig.~\ref{fig:axion_N=3and4} in the axion model with $N=3$. Not only the background solution, the corresponding fermion zero modes given in Figs.~\ref{fig:axion_zeromode_N=3and4}(a) [see also \ref{fig:axion_zeromode_N=3_2}(a)] and \ref{fig:fermion_susy_nf3}(a) are qualitatively very similar to each other. But we can only say the latter is numerically zero mode whereas the former is exactly zero mode.
In contrast, the non-equiangular three-pronged junction in Fig.~\ref{fig:susy_dwj_n3_2} is peculiar to the SUSY motivated model whose counterpart does not exist in the axion model. As shown in Fig.~\ref{fig:fermion_susy_nf3}(b), shape of the fermion zero mode function is deformed according to the background solution.

From Fig.~\ref{fig:fermion_susy_nf3} together with Figs.~\ref{fig:susy_dwj_n3_1}(a) and \ref{fig:susy_dwj_n3_2}(a), we see the fermion zero mode localizes not on the domain walls but around the strings as in the axion model.
This can be again explained by the position-dependent fermion mass $m_{\rm f} = h\varphi$. As mentioned above, the fermion zero mode function peaks at the point $(x^1,x^2)$ where $m_{\rm f}(x^1,x^2)$ vanishes.
As is seen by both Figs.~\ref{fig:susy_dwj_n3_1}(b) and \ref{fig:susy_dwj_n3_2}(b), the vanishing point $\vec\varphi = 0$ indeed exists inside the red triangles. More precisely, it occurs on an interior point (neither on an edge nor on a vertex) of the red triangles which correspond to the string.
This is the reason why the fermions localize on the strings.

\begin{figure}[hbtp]
\centering
\begin{minipage}[b]{0.45\linewidth}
\centering
\includegraphics[keepaspectratio,scale=0.4]{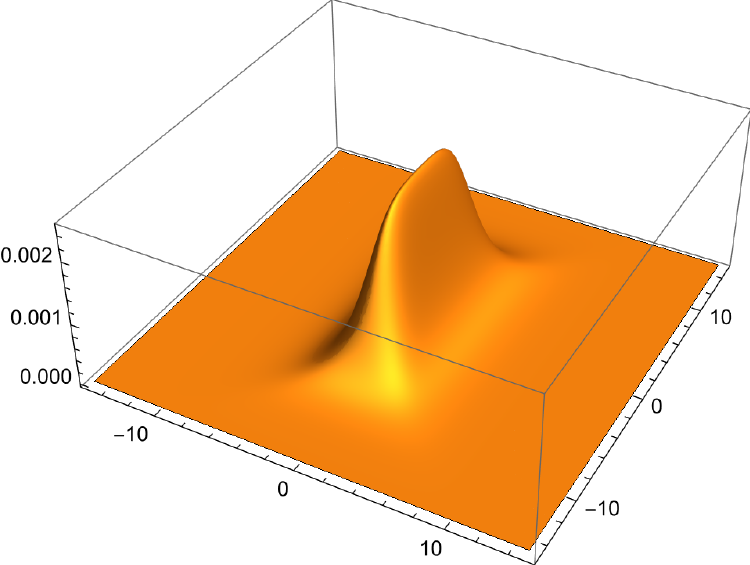}
\subcaption{}
\end{minipage}
\begin{minipage}[b]{0.45\linewidth}
\centering
\includegraphics[keepaspectratio,scale=0.4]{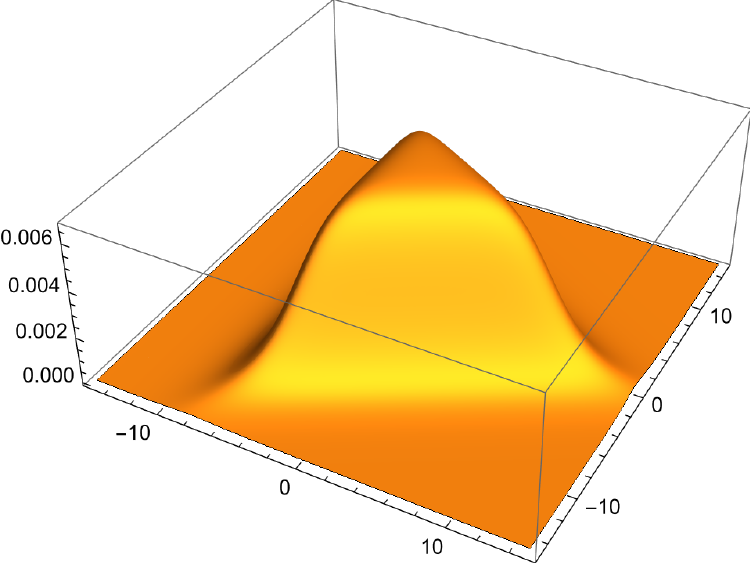}
\subcaption{}
\end{minipage}
\caption{The fermion zero mode functions on the BPS string-wall networks: (a) On the inner domain wall in Fig.~\ref{fig:susy_dwj_Nf4}(a--d). (b) Inside the inner vacuum in Fig.~\ref{fig:susy_dwj_Nf4}(e--h).}
\label{fig:fermion_susy_nf4}
\end{figure}

The fermion zero modes for the string-wall networks with $N_{\rm F} > 3$ are also readily obtained in the similar manner as explained above.
New aspects peculiar to the network are the finite substructures such as an inner domain wall between two strings as shown in Fig.~\ref{fig:susy_dwj_Nf4}(a-d) and an inner vacuum surrounded by the loop of domain walls as shown in Fig.~\ref{fig:susy_dwj_Nf4}(e-h).
For the former, we take the mass polygon as $\{(m,0), (0,m), (-m,0), (0,-m)\}$, and $m_{\rm f}  = 0$ occurs on the inner horizontal edge, see Fig.~\ref{fig:susy_dwj_Nf4}(b). Accordingly, the fermion zero mode function given in Fig.~\ref{fig:fermion_susy_nf4}(a) is localized on the inner vertical domain wall of finite length [Fig.~\ref{fig:susy_dwj_Nf4}(a)]. 
The fermion zero mode function is not point-like but linearly spreads on the inner domain wall. It is normalizable since the inner domain wall has a finite length. Thus the massless fermion lives on the domain wall band whose volume is $\mathbb{R}^1 \times I$ where $I$ stands for the one-dimensional finite line segment.
Similarly, for the latter, the mass polygon consists of four points, the three vertices of the triangle $m (-1,0)$, $m(-\cos\frac{2\pi}{3},\sin\frac{2\pi}{3})$, and $m(-\cos\frac{2\pi}{3},-\sin\frac{2\pi}{3})$ with the  inner vertex at $(0,0)$. As is seen in Fig.~\ref{fig:susy_dwj_Nf4}(f) $m_{\rm f} = 0$ takes place onto the inner vertex at $(0,0)$. Therefore, the fermion zero mode function is localized neither on the string nor on the domain walls. It is localized and spreads over whole the inner vacuum as shown in Fig.~\ref{fig:fermion_susy_nf4}(b). Hence, the fermion lives on the triangular prism $\mathbb{R}^1 \times \triangle$ where $\triangle$ stands for the inner vacuum of the finite size.

\subsubsection{A constant shift of the bosonic mass}
\label{sec:bosonic_mass_susy}

Another peculiar point in the BPS string-wall composite is related to the overall constant shift of the mass vectors, briefly mentioned in Sec.~\ref{sec:shift}. 
The constant shift $\vec m_A \to \vec m_A + \vec m_0$ ($A=1,2,\cdots,N_{\rm F}$) does not affect the physical observables such as energy density and the topological charge density, and therefore it has been overlooked in the literature. 
However, the fermions $\xi$ and $\chi$ directly couple to $\vec\varphi$ via the Yukawa term in Eq.~(\ref{eq:L_fermion}), so that the fermionic mode functions are sensitive to the constant shift.
From Eqs.~(\ref{eq:master}) and (\ref{eq:Omega0}) the constant mass shift $\vec m_0$ yields the additional factor to the background solution $\Omega$ as $\Omega \to \Omega' = e^{2\vec m_0\cdot \vec x} \Omega$. Therefore, the scalar field is shifted by $\vec m_0$ as
\be
\vec \varphi \to \vec\varphi'  
= \frac{1}{2}\vec\p \log \Omega'
=  \frac{1}{2}\vec\p \log \Omega + \vec m_0 = \vec\varphi + \vec m_0 \,,
\ee
and this immediately leads to change in the zero mode function
 \be
 f_1^{(0)}  \to f_1^{(0)}{}' = (\Omega')^{-\frac{h}{2}} = e^{- h \vec m_0 \cdot \vec x} \Omega^{-\frac{h}{2}} 
 = e^{- h \vec m_0 \cdot \vec x} f_1^{(0)}\,.
 \ee
The additional factor $e^{- h \vec m_0 \cdot \vec x}$ obviously affects both localization and normalizability of the zero mode function. 
This can be clearly understood with respect to the position-dependent fermion mass in Eq.~(\ref{eq:mf}).
The bosonic mass shift effectively transmuted into the shift  of fermion mass as
\be
m_{\rm f} \to  h \varphi' = h(\varphi + m_0)\,, \quad m_0 \equiv m_{0,1} + i m_{0,2}\,.
\ee
Hence, the vanishing point of  $m_f$ moves in accordance with the change from $\varphi=0$ to $\varphi + m_0 = 0$. 
Then the normalizability is also affected by $m_0$.
Let us assume that the initial mass polygon $\{\vec m_1,\cdots, \vec m_{N_{\rm F}}\}$ includes $\vec\varphi = 0$ inside the mass polygon. After the shift,
the polygon consists of the points $\{\vec m_1 + \vec m_0,\cdots, \vec m_{N_{\rm F}} + \vec m_0\}$.
If the shift $|\vec m_0|$ is too large, the shifted mass polygon no longer includes $\vec\varphi = 0$. If this is the case, the fermion zero mode function does not peak at any points in $x^1x^2$ plane. Namely, it continuously brows up along a spatial direction, and therefore the zero mode is non-normalizable and is unphysical.
\begin{figure}[h]
\centering
\begin{minipage}[b]{0.32\linewidth}
\centering
\includegraphics[keepaspectratio,scale=0.4]{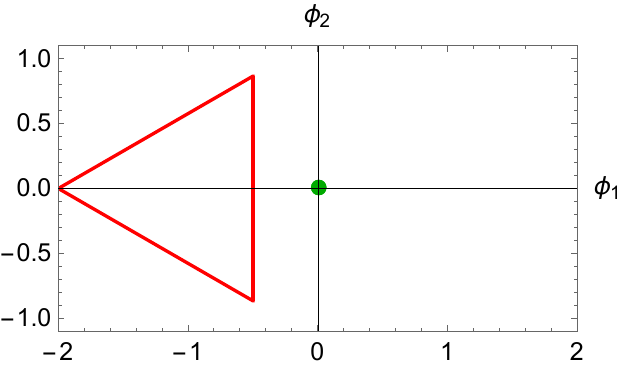}
\end{minipage}
\begin{minipage}[b]{0.32\linewidth}
\centering
\includegraphics[keepaspectratio,scale=0.4]{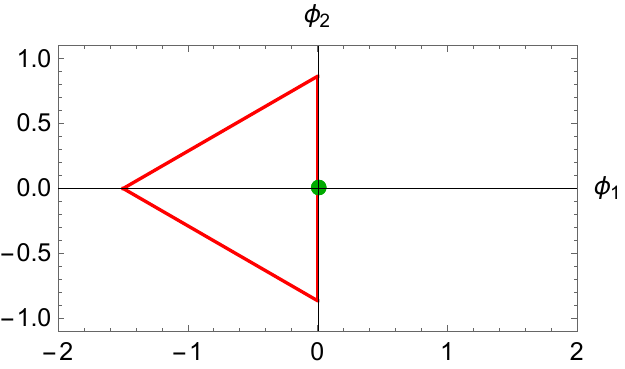}
\end{minipage}
\begin{minipage}[b]{0.32\linewidth}
\centering
\includegraphics[keepaspectratio,scale=0.4]{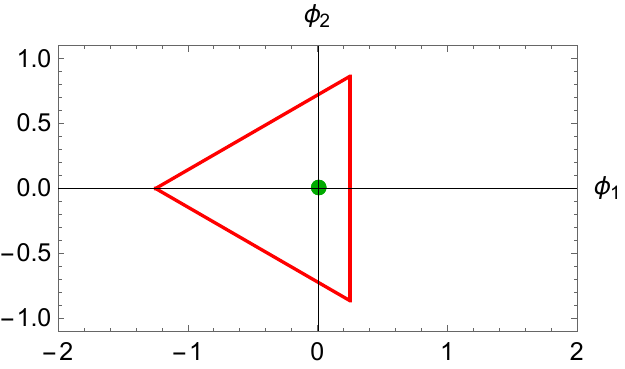}
\end{minipage}
\begin{minipage}[b]{0.32\linewidth}
\centering
\includegraphics[keepaspectratio,scale=0.3]{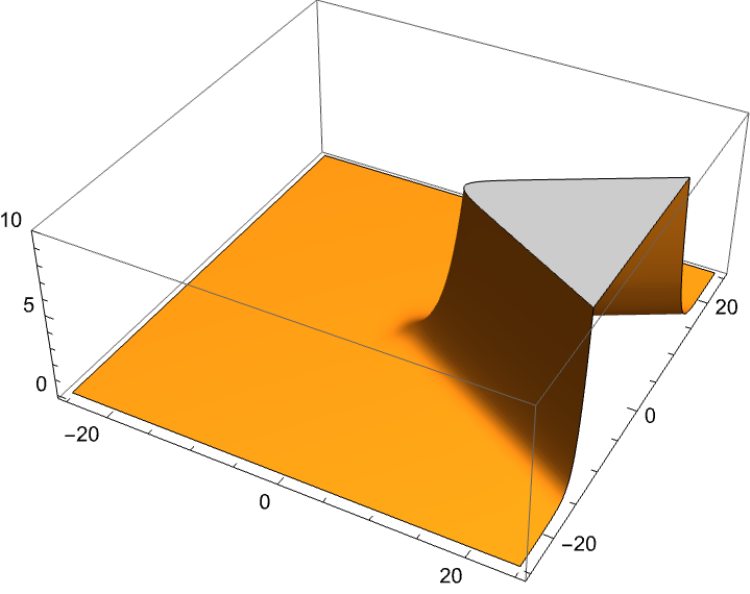}
\subcaption{$(-m,0)$}
\vspace*{.2cm}
\end{minipage}
\begin{minipage}[b]{0.32\linewidth}
\centering
\includegraphics[keepaspectratio,scale=0.3]{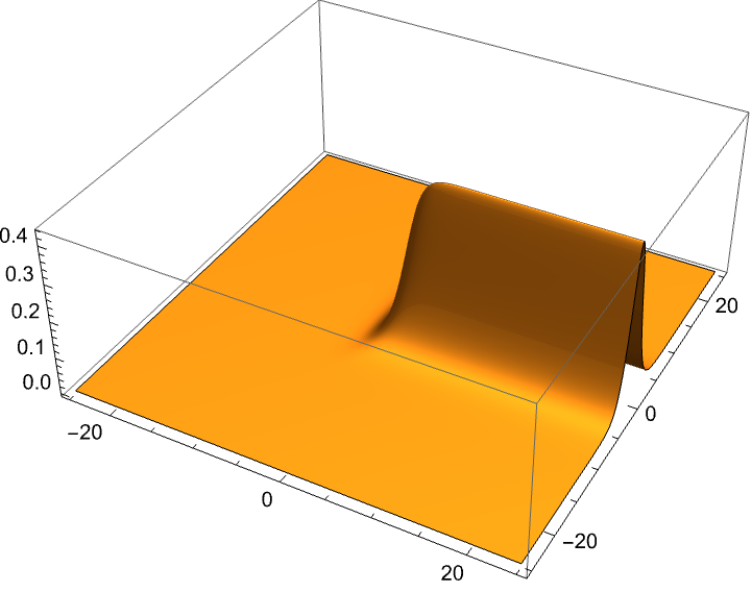}
\subcaption{$(-m/2,0)$}
\vspace*{.2cm}
\end{minipage}
\begin{minipage}[b]{0.32\linewidth}
\centering
\includegraphics[keepaspectratio,scale=0.3]{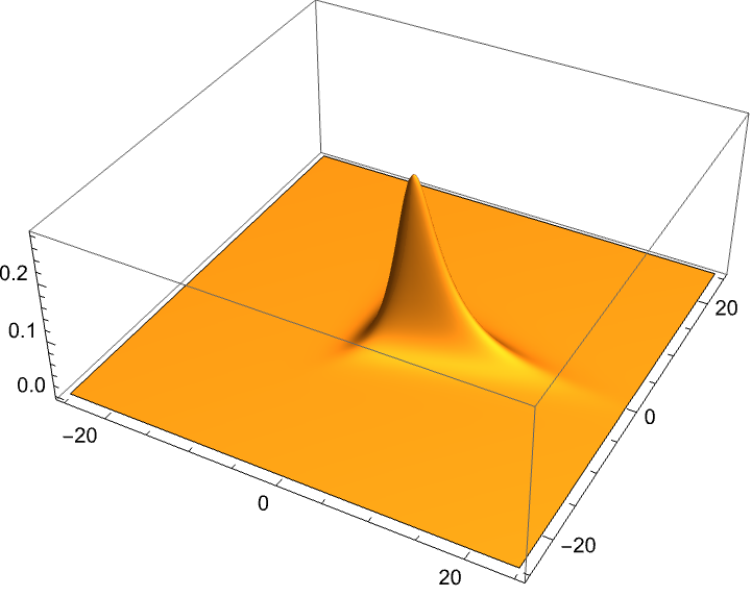}
\subcaption{$(-m/4,0)$}
\vspace*{.2cm}
\end{minipage}
\begin{minipage}[b]{0.32\linewidth}
\centering
\includegraphics[keepaspectratio,scale=0.4]{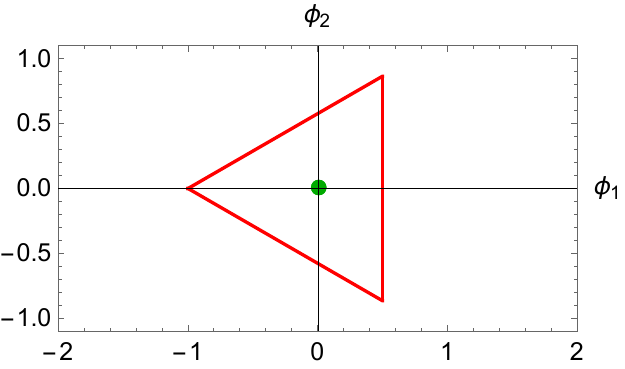}
\end{minipage}
\begin{minipage}[b]{0.32\linewidth}
\centering
\includegraphics[keepaspectratio,scale=0.4]{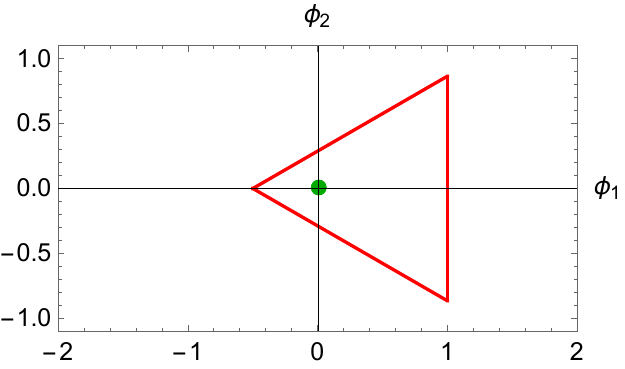}
\end{minipage}
\begin{minipage}[b]{0.32\linewidth}
\centering
\includegraphics[keepaspectratio,scale=0.4]{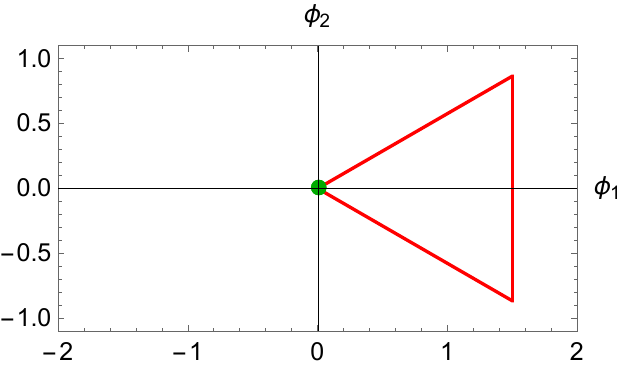}
\end{minipage}
\begin{minipage}[b]{0.32\linewidth}
\centering
\includegraphics[keepaspectratio,scale=0.3]{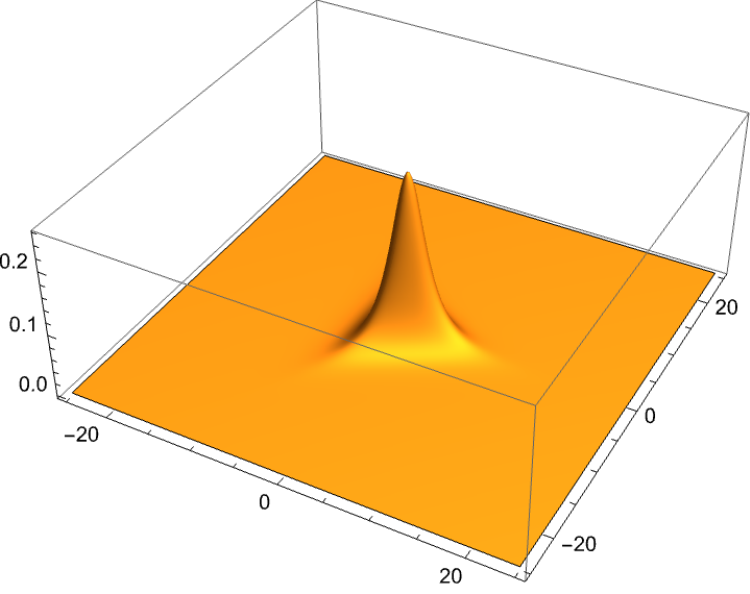}
\subcaption{$(0,0)$}
\vspace*{.2cm}
\end{minipage}
\begin{minipage}[b]{0.32\linewidth}
\centering
\includegraphics[keepaspectratio,scale=0.3]{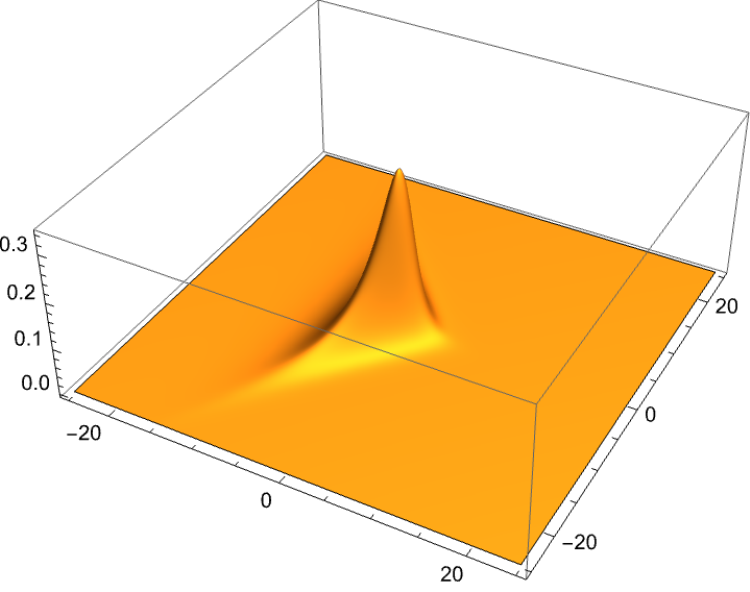}
\subcaption{$(m/2,0)$}
\vspace*{.2cm}
\end{minipage}
\begin{minipage}[b]{0.32\linewidth}
\centering
\includegraphics[keepaspectratio,scale=0.3]{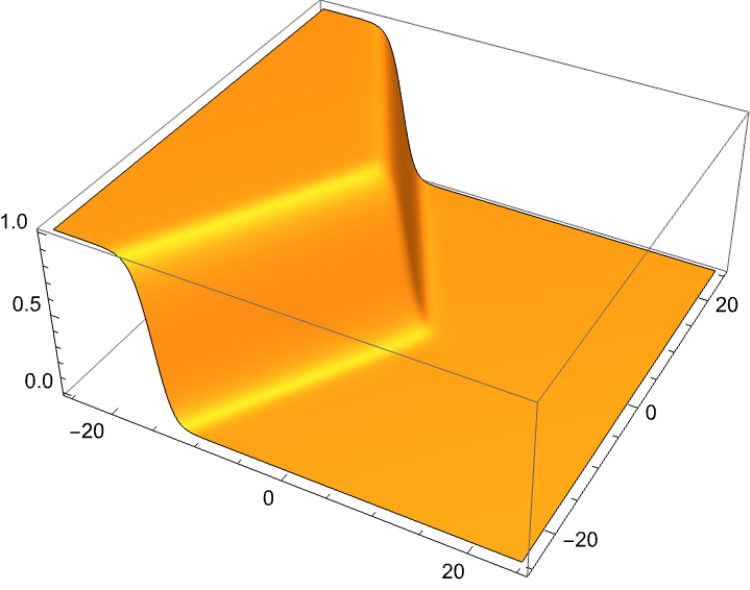}
\subcaption{$(m,0)$}
\vspace*{.2cm}
\end{minipage}
\begin{minipage}[b]{0.32\linewidth}
\centering
\includegraphics[keepaspectratio,scale=0.4]{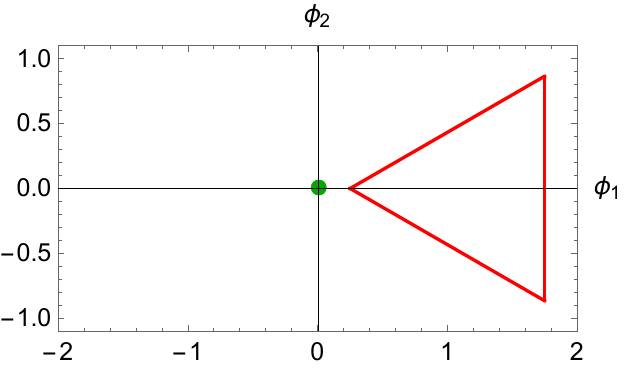}
\end{minipage}
\begin{minipage}[b]{0.32\linewidth}
\centering
\phantom{a}
\vspace*{.2cm}
\end{minipage}
\begin{minipage}[b]{0.32\linewidth}
\centering
\phantom{a}
\vspace*{.2cm}
\end{minipage}
\begin{minipage}[b]{0.32\linewidth}
\centering
\includegraphics[keepaspectratio,scale=0.3]{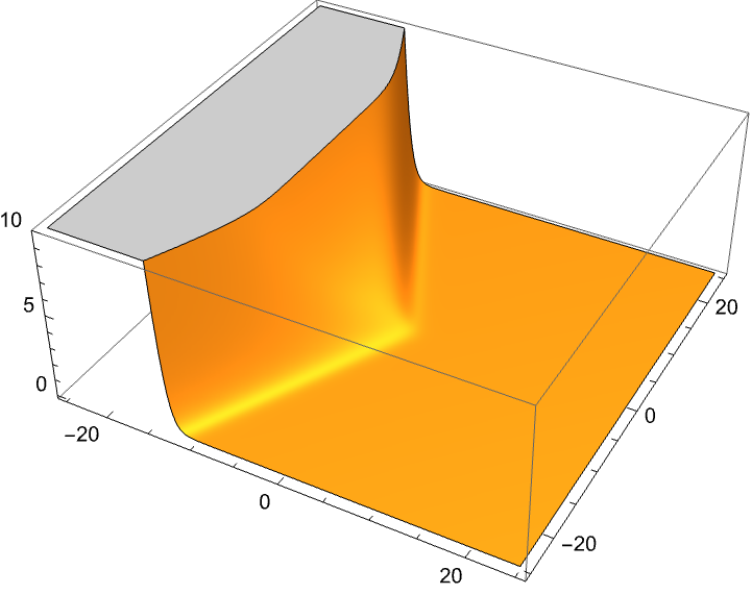}
\subcaption{$(5m/4,0)$}
\vspace*{.2cm}
\end{minipage}
\begin{minipage}[b]{0.32\linewidth}
\centering
\phantom{a}
\vspace*{.2cm}
\end{minipage}
\begin{minipage}[b]{0.32\linewidth}
\centering
\phantom{a}
\vspace*{.2cm}
\end{minipage}
\caption{The fermion zero mode functions for the equilateral triangular case shown in Fig.~\ref{fig:susy_dwj_n3_1}(a) with the mass shift $\vec m_0$. The subcaptions show the value of $\vec m_0$. The panel (d) is unshifted one which is identical to Fig.~\ref{fig:fermion_susy_nf3}(a).}
\label{fig:shift}
\end{figure}
In Fig.~\ref{fig:shift} we show how the fermion zero mode function changes in accordance with the mass shift $\vec m_0$ for the equilateral triangular case of Fig.~\ref{fig:susy_dwj_n3_1}(a). When $|\vec m_0|$ is small such that the shifted triangle includes $\vec\varphi=0$ on its inside, the fermion zero mode localized on the string, see Fig.~\ref{fig:shift}(c), (d), and (e).
There are two special cases: One is that $\vec\varphi=0$ occurs on an edge of the triangle. The zero mode function spreads over the corresponding domain wall as Fig.~\ref{fig:shift}(b). 
This implies that the zero mode freely propagates along the semi-infinite domain wall.
The other is that $\vec\varphi$ vanishes at a vertex as Fig.~\ref{fig:shift}(f). The zero mode function has a flat support expanded over the corresponding vacuum region. Namely, the fermion zero mode freely propagates only in the corresponding vacuum.
When the triangle does not include the zero as Fig.~\ref{fig:shift}(a) and (g), the fermion zero mode function diverges exponentially fast. So they are unphysical states. Hence, the fermion mass spectrum does not have zero modes and includes massive modes only with a mass gap which is of order of the distance between the vanishing point of $\vec\varphi$ and the triangle in the $\varphi_1\varphi_2$ plane.

\subsubsection{Fully analytic fermion zero mode functions at $g^2 \to \infty$}

As we explained in Sec.~\ref{sec:strong}, the largest advantage in the strong gauge coupling limit $g^2\to\infty$ is that all the analytic solutions $\Omega$ to the master equation are available as given in Eq.~(\ref{eq:Omega_infty}).
Together with our finding in Eq.~(\ref{eq:susy_fermion_zero_mode}), all the fermion zero mode functions can also be given fully analytically as
\be
f_1^{(0)} = \Omega_0^{-\frac{h}{2}} = \left(\sum_{A=1}^{N_{\rm F}} |H_0^A|^2 e^{2 \vec m_A\cdot \vec x}\right)^{-\frac{h}{2}}\,,\quad
f_2^{(0)} = 0\,.
\label{eq:fermion_strong}
\ee
Thus, by giving the moduli matrix $\{H_0^A\}$, one can immediately produce not only the bosonic background solution by Eq.~(\ref{eq:2.9}) but also the fermionic zero mode function by Eq.~(\ref{eq:fermion_strong}).

\begin{figure}[h]
\centering
\begin{minipage}[t]{0.32\linewidth}
\centering
\includegraphics[keepaspectratio,scale=0.35]{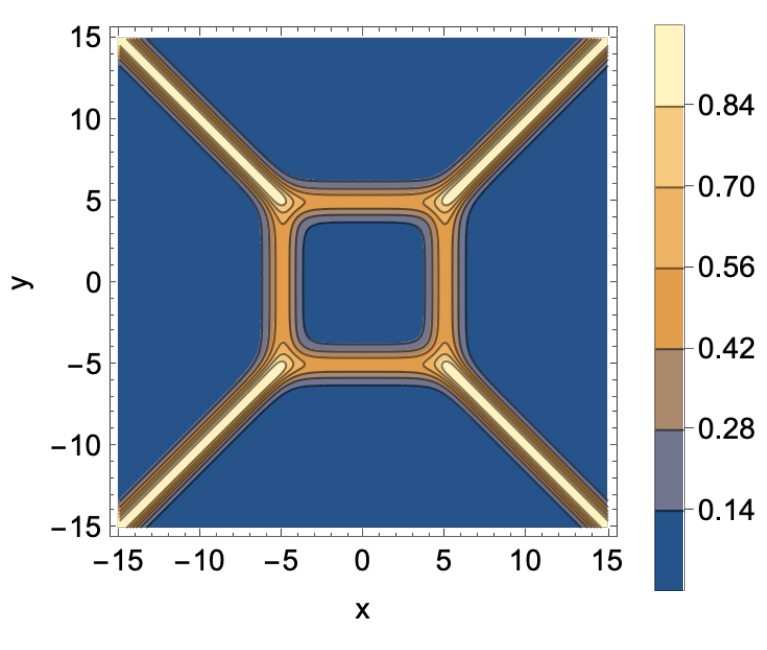}
\end{minipage}
\begin{minipage}[t]{0.32\linewidth}
\centering
\includegraphics[keepaspectratio,scale=0.35]{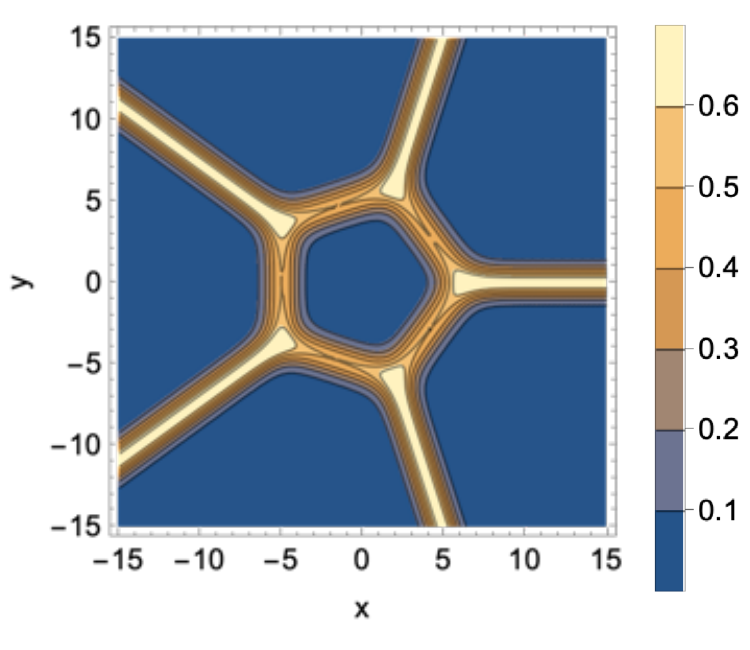}
\end{minipage}
\begin{minipage}[t]{0.32\linewidth}
\centering
\includegraphics[keepaspectratio,scale=0.35]{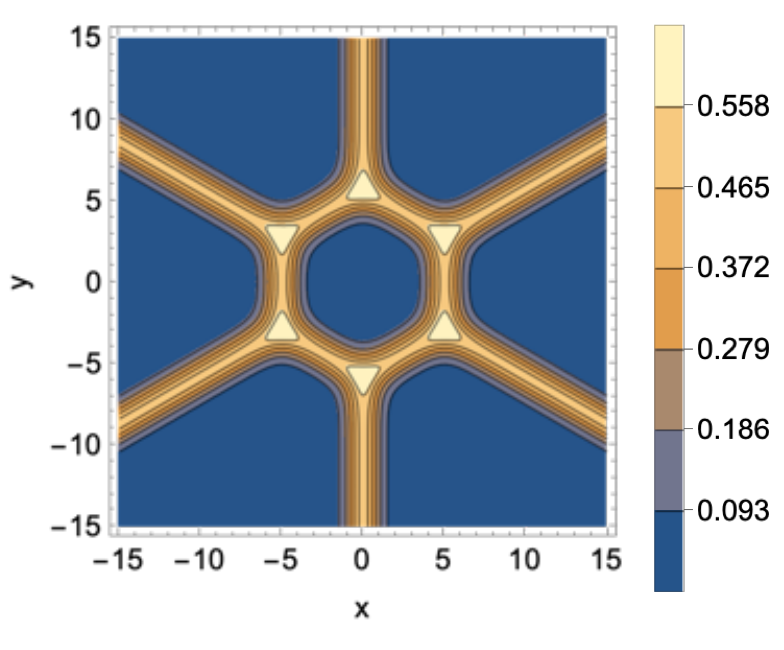}
\end{minipage}
\begin{minipage}[t]{0.32\linewidth}
\centering
\includegraphics[keepaspectratio,scale=0.35]{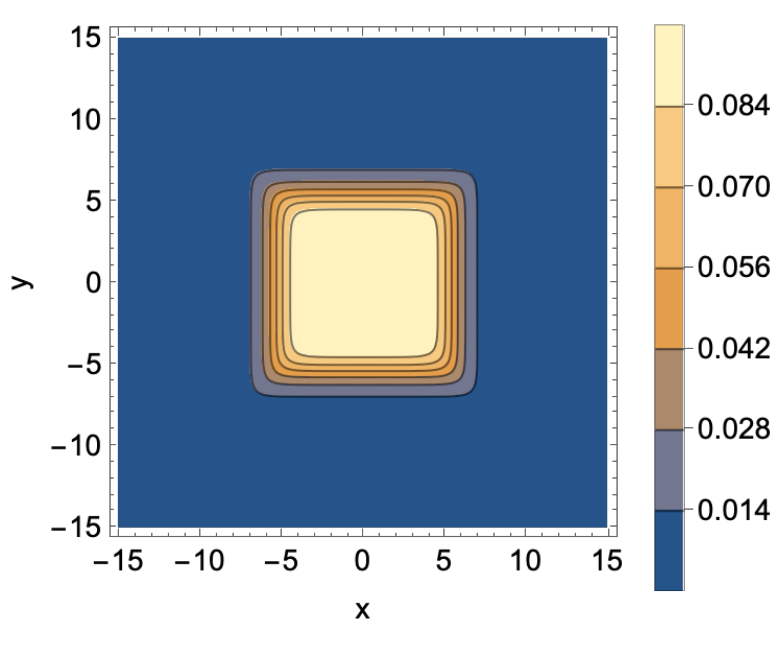}
\subcaption{$N_{\rm F}=5$}
\end{minipage}
\begin{minipage}[t]{0.32\linewidth}
\centering
\includegraphics[keepaspectratio,scale=0.35]{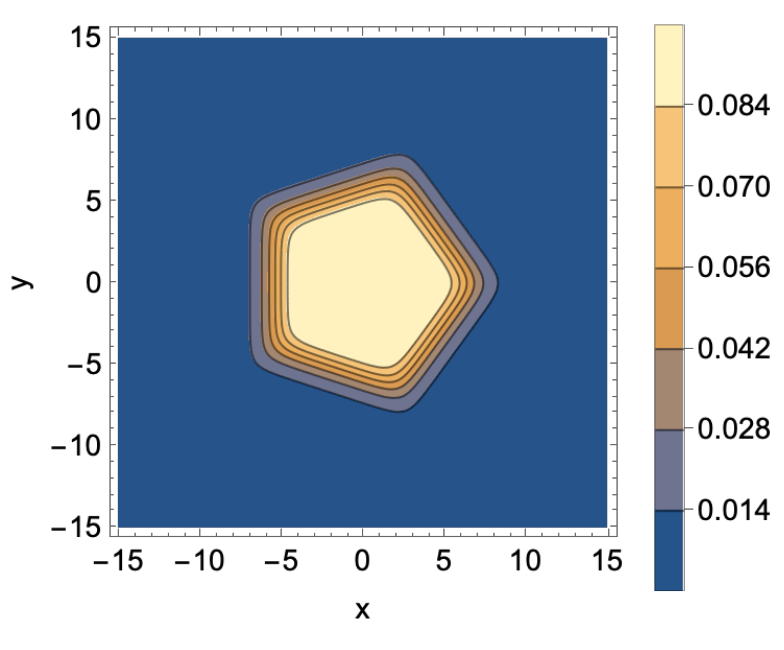}
\subcaption{$N_{\rm F}=6$}
\end{minipage}
\begin{minipage}[t]{0.32\linewidth}
\centering
\includegraphics[keepaspectratio,scale=0.35]{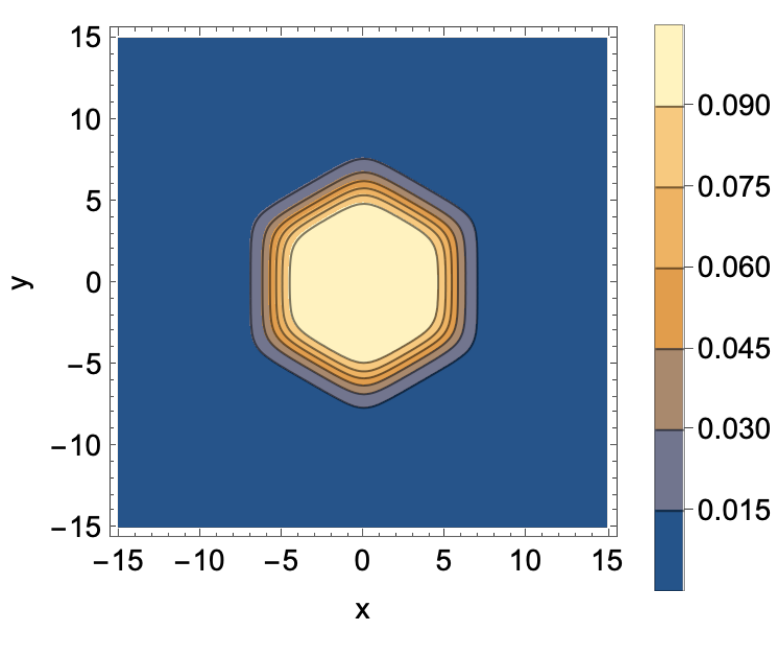}
\subcaption{$N_{\rm F}=7$}
\end{minipage}
\caption{The topological charge densities ${\cal Z}_1 + {\cal Z}_2$ (top row) and the fermion zero mode functions (bottom row) for $N_{\rm F} = 5,6,7$ in the strong gauge coupling limit $g^2 \to \infty$. The fermion zero mode is localized in the inner vacuum surrounded by the domain wall loop. }
\label{fig:strong}
\end{figure}

In order to demonstrate tractability of the exact formulae given in Eqs.~(\ref{eq:fermion_strong}) and  (\ref{eq:fermion_strong}), we show three examples for $N_{\rm F} = 5,6,7$ in Fig.~\ref{fig:strong}.

\section{Effects of fermion bulk mass in axion-like models}
\label{sec:bulkmass}

Let us add a fermionic bulk mass $\mu$ to the fermionic Lagrangian as
\be
\mathcal{L}_\mathrm{F} &=& i \bar \chi \bar \sigma^\mu D_\mu \chi + i \xi \sigma^\mu D_\mu \bar \xi - h \varphi \chi \xi - h \varphi^*  \bar\chi \bar\xi
+ \mu \chi\xi + \mu^*\bar\chi\bar\xi \nonumber\\
&=& i \bar\psi \gamma^\mu D_\mu \psi- h \bar\psi\left(\varphi_1 + i \gamma_5 \varphi_2\right)\psi  +  \bar\psi (\mu_1 + i \gamma_5\mu_2) \psi  \,,
\label{eq:L_fermion_2}
\ee
where $\mu$ is a complex as $\mu = \mu_1 + i \mu_2$, and we will also use the vector notation $\vec \mu = (\mu_1,\mu_2)$.
The axial $U(1)_{\rm A}$ symmetry (\ref{eq:U(1)A}) is explicitly broken independently of the bosonic potential in Eq.~(\ref{eq:Lag_axion}). 
Thus, when we combine the axion Lagrangian (\ref{eq:Lag_axion}) and the fermion Lagrangian (\ref{eq:L_fermion_2}), the phase $\zeta = {\rm arg}\,\varphi$ cannot be regarded as the QCD axion. However, there are still rooms for $\zeta$ to play some roles as axion-like particles which has been intensively studied in high energy astrophysics.

Note that the fermionic bulk mass in Eq.~(\ref{eq:L_fermion_2}) can be formally considered as just a constant shift
\be
\varphi \to \varphi - \frac{\mu}{h}\,.
\label{eq:phi_shift}
\ee
In the SUSY-like model the constant shift of $\varphi$ is already installed in the bosonic Lagrangian (\ref{eq:susy_pot}) 
as the constant shift in the bosonic mass vector $\vec m_A \to \vec m_A + \vec \mu/h$. 
Since we have already examined the effects of the bosonic mass shift in the SUSY-like models in Sec.~\ref{sec:bosonic_mass_susy} in details, we do not repeat it and we will concentrate on the axion-like model below.

\subsection{$N=0$:  Analytic zero mode function of axion-like strings}
\label{sec:delocalization}

Let us study the effect of fermionic bulk mass in the axion model with $N=0$. Namely, the $U(1)_{\rm A}$ is explicitly broken only by the fermion bulk mass. This was recently investigated in Ref.~\cite{Bagherian:2023jxy}. The following results in this section partially overlap with \cite{Bagherian:2023jxy} but also partially improve it.
As mentioned above, the $\mu \neq 0$ in Eq.~(\ref{eq:L_fermion_2}) is formally mere constant shift for $\varphi$ in the Yukawa term.
Therefore, all the calculations for deriving the formula for $\mu=0$ in Sec.~\ref{sec:DWSF} obviously remain correct without any changes except for the following modification to the potential $U$ given in Eq.~(\ref{eq:U_axion}) as
\be
U(x) &=&  v \int^\rho F(\sqrt{\lambda}v\rho')\, d\rho' - \frac{\vec\mu\cdot \vec x}{h}\,,\\
\varphi_a(x) &=& \p_a U(x) = v F(\sqrt{\lambda}v \rho) \frac{x_a}{\rho} - \frac{\mu_a}{h}\,.
\label{eq:varphi_dU}
\ee
By substituting this into Eq.~(\ref{eq:f10}), we find the fermionic zero mode function under the presence of the fermionic bulk mass
\be
f_1^{(0)}(x) &=& A \exp\left( -h v \int^\rho F(\sqrt{\lambda} v \rho')d\rho' + \vec \mu\cdot \vec x\right)\nonumber\\
&=& A \exp \left\{ - \frac{h}{\sqrt{\lambda}} \left(\int^{\tilde\rho} F(\tilde \rho')\, d\tilde\rho' - \frac{\vec \mu}{hv} \cdot \vec{\tilde x} \right)\right\}\,,
\label{eq:fermion_axion_string_shift}
\ee
with $f_2^{(0)} = 0$. 
We emphasize that this is the analytic solution. Therefore, the eigenvalue is exactly zero, and the existence of the zero mode is proved by this analytic solution.

\begin{wrapfigure}{r}[0pt]{0.4\textwidth}
\vspace*{-.2cm}
 \centering
 \includegraphics[width=0.4\textwidth]{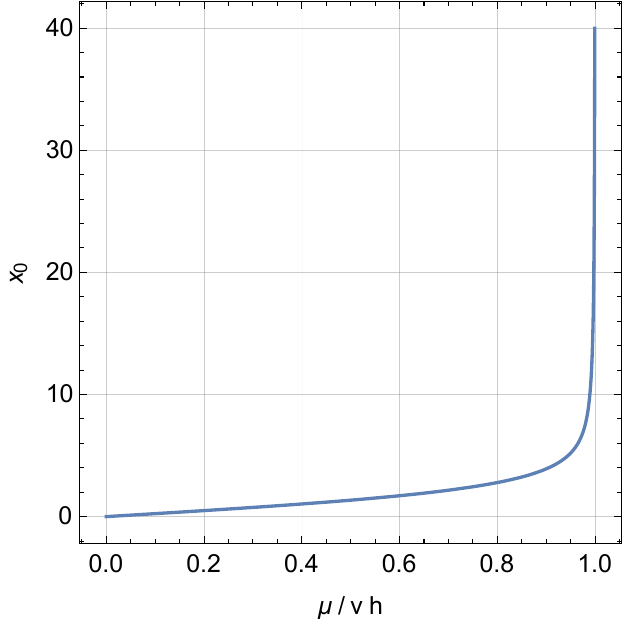}
 \caption{The peak coordinate $\tilde x_0$ of the fermionic zero mode as the function of $\mu_1/vh$.}
 \label{fig:x0}
\end{wrapfigure}
As we explained above, the localization point of the fermion zero mode corresponds to the point at which $\vec\varphi=0$ holds. Due to the axial symmetry of the background configuration, we can assume $\vec\mu = (\mu_1,0)$ and $\mu_1 \ge 0$ without loss of generality. Then the localization point is on the positive region of the $x$ axis, and from Eq.~(\ref{eq:varphi_dU}) its coordinate $(x,y) = (x_0,0)$ is determined by
\be
F(\tilde x_0) = \frac{\mu_1}{vh}\,,
\ee
with $\tilde x_0 = \sqrt{\lambda} v x_0$.
We show the relation between $\tilde x_0$ and $\mu_1/vh$ in Fig.~\ref{fig:x0}.
Since $F(\tilde \rho)$ is the monotonically increasing function between $F(0) = 0$ and $F(\infty) = 1$,
the fermion zero mode is normalizable only when
\be
0 \le \frac{\mu_1}{vh} \le 1\,.
\ee

This can be also understood by means of  the position-dependent fermion mass
\be
m_{\rm f}(x) = h \varphi(x) - \mu\,.
\ee 
By regarding the axisymmetric string configuration $\varphi(x)$ as the map from the real plane $\mathbb{R}^2$ to the internal space $\varphi \in \mathbb{C}$, the image of the map is onto the disk of radius $v$ centered at the origin as shown in Fig.~\ref{fig:axion_N=0_BG}(b). Therefore, $m_{\rm f}(x)$ can also be thought of as the map from $\mathbb{R}^2$ to $\mathbb{C}$ whose image is the disk of radius $hv$ centered at $m_{\rm f} = -\mu$. Hence in order for the $m_{\rm f}$ vanishing point to exist, the fermion  bulk mass $\mu$ should satisfy 
\be
|\mu| \le hv\,.
\label{eq:cond_normalizable}
\ee

\begin{figure}[htbp]
\centering
\begin{minipage}[b]{0.45\linewidth}
\centering
\includegraphics[keepaspectratio,scale=0.45]{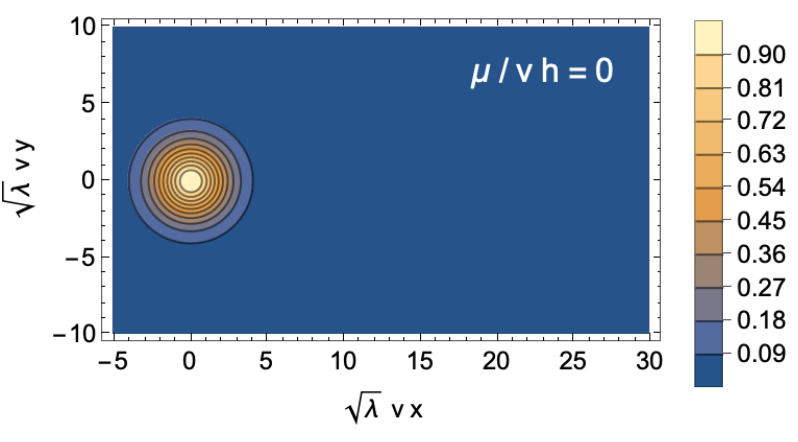}
\vspace*{.5cm}
\end{minipage}
\begin{minipage}[b]{0.45\linewidth}
\centering
\includegraphics[keepaspectratio,scale=0.45]{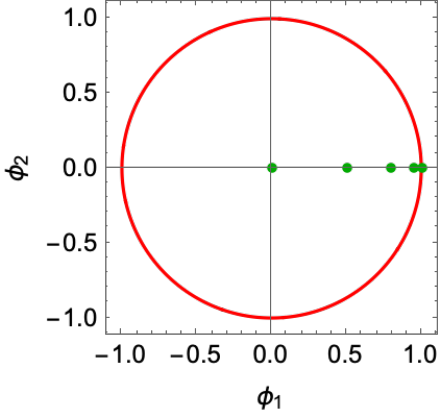}
\vspace*{.5cm}
\end{minipage}
\begin{minipage}[b]{0.45\linewidth}
\centering
\includegraphics[keepaspectratio,scale=0.45]{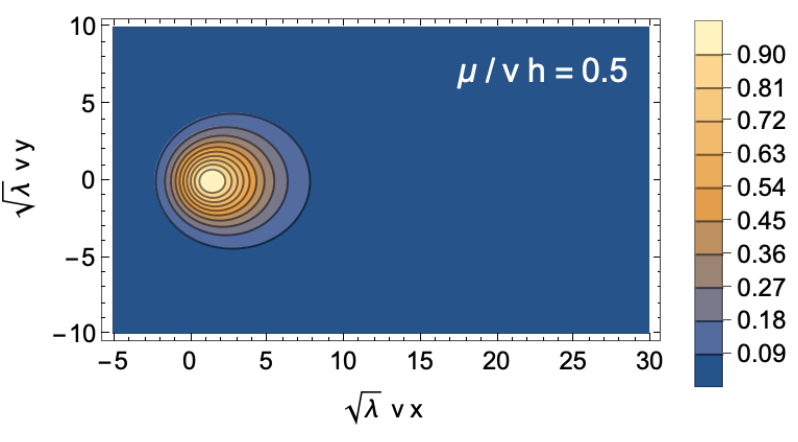}
\vspace*{.5cm}
\end{minipage}
\begin{minipage}[b]{0.45\linewidth}
\centering
\includegraphics[keepaspectratio,scale=0.45]{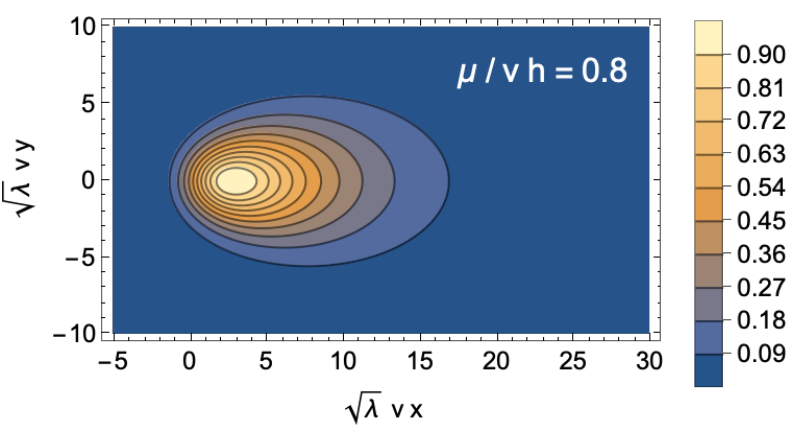}
\vspace*{.5cm}
\end{minipage}
\begin{minipage}[b]{0.45\linewidth}
\centering
\includegraphics[keepaspectratio,scale=0.45]{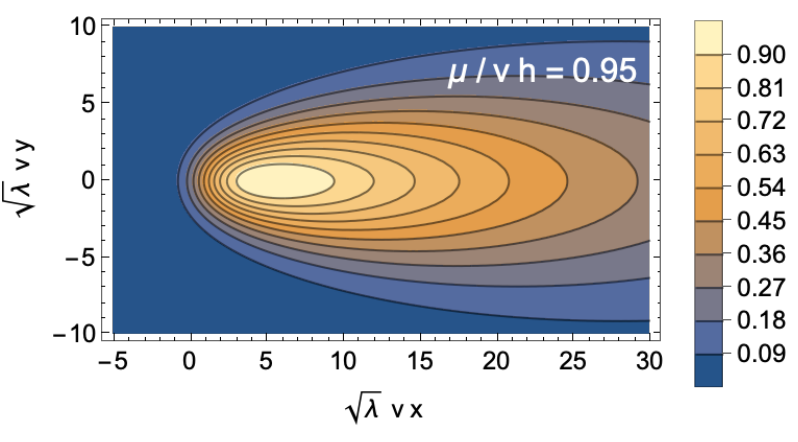}
\vspace*{.5cm}
\end{minipage}
\begin{minipage}[b]{0.45\linewidth}
\centering
\includegraphics[keepaspectratio,scale=0.45]{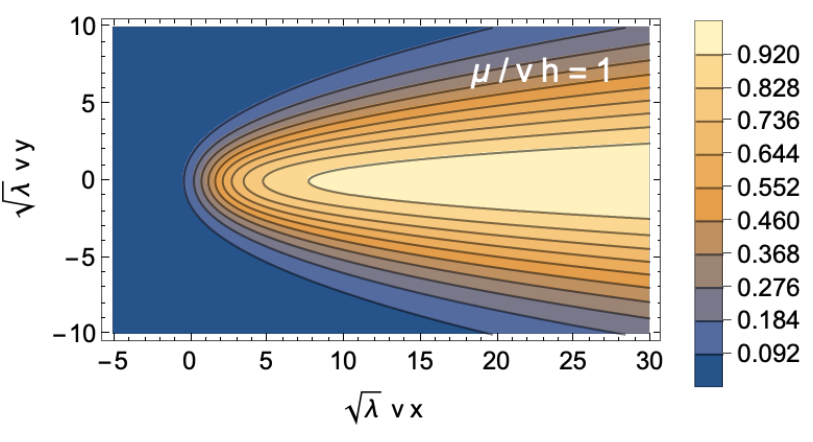}
\vspace*{.5cm}
\end{minipage}
\caption{The analytic fermion zero mode functions $f_1^{(0)}$ given in Eq.~(\ref{eq:fermion_axion_string_shift}) for the axially symmetric axion string given in Fig.~\ref{fig:axion_N=0_BG}. We choose $\mu/vh = \{0,\, 0.5,\, 0.8,\, 0.95,\, 1\}$ and we set $h = \sqrt{\lambda}$.}
\label{fig:axion_string_shift}
\end{figure}

We show the analytic zero mode functions for $\mu/vh = \{0,\, 0.5,\, 0.8,\, 0.95,\, 1\}$ in Fig.~\ref{fig:axion_string_shift} and $h = \sqrt{\lambda}$.
In the top-right panel of Fig.~\ref{fig:axion_string_shift} the red circle is the image of the map $\phi(x) = \varphi(x)/v$ at the spatial infinity, and the green dots show the points at which $m_{\rm f}$ vanishes for the five different $\mu$'s given above. The mode functions are deformed according to $\mu$ and its peaks can be read from Fig.~\ref{fig:x0}.
One generic feature is that the closer the green dot gets to the red circle, the longer tail of the zero mode function is.
At the critical limit where the green dot is on the red circle, the tail becomes infinitely long.
If once the green dot gets out from the red circle violating the condition (\ref{eq:cond_normalizable}), the zero mode function diverges exponentially fast. Our analytic solutions agree well with the results obtained independently in \cite{Bagherian:2023jxy} where the Dirac equation is solved by numerical computations and an analytical approximation for an asymptotic region.
We would like to point out that our analytic formula is advantageous in obtaining the fermionic mode function especially for the cases with the green dots close to or even on the red circle in Fig.~\ref{fig:axion_string_shift}. In those cases, the fermion wave functions have very long tails (indeed it is infinitely long when the green dot is onto the red circle). Hence, we need to introduce a cut off to make the system finite when we solve the problem numerically.

Normalizability at the upper limit $|\mu| = vh$ is marginal since the zero mode function is asymptotically flat, see the bottom-right panel of Fig.~\ref{fig:axion_string_shift}. However the mode function does not diverge, and therefore we understand it as a physical zero mode. In contrast, if $|\mu| > vh$, the zero mode function diverges exponentially fast, so it is unphysical.

\subsection{$N\neq0$: Numerical zero mode function of axion-like string-wall composites}

Next we study effect of the fermion bulk mass to the string-wall composite in the axion-like model with $N \neq 0$. 
The $U(1)_{\rm A}$ is already explicitly broken in the bosonic Lagrangian. In addition, the fermionic bulk mass further breaks $U(1)_{\rm A}$ explicitly. 

As explained in Sec.~\ref{sec:axion_N>=1}, the background solutions for the axion model with $N\neq0$ in general do not satisfy the rotation-free condition $\epsilon_{ab}\p_a\varphi_b = 0$. Thus, we cannot make use of the analytic formula (\ref{eq:f10}), so we go back to the matrix equation (\ref{eq:DDm}) and solve it numerically.

Here we take $N=2$ case as an example. The background solution is the static composite soliton consisting of the string attached by the two domain walls. The numerical solution of the background configuration is given in Fig.~\ref{fig:axion_N=2_a=0.1_BG} and the fermionic zero mode without the bulk mass ($\mu=0$) is given in Fig.~\ref{fig:axion_zeromode_N=2}.

Since we do not have an analytic solution like the one in Eq.~(\ref{eq:f10}), it is hard to give the condition for the fermionic zero mode to exist in an analytic formula as Eq.~(\ref{eq:cond_normalizable}) for $N = 0$.
Therefore, we numerically examine the fermionic zero mode one by one with different $\mu$'s.
Note that since the $U(1)$ global symmetry is explicitly broken, we should turn on both real and imaginary component of the bulk mass $\mu = \mu_1 + i \mu_2$. Then we replace $\varphi$ by $\varphi - \mu/h$ in the Hermitian operator $D^\dag D$ in Eq.~(\ref{eq:DdD}), and numerically obtain its zero mode.
\begin{figure}[htbp]
\centering
\begin{minipage}[b]{0.32\linewidth}
\centering
\includegraphics[keepaspectratio,scale=0.35]{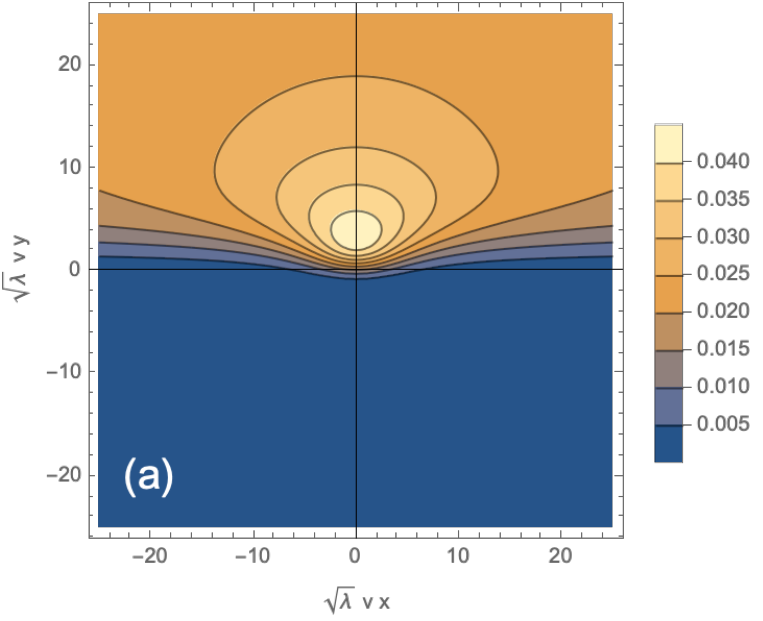}
\vspace*{.5cm}
\end{minipage}
\begin{minipage}[b]{0.32\linewidth}
\centering
\includegraphics[keepaspectratio,scale=0.35]{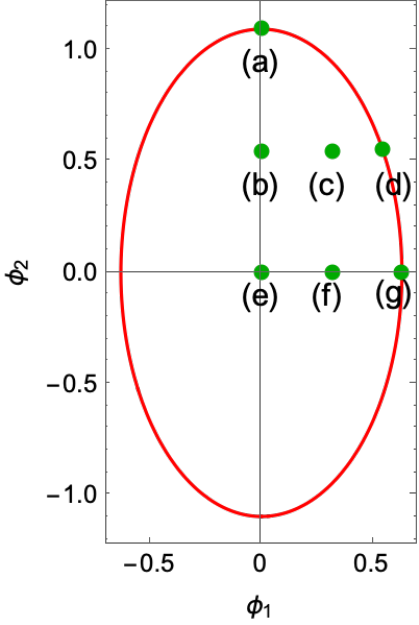}
\vspace*{.5cm}
\end{minipage}
\begin{minipage}[b]{0.32\linewidth}
\centering
\phantom{a}
\vspace*{.5cm}
\end{minipage}
\begin{minipage}[b]{0.32\linewidth}
\centering
\includegraphics[keepaspectratio,scale=0.35]{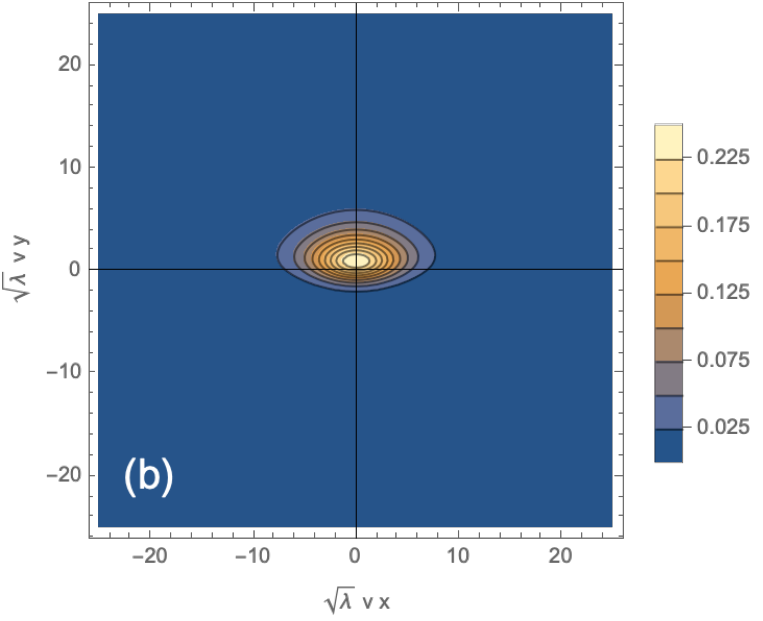}
\vspace*{.5cm}
\end{minipage}
\begin{minipage}[b]{0.32\linewidth}
\centering
\includegraphics[keepaspectratio,scale=0.35]{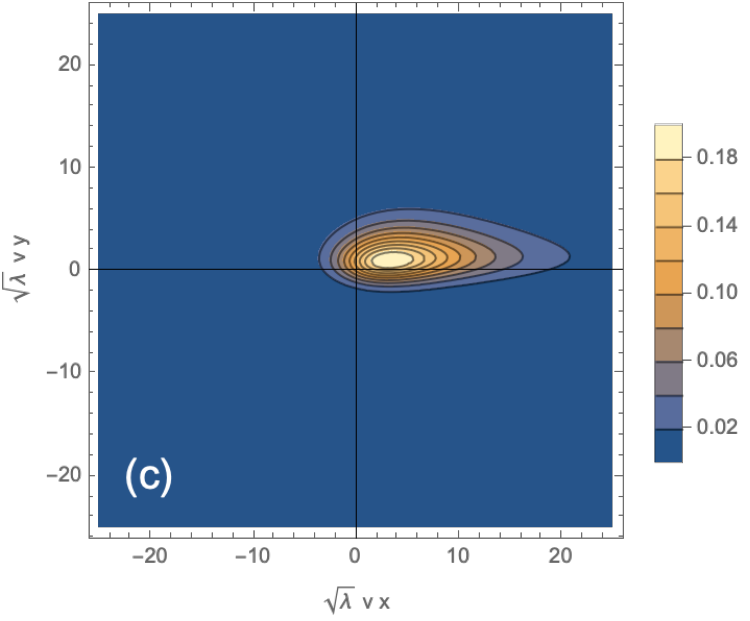}
\vspace*{.5cm}
\end{minipage}
\begin{minipage}[b]{0.32\linewidth}
\centering
\includegraphics[keepaspectratio,scale=0.35]{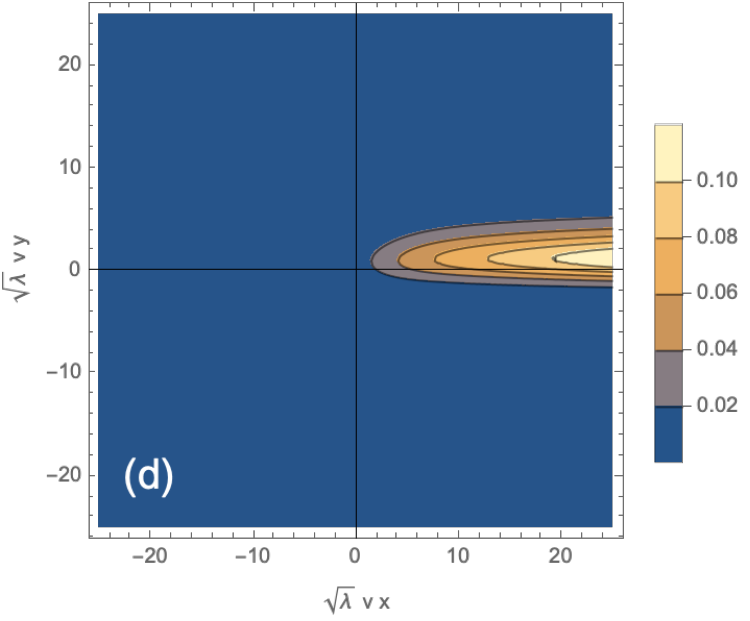}
\vspace*{.5cm}
\end{minipage}
\begin{minipage}[b]{0.32\linewidth}
\centering
\includegraphics[keepaspectratio,scale=0.35]{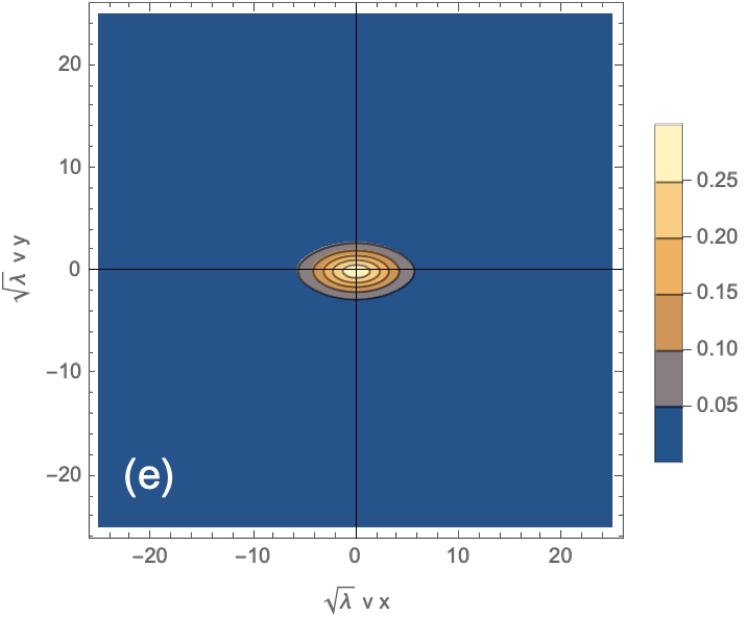}
\vspace*{.5cm}
\end{minipage}
\begin{minipage}[b]{0.32\linewidth}
\centering
\includegraphics[keepaspectratio,scale=0.35]{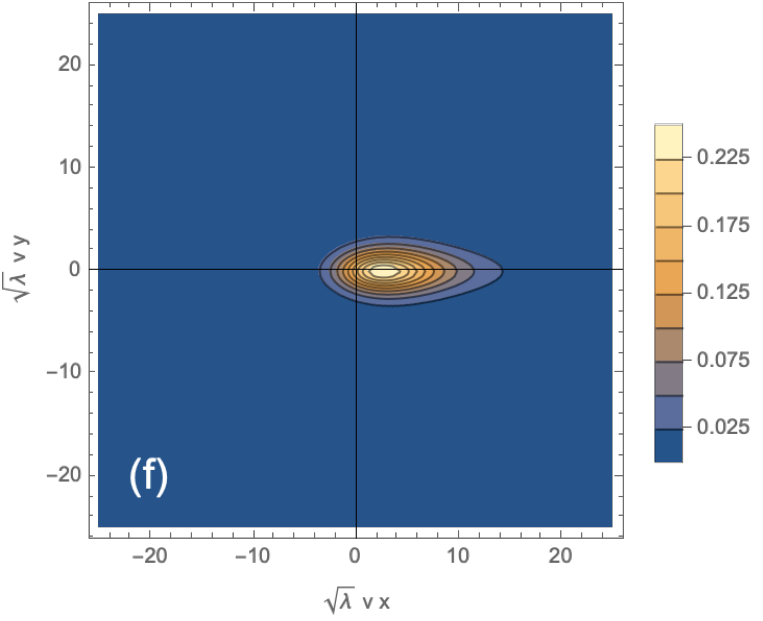}
\vspace*{.5cm}
\end{minipage}
\begin{minipage}[b]{0.32\linewidth}
\centering
\includegraphics[keepaspectratio,scale=0.35]{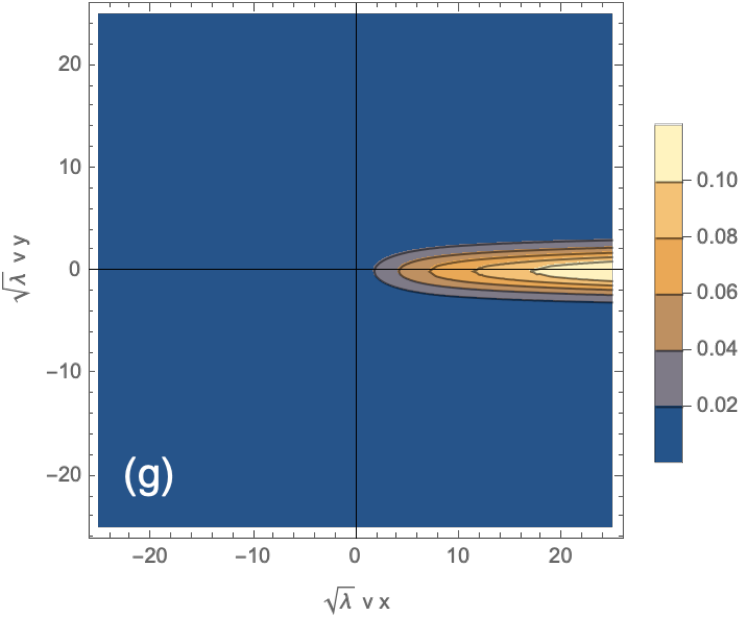}
\vspace*{.5cm}
\end{minipage}
\caption{The numerical fermion zero mode function $|f_1^{(0)}+if_2^{(0)}|$ for the $N=2$ axion string-wall composite given in Fig.~\ref{fig:axion_N=2_a=0.1_BG}. We put the fermion bulk mass $\mu$ shown by the green dots in the top-middle panel.
(a) the fermion zero mode spreads over whole upper half plane, (b,c,e,f) it localizes around the string center, and (d,g) it appears on the domain wall attaching to the string from the right hand side.}
\label{fig:shift_n2}
\end{figure}
The results are shown in Fig.~\ref{fig:shift_n2}.
As before we regard $\varphi$ as the map from $\mathbb{R}^2$ to $\mathbb{C}$. The image of $\mathbb{R}^2$ is the red oval and its interior shown in the top-middle panel of Fig.~\ref{fig:shift_n2}. As we showed in Fig.~\ref{fig:axion_N=2_a=0.1_BG}(b), the left and right half ovals of the string-wall composite well coincide with the images of the domain walls given by Eq.~(\ref{eq:wall_n2}). Therefore, the oval is expressed by
\be
\left(\frac{\phi_1}{\sqrt{1-12\alpha}}\right)^2 + \left(\frac{\phi_2}{\sqrt{1+4\alpha}}\right)^2 = 1\,,
\ee
namely the major diameter is $\sqrt{1+4\alpha} \to \sqrt{6/5}$ and the minor one is $\sqrt{1-12\alpha} \to \sqrt{2/5}$ for $\alpha = 1/20$.
We take seven different masses which are represented by the green dots (a)--(g) in the top-middle panel of Fig.~\ref{fig:shift_n2}
(The mass $(\mu_1,\mu_2)$ corresponds to the coordinate of the green dot).
The case (e) corresponds to $\mu=0$, which is identical to Fig.~\ref{fig:axion_zeromode_N=2}.
The fermion zero mode is localized precisely on the string center.
When we slightly shift the green dot from the origin as the cases (b), (c) and (f), the fermion zero mode shifts accordingly, and its peaks are slightly dislocated from the string center. We numerically confirm the peaks appear at the points where the fermion mass $m_{\rm f}$ vanishes.
The closer the green dot to the red oval, the wider/longer the spread of the zero mode function, see the case (c).
When the green dot is put at the northern vacuum, namely in the case (a), the fermion zero mode spreads over whole upper half plane.
It is not normalizable but the zero mode function does not diverge, and we interpret it as massless fermion living in the upper half of the $xy$ plane. In the case (g) where the green dot is put on the middle point on the oval boarder, the fermion zero mode function one-dimensionally extends over the domain wall to the right of the string. That is, the massless fermion lives on a half-line terminated at the origin by the string.
When the green dot is placed on the oval's boarder, except for the north and south poles which correspond to the vacua, the massless fermion appears on either the left or right domain wall as the case (d). When the green dot is placed outside the oval, the fermion zero mode diverges.
Therefore, it is numerically confirmed that the condition for the existence of a massless fermion is whether there exists a point where the fermion mass $m_{\rm f}$ vanishes, as expected from the analytically solvable cases.

\begin{figure}[hbtp]
\centering
\includegraphics[width=15cm]{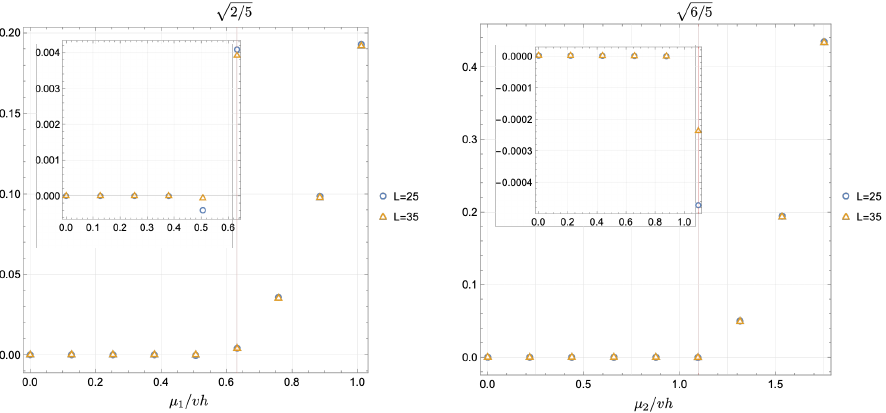}
\caption{The left panel shows the lowest eigenvalue of Eq.~(\ref{eq:DDm}) as the function of real bulk mass shift $\mu_1$ ($\mu_2 = 0$)
for the $N=2$ axion string-wall composite given in Fig.~\ref{fig:axion_N=2_a=0.1_BG}.
The right panel shows the same one but as the function of the imaginary bulk mass shift $\mu_2$ ($\mu_1 = 0$).
The blue circles are obtained by the numerical computation of the box size $50$ whereas the orange triangles are those in the box size $70$.
The red lines correspond to the critical values for $\alpha = 1/20$.}
\label{fig:fermion_mass_shift_n2_detail}
\end{figure}

Finally, we make a comment on a finite volume effect in our numerical analysis.
As can be seen in Fig.~\ref{fig:axion_N=2_a=0.1_BG}(c), the background configuration well fits into a square box with the side length 40 in the unit of $(\sqrt{\lambda} v)^{-1}$. Then we have performed two numerical calculations for solving Eq.~(\ref{eq:DDm}) by setting the computation box size 50 and 70. The results are shown in Fig.~\ref{fig:fermion_mass_shift_n2_detail}.
In the left panel we show the lowest eigenvalues for the different fermion bulk mass shift ($\mu = \mu_1 + i 0$) along the minor axes of the oval.
The lowest eigenvalues stay at very close to zero within the numerical accuracy for small $\mu_1$, but it is slightly off the zero as the shift $\mu_1$ approaching the critical value $\sqrt{2/5} vh$ from below. We compares two eigenvalues for the box size 50 and 70, and find the absolute value of the former is universally larger than that of the latter. This discrepancy clearly shows the finite volume effect.
Above the critical value (the red vertical line in Fig.~\ref{fig:fermion_mass_shift_n2_detail}) the lowest eigenvalues linearly grow up with $\mu_1$. The similar observation was obtained for the mass shift along the major axes of the oval ($\mu = 0 + i \mu_2$) as is shown in the right panel of Fig.~\ref{fig:fermion_mass_shift_n2_detail}. Again the lowest eigenvalues stay at very close to zero below the critical value $\sqrt{6/5} vh$,
and it blows up above the critical value. It seems the finite volume effect is slightly milder in this direction. 

For the cases of $N = 3$ and $4$, we obtained qualitatively similar results. Namely, the fermion zero mode is localized around the string center
if the fermion bulk mass is included in the interior of the rounded triangle or square shown in Fig.~\ref{fig:axion_N=3and4}(b) and (e).
The fermion zero mode appears in one of the vacua when the bulk mass is located precisely at one of the vertices of the triangle and square.
If the mass is on the rounded edge, the fermion zero mode appears on the corresponding domain wall.

\section{Fermionic superconducting current and anomaly inflow}
\label{sec:supercurrrent}

The massless fermion on the composite solitons is right-moving as we explained in Sec.~\ref{sec:chiral_fermion}.
If the fermion carries electromagnetic charge, and they are put in a static  electromagnetic field along the $z$-direction,
the superconducting current will build up in the composite soliton \`a la Witten \cite{WITTEN1985557}.

Let us first consider the axion model without bulk fermion mass ($\mu=0$) studied in Sec.~\ref{sec:axion_N>=1}.
The string is attached by $N$ domain walls.
The massless fermion is localized on the string, and its mode function is not axially symmetric but has a $N$-gonal shape as shown
in Figs.~\ref{fig:axion_zeromode_N=2}, \ref{fig:axion_zeromode_N=3and4} and \ref{fig:axion_zeromode_N=3_2}.
The electromagnetic current $J^\mu = e \bar\psi \gamma^\mu \psi$ with $\psi$ given in Eq.~(\ref{eq:zeromode}) reads
\be
J^0 = J^3 = 2e |f^{(0)}|^2\,,\quad J^1 = J^2 = 0\,,
\ee
implying indeed the electromagnetic current is confined inside the region where $|f^{(0)}|^2$ is nonzero.
Although the current is not axially symmetric, the zero mode is localized on the string. Therefore, the massless fermion
should be well described by a $1+1$ dimensional effective theory. The superconducting current builds up in the electromagnetic field is \cite{WITTEN1985557}
\be
J^3 = \frac{e^2Et}{2\pi}\,.
\ee

If $\psi$ is the only fermion field in the theory, the $1+1$ dimensional effective theory on the string is anomalous and the electric charge conservation low is violated. However, when an electric field is applied in the presence of a varying NG boson, an anomalous electric current is induced \cite{CALLAN1985427}
\be
J_\mu = \frac{e}{8\pi^2}\epsilon_{\mu\nu\rho\lambda}  {\cal J}^\nu F^{\lambda\rho}\,,\quad
{\cal J}^\nu = \frac{- i}{2} \frac{\varphi^*\p^\nu\varphi - \varphi\p^\nu\varphi^*}{|\varphi|^2}\,.
\label{eq:anomaly_inflow}
\ee
For $\varphi = \varphi(x,y)$ and $F^{30}=E$, this reduces to
\be
J_a = - \frac{e}{4\pi^2}\epsilon_{ab} {\cal J}^b E\,,\quad(a=1,2)\,.
\ee
For the usual axisymmetric strings, this current flows radially inward and exactly compensate the anomalous charge imbalance on the strings. 

\begin{figure}[htbp]
\centering
\begin{minipage}[b]{0.32\linewidth}
\centering
\includegraphics[keepaspectratio,scale=0.35]{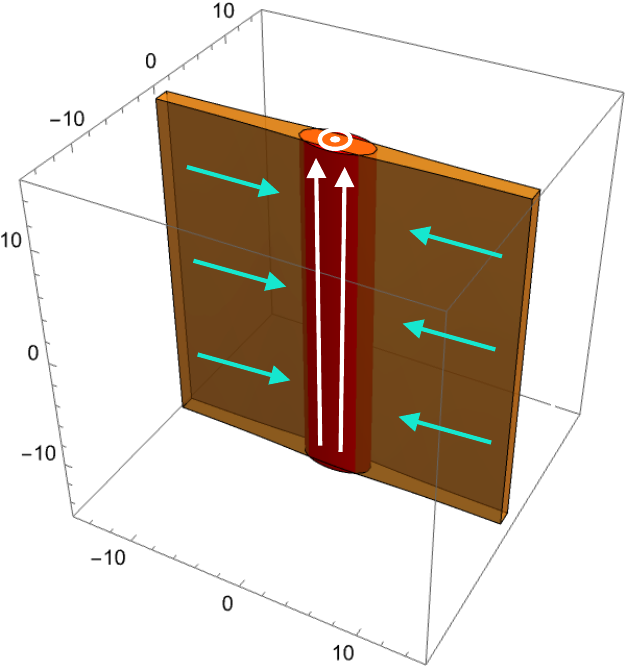}
\vspace*{.5cm}
\end{minipage}
\begin{minipage}[b]{0.32\linewidth}
\centering
\includegraphics[keepaspectratio,scale=0.35]{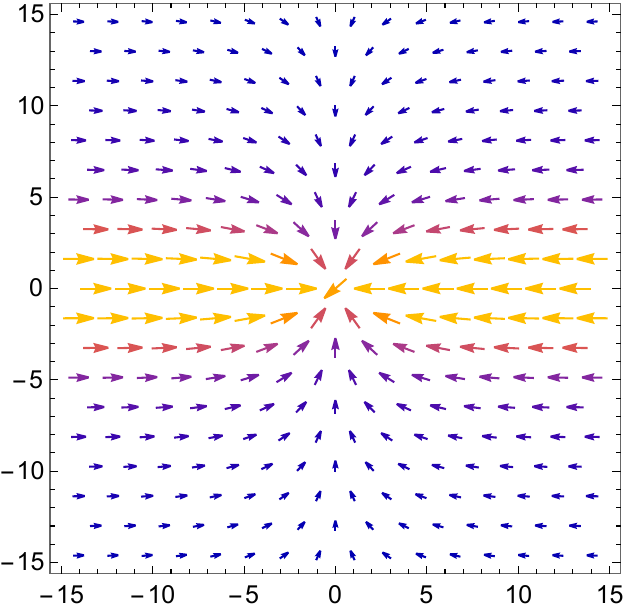}
\vspace*{.5cm}
\end{minipage}
\begin{minipage}[b]{0.32\linewidth}
\centering
\includegraphics[keepaspectratio,scale=0.35]{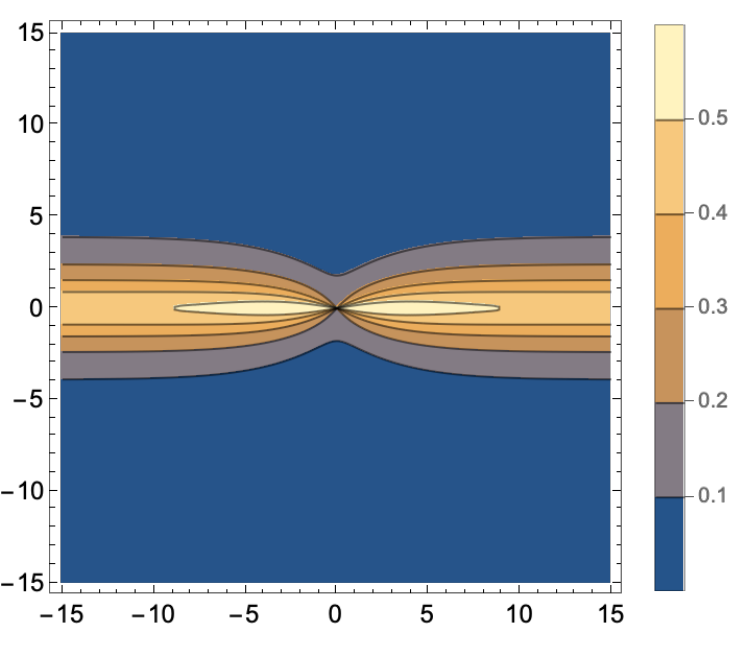}
\vspace*{.5cm}
\end{minipage}
\begin{minipage}[b]{0.32\linewidth}
\centering
\includegraphics[keepaspectratio,scale=0.35]{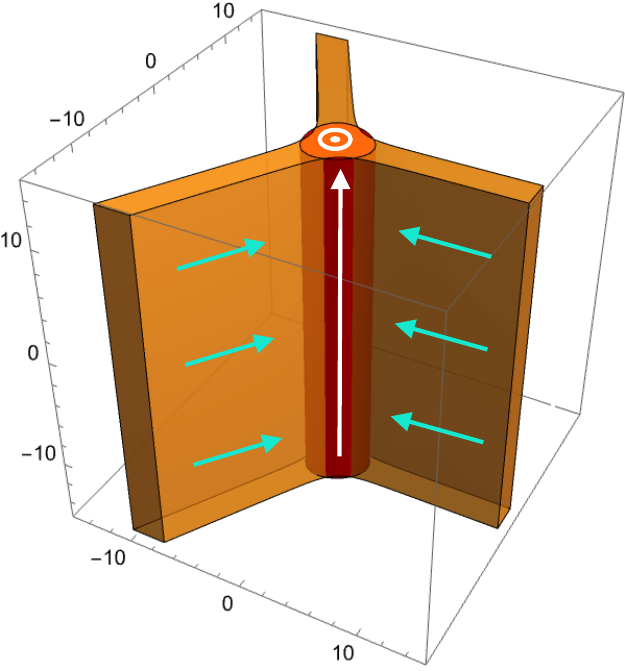}
\vspace*{.5cm}
\end{minipage}
\begin{minipage}[b]{0.32\linewidth}
\centering
\includegraphics[keepaspectratio,scale=0.35]{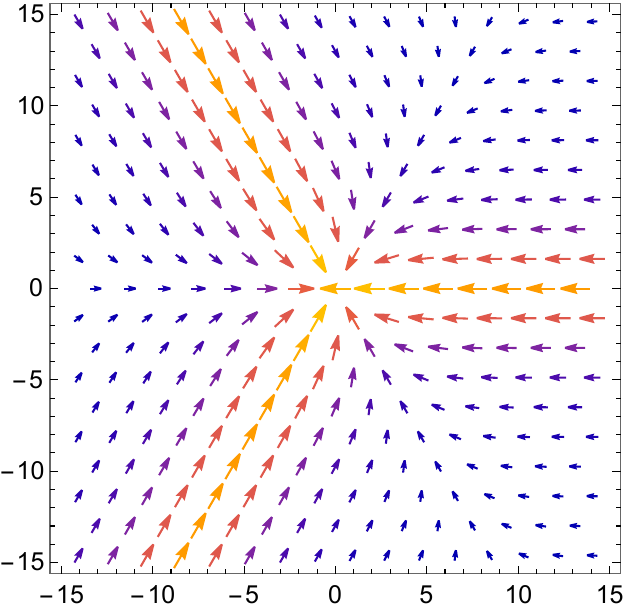}
\vspace*{.5cm}
\end{minipage}
\begin{minipage}[b]{0.32\linewidth}
\centering
\includegraphics[keepaspectratio,scale=0.35]{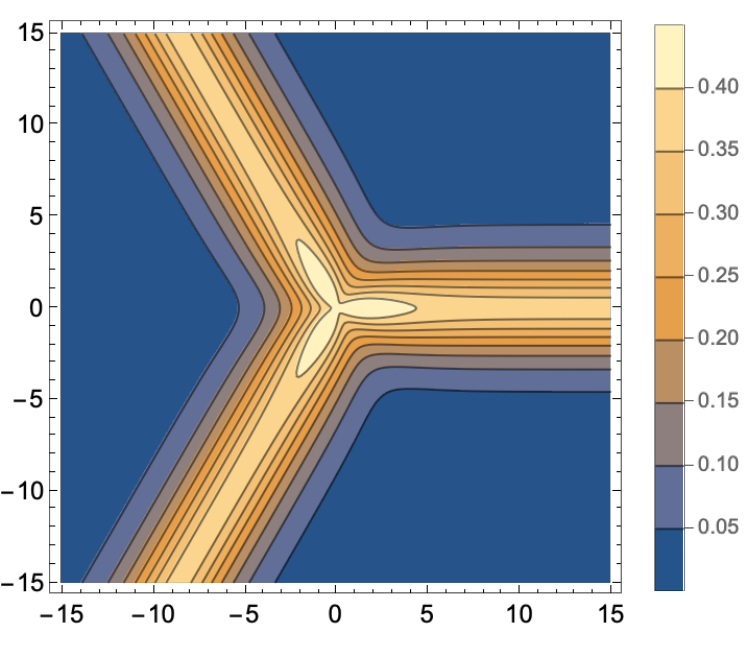}
\vspace*{.5cm}
\end{minipage}
\caption{The top (bottom) row corresponds to the $N=2$ ($N=3$) case given in Figs.~\ref{fig:axion_N=2_a=0.1_BG} and \ref{fig:axion_zeromode_N=2} (Figs.~\ref{fig:axion_N=3and4} and \ref{fig:axion_zeromode_N=3and4}). In the left-most panels, the orange surface shows an isosurface of energy density, and the red one corresponds to an isosurface of $|\psi_0|^2$. The middle and the right-most panels show the anomalous electric current $(J_1,J_2)$ and $|J_1+iJ_2|$, respectively.}
\label{fig:anomaly_inflow_n2}
\end{figure}

Under the presence of domain walls attaching to the strings, the NG boson is not radially symmetric, so that the anomalous currents flow in on the domain walls \cite{Vilenkin:2000jqa}.
We can now verify this by our analytic solution. The background solution and the fermion zero mode function for $N=2$ are shown in Figs.~\ref{fig:axion_N=2_a=0.1_BG} and \ref{fig:axion_zeromode_N=2}, respectively.
Since the massless fermion is localized in an elliptical shape on the string, the supercurrent flowing over the string also has an elliptical shape. The anomalous electric currents radially flow in on the two domain walls as shown in the panels at the top row of Fig.~\ref{fig:anomaly_inflow_n2}.
When $N > 2$, almost the same phenomenon occurs as when $N=2$. The only difference is the number of domain walls and the shape of the supercurrent on the string. For example, see the bottom row of Fig.~\ref{fig:anomaly_inflow_n2} for $N=3$.
The shape of supercurrent along $z$-axis is like a triangle prism and the anomalous inflows come from three domain walls.

\begin{figure}[htbp]
\centering
\begin{minipage}[b]{0.32\linewidth}
\centering
\includegraphics[keepaspectratio,scale=0.35]{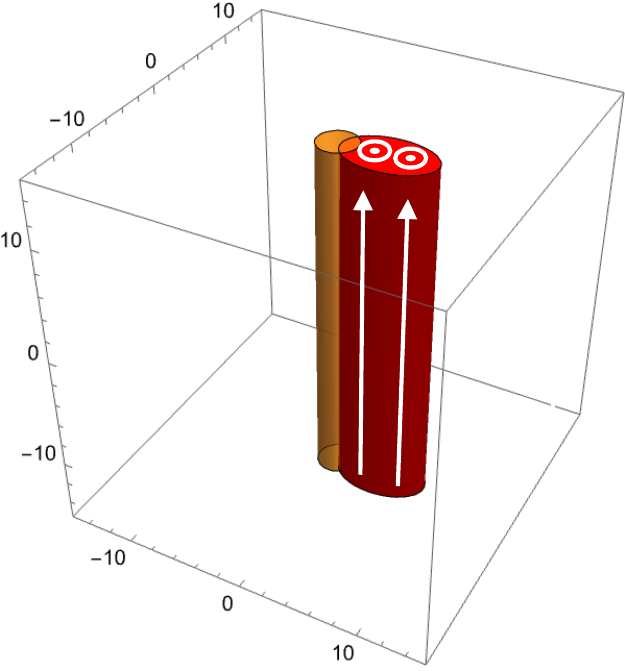}
\vspace*{.5cm}
\end{minipage}
\begin{minipage}[b]{0.32\linewidth}
\centering
\includegraphics[keepaspectratio,scale=0.35]{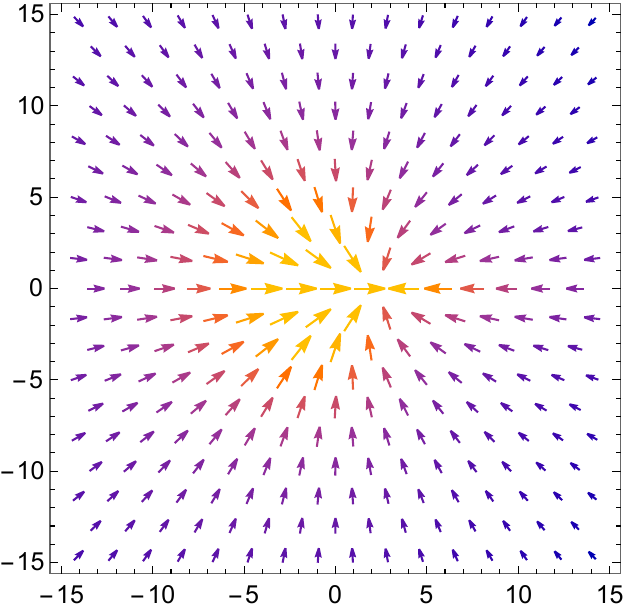}
\vspace*{.5cm}
\end{minipage}
\begin{minipage}[b]{0.32\linewidth}
\centering
\includegraphics[keepaspectratio,scale=0.35]{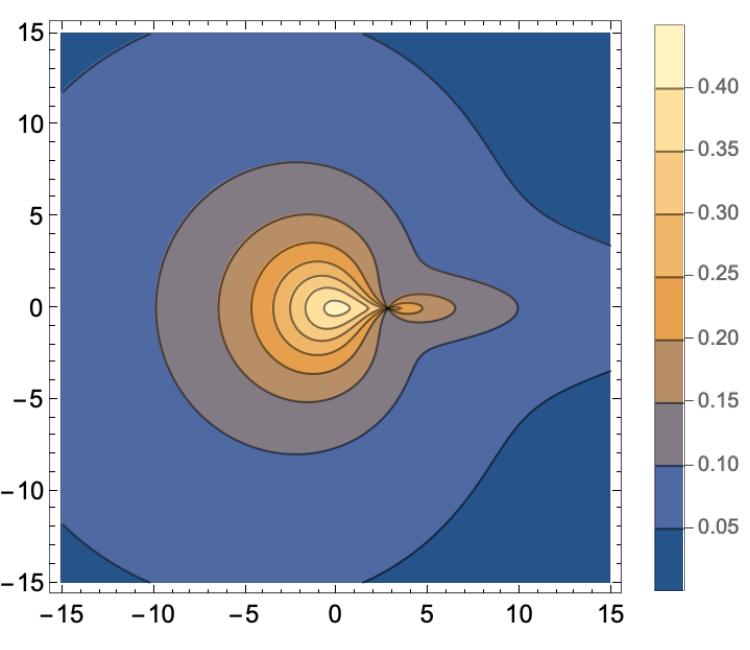}
\vspace*{.5cm}
\end{minipage}
\begin{minipage}[b]{0.32\linewidth}
\centering
\includegraphics[keepaspectratio,scale=0.35]{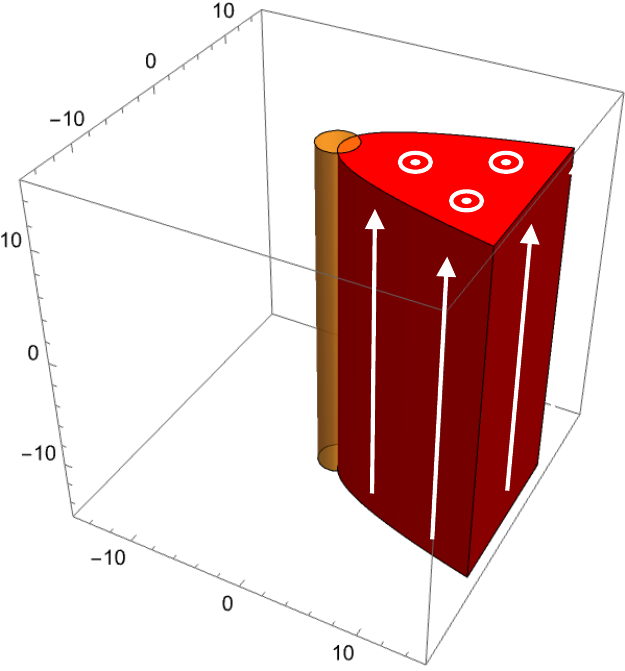}
\vspace*{.5cm}
\end{minipage}
\begin{minipage}[b]{0.32\linewidth}
\centering
\includegraphics[keepaspectratio,scale=0.35]{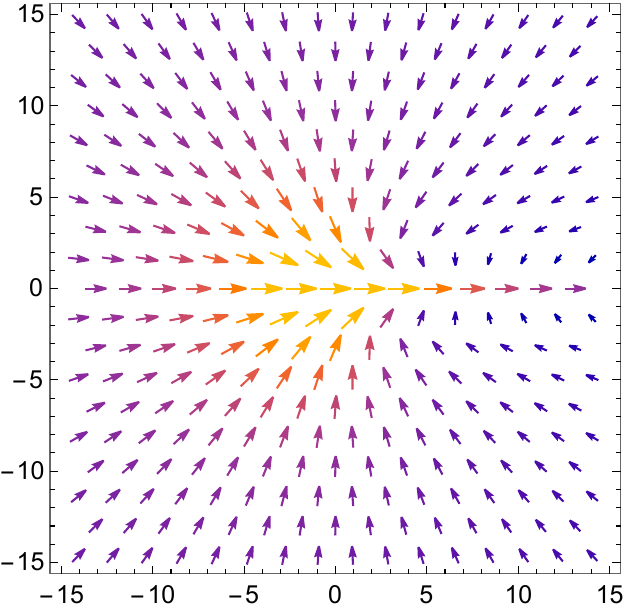}
\vspace*{.5cm}
\end{minipage}
\begin{minipage}[b]{0.32\linewidth}
\centering
\includegraphics[keepaspectratio,scale=0.35]{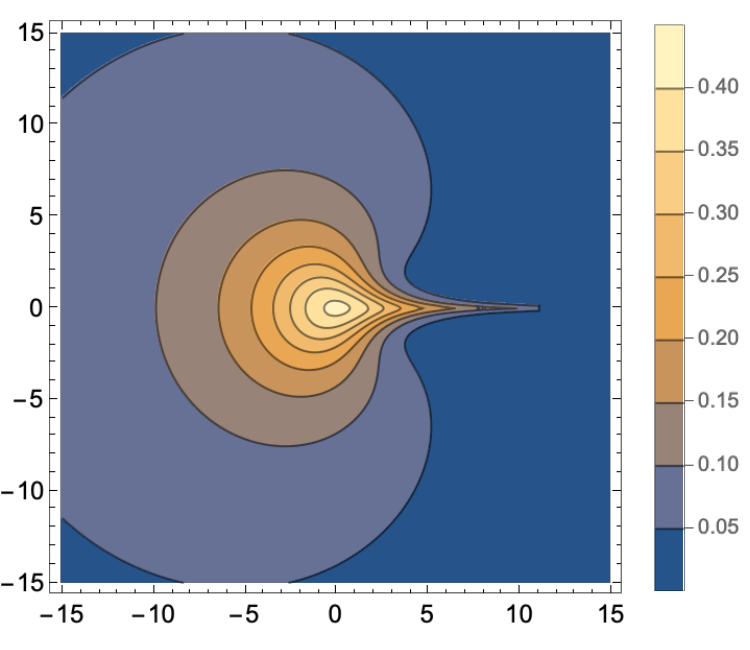}
\vspace*{.5cm}
\end{minipage}
\caption{The supercurrent and anomalous electric current for the $N=0$ axisymmetric axion string given in Fig.~\ref{fig:axion_N=0_BG}. The top (bottom) row corresponds to $\mu/vh = 0.8$ ($\mu/vh=1$), respectively. See the caption of Fig.~\ref{fig:anomaly_inflow_n2} for a detail explanation.}
\label{fig:anomaly_inflow_n0_asym}
\end{figure}

\begin{figure}[htbp]
\centering
\begin{minipage}[b]{0.32\linewidth}
\centering
\includegraphics[keepaspectratio,scale=0.35]{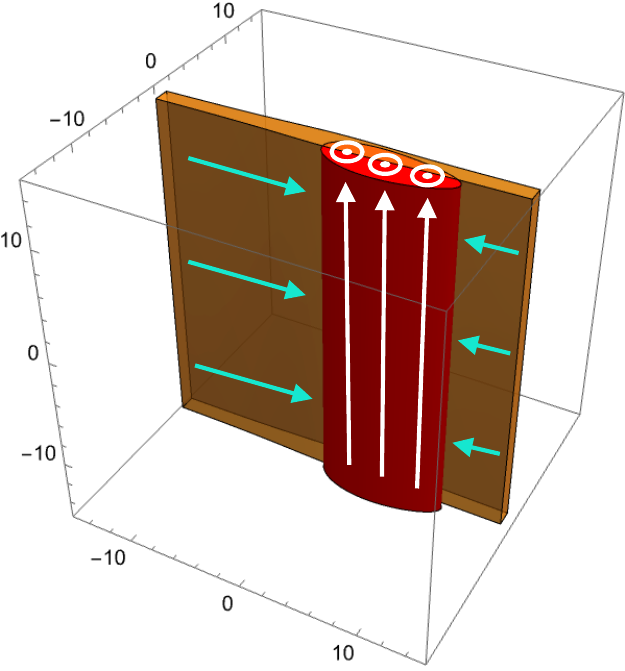}
\vspace*{.5cm}
\end{minipage}
\begin{minipage}[b]{0.32\linewidth}
\centering
\includegraphics[keepaspectratio,scale=0.35]{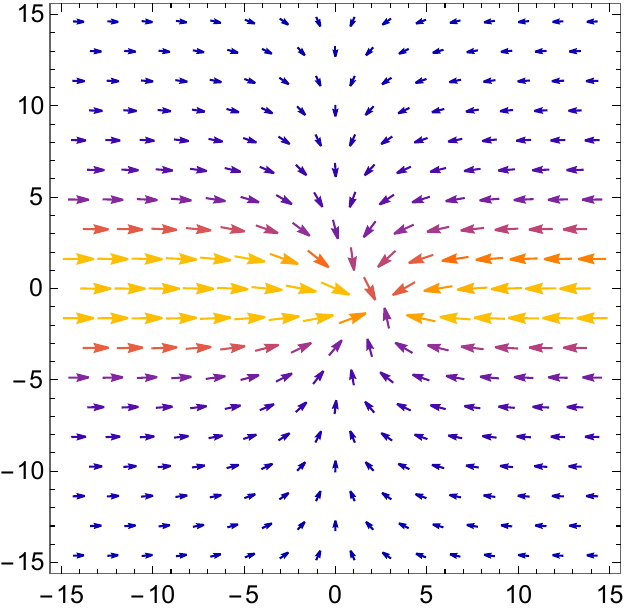}
\vspace*{.5cm}
\end{minipage}
\begin{minipage}[b]{0.32\linewidth}
\centering
\includegraphics[keepaspectratio,scale=0.35]{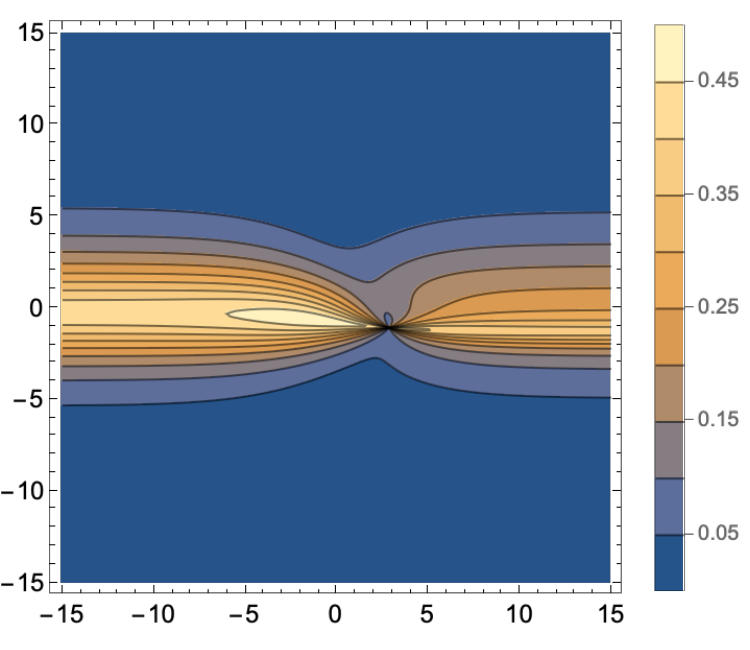}
\vspace*{.5cm}
\end{minipage}
\begin{minipage}[b]{0.32\linewidth}
\centering
\includegraphics[keepaspectratio,scale=0.35]{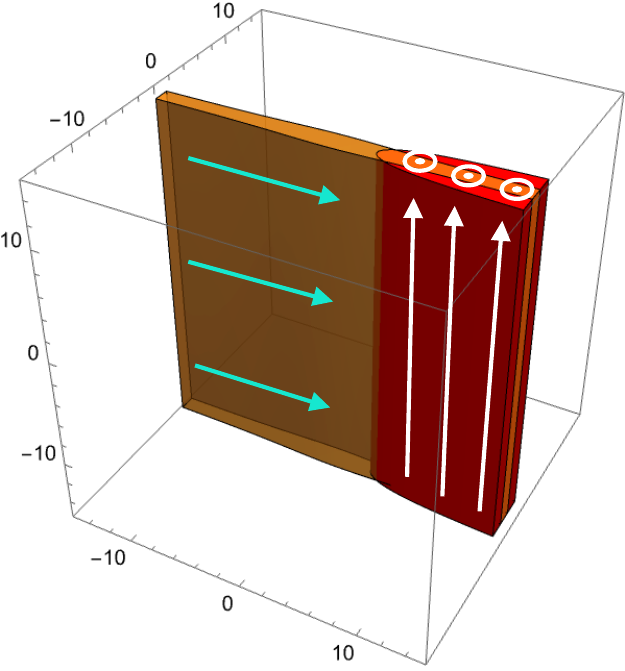}
\vspace*{.5cm}
\end{minipage}
\begin{minipage}[b]{0.32\linewidth}
\centering
\includegraphics[keepaspectratio,scale=0.35]{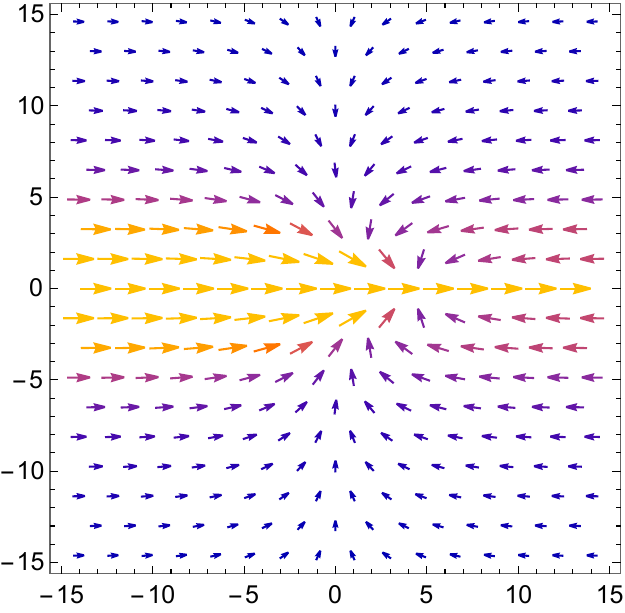}
\vspace*{.5cm}
\end{minipage}
\begin{minipage}[b]{0.32\linewidth}
\centering
\includegraphics[keepaspectratio,scale=0.35]{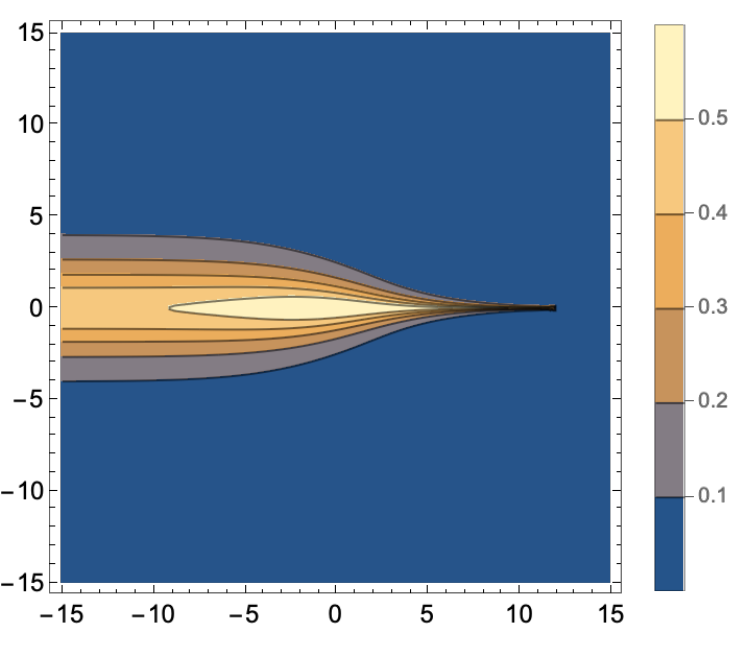}
\vspace*{.5cm}
\end{minipage}
\begin{minipage}[b]{0.32\linewidth}
\centering
\includegraphics[keepaspectratio,scale=0.35]{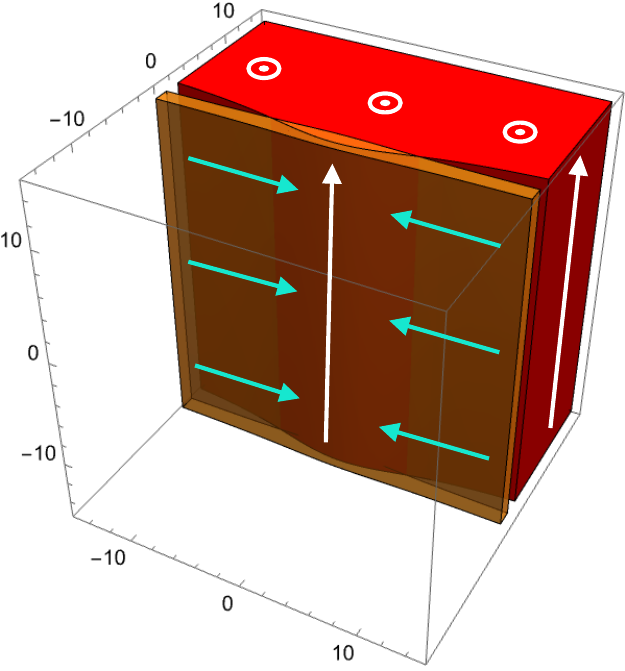}
\vspace*{.5cm}
\end{minipage}
\begin{minipage}[b]{0.32\linewidth}
\centering
\includegraphics[keepaspectratio,scale=0.35]{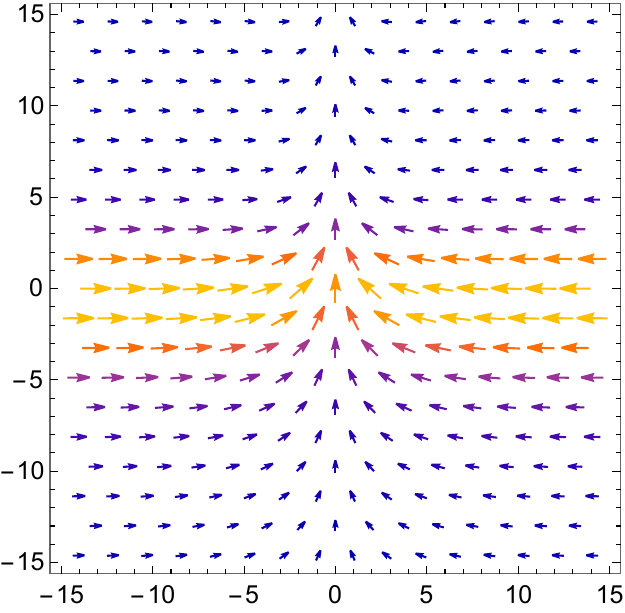}
\vspace*{.5cm}
\end{minipage}
\begin{minipage}[b]{0.32\linewidth}
\centering
\includegraphics[keepaspectratio,scale=0.35]{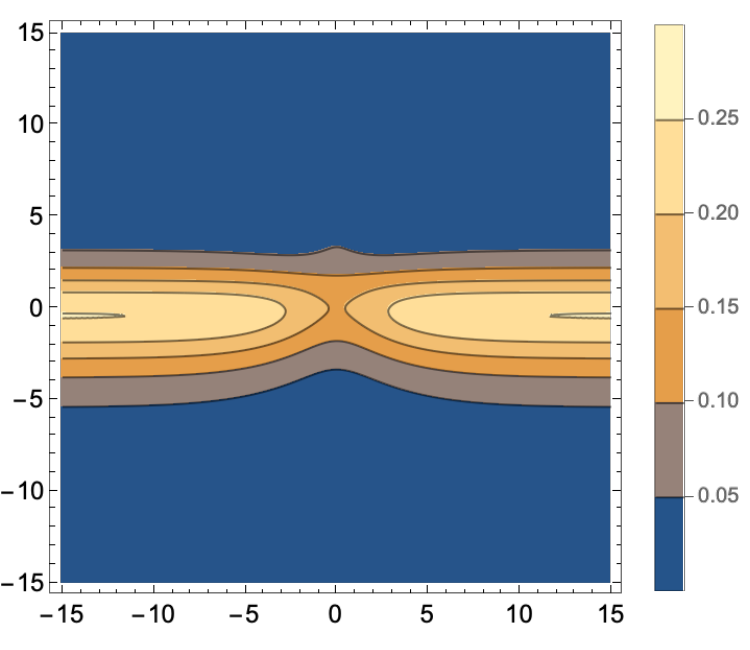}
\vspace*{.5cm}
\end{minipage}
\caption{The supercurrents and anomalous electric currents on the $N=2$ string-wall composites in the axion models shown in Fig.~\ref{fig:axion_N=2_a=0.1_BG}.
Each row from top to bottom corresponds to panels (c), (g), and (a) in Fig.~\ref{fig:shift_n2}, respectively. See the caption of Fig.~\ref{fig:anomaly_inflow_n2} for a detail explanation.}
\label{fig:anomaly_inflow_n2_asym}
\end{figure}

Let us next see effects of the bulk fermion mass $\mu = \mu_1 + i \mu_2$ in Eq.~(\ref{eq:L_fermion_2}).
As we mentioned in Eq.~(\ref{eq:phi_shift}), including bulk mass can be interpreted as the shift of the scalar field $\varphi \to \varphi - \mu/h$.
Accordingly, we replace $\varphi$ with $\varphi-\mu/h$ in the current ${\cal J}^\nu$ in Eq.~(\ref{eq:anomaly_inflow}).
The first example is the $N=0$ case. The background soliton is the axisymmetric string given in Fig.~\ref{fig:axion_N=0_BG}, and the non-axisymmetric massless mode functions for various $\mu$ are given in Fig.~\ref{fig:axion_string_shift}.
In Fig.~\ref{fig:anomaly_inflow_n0_asym} we show the supercurrent and anomalous electric current for $\mu/vh = 0.8$ and $1$.
In the former case, the elliptical supercurrent is off the center of the string, and the anomalous current flows in is not radially symmetric.
In the latter case, the supercurrent is not confined in a compact region but spreads infinitely in the positive $x$ direction, and the anomalous current flows in from the opposite side. 
Next let us see the second examples of $N=2$.
The $N=2$ composite is shown in Fig.~\ref{fig:axion_N=2_a=0.1_BG} and the zero mode functions for various mass shifts are given in Fig.~\ref{fig:shift_n2}. We take three concrete masses $(\mu_1,\mu_2)/vh = (\sqrt{1/10},-\sqrt{3/10})$, $(\sqrt{2/5},0)$, and $(0,\sqrt{6/5})$. The currents for the case of $(\mu_1,\mu_2) = (0,0)$ is shown in the top row of Fig.~\ref{fig:anomaly_inflow_n2}, and changes from this are shown in Fig.~\ref{fig:anomaly_inflow_n2_asym}. The first mass choice only shifts the supercurrent and anomalous current as shown in the top row of Fig.~\ref{fig:anomaly_inflow_n2_asym}. By contrast, the second and third choices lead to drastic changes. For the second choice, there are two walls, but the massless fermion only lives in one. Hence, the supercurrent only flows on that domain wall, and the anomalous current flows in on the other domain wall, see the middle row of Fig.~\ref{fig:anomaly_inflow_n2_asym}. 
Note that the two cases, one in the bottom row of Fig.~\ref{fig:anomaly_inflow_n0_asym} and the other in the middle row of Fig.~\ref{fig:anomaly_inflow_n2_asym}, look similar to one another but they are qualitatively different.
For the third parameter choice, the fermion zero mode is restricted to live in a half volume as the bottom row in Fig.~\ref{fig:anomaly_inflow_n2_asym}. The supercurrent flows along the $z$ axis through the entire half of the volume whereas the anomalous current flows through both domain walls toward the string and flows out into the half of the volume.

\begin{figure}[htbp]
\centering
\begin{minipage}[b]{0.32\linewidth}
\centering
\includegraphics[keepaspectratio,scale=0.35]{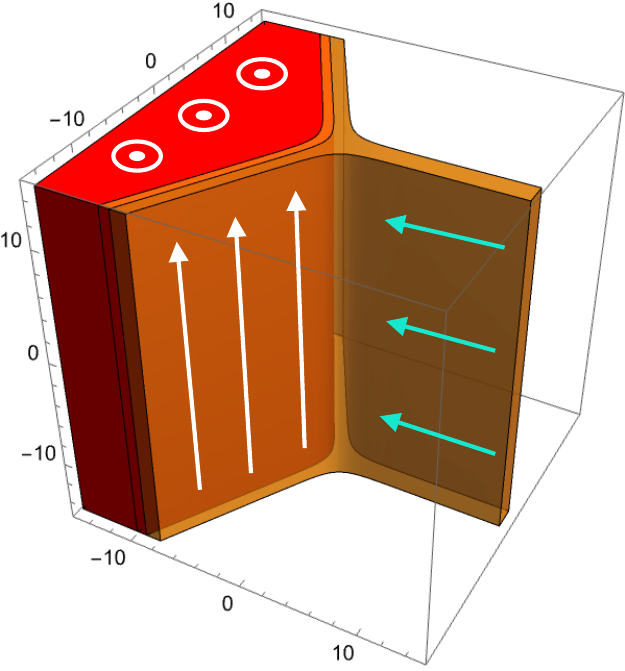}
\vspace*{.5cm}
\end{minipage}
\begin{minipage}[b]{0.32\linewidth}
\centering
\includegraphics[keepaspectratio,scale=0.35]{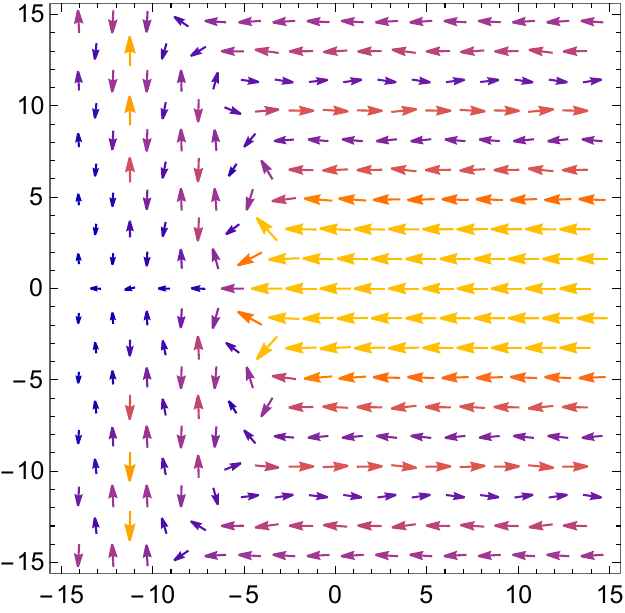}
\vspace*{.5cm}
\end{minipage}
\begin{minipage}[b]{0.32\linewidth}
\centering
\includegraphics[keepaspectratio,scale=0.35]{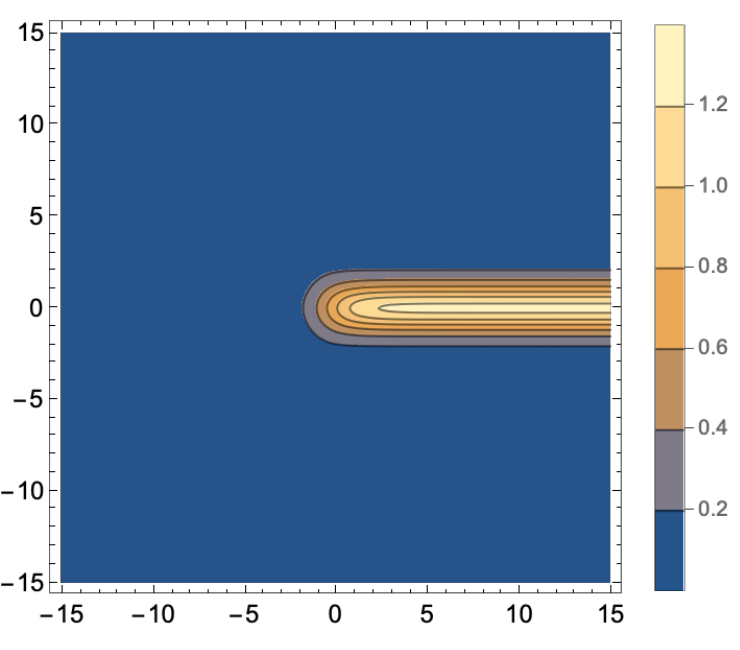}
\vspace*{.5cm}
\end{minipage}
\begin{minipage}[b]{0.32\linewidth}
\centering
\includegraphics[keepaspectratio,scale=0.35]{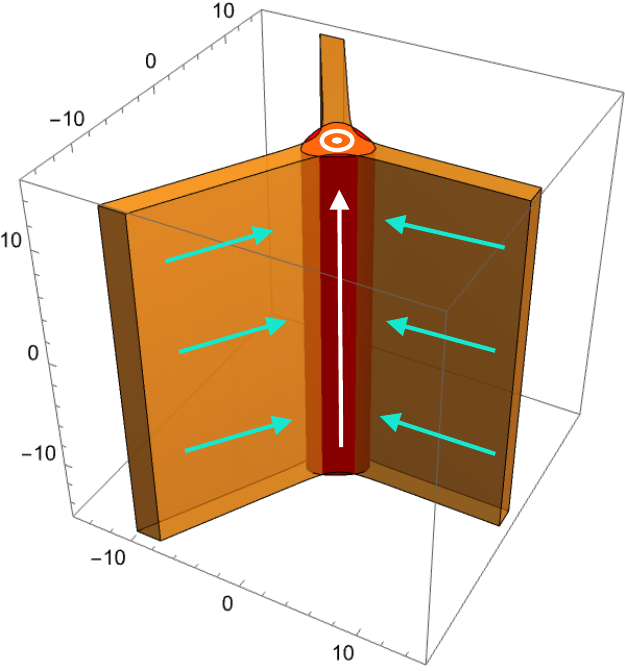}
\vspace*{.5cm}
\end{minipage}
\begin{minipage}[b]{0.32\linewidth}
\centering
\includegraphics[keepaspectratio,scale=0.35]{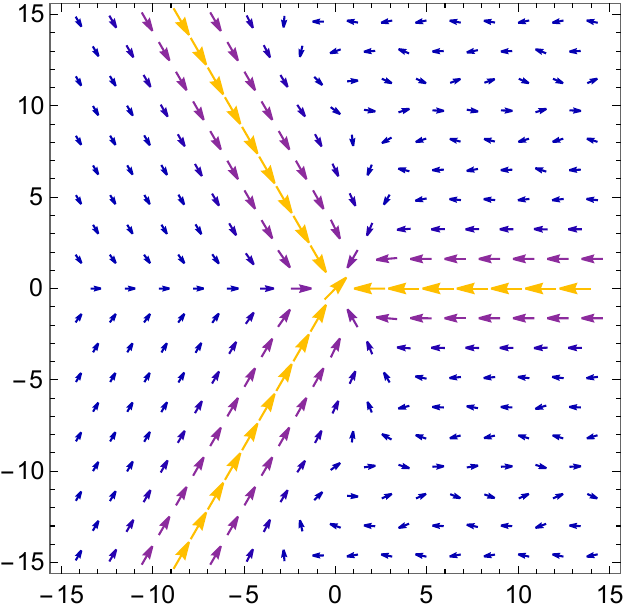}
\vspace*{.5cm}
\end{minipage}
\begin{minipage}[b]{0.32\linewidth}
\centering
\includegraphics[keepaspectratio,scale=0.35]{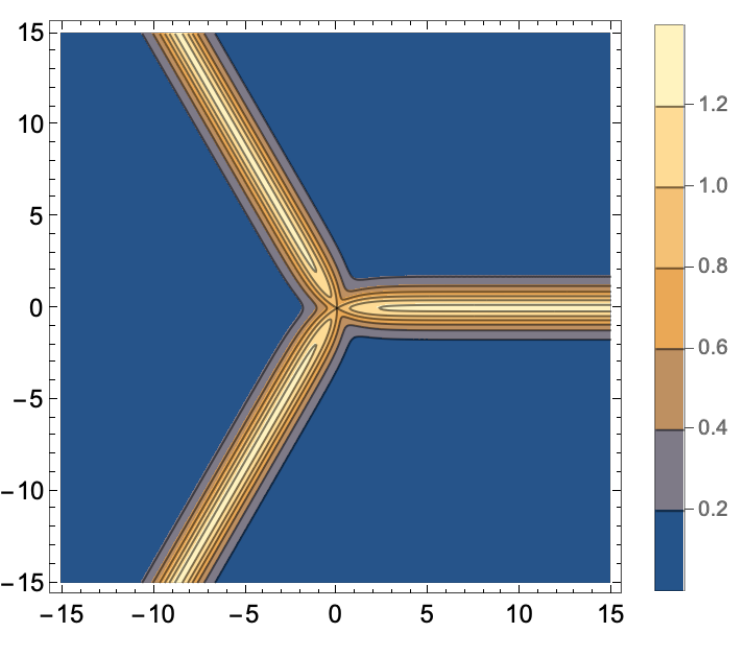}
\vspace*{.5cm}
\end{minipage}
\begin{minipage}[b]{0.32\linewidth}
\centering
\includegraphics[keepaspectratio,scale=0.35]{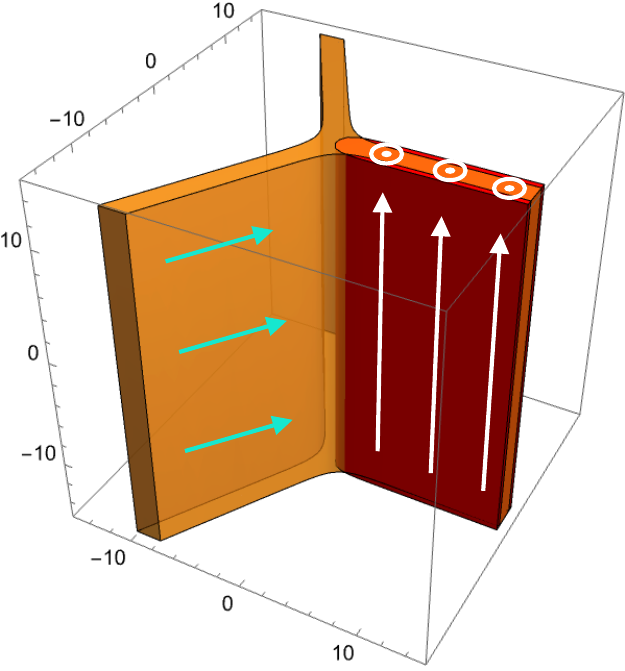}
\vspace*{.5cm}
\end{minipage}
\begin{minipage}[b]{0.32\linewidth}
\centering
\includegraphics[keepaspectratio,scale=0.35]{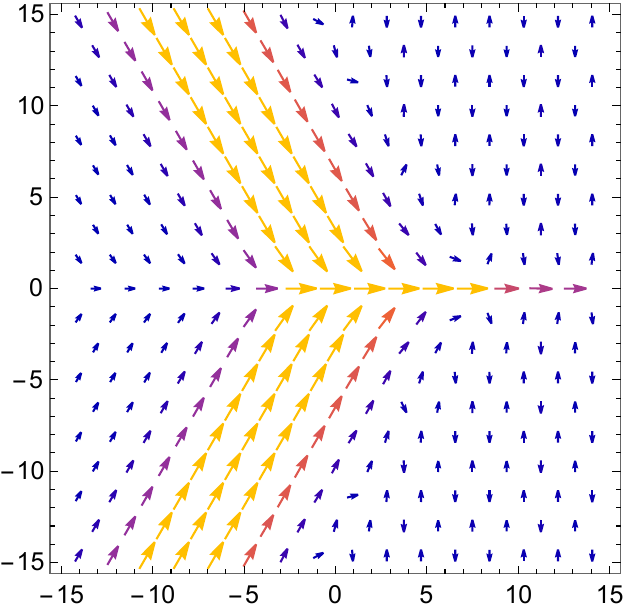}
\vspace*{.5cm}
\end{minipage}
\begin{minipage}[b]{0.32\linewidth}
\centering
\includegraphics[keepaspectratio,scale=0.35]{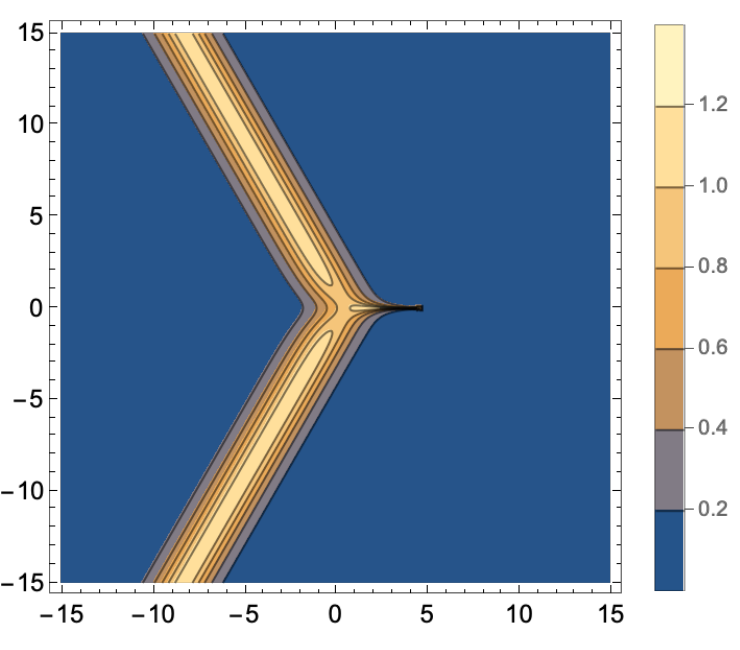}
\vspace*{.5cm}
\end{minipage}
\caption{The supercurrents and anomalous electric currents on the BPS string-wall composites in the SUSY-like models with $N_F=3$. 
Each row from top to bottom corresponds to panels (f), (d), and (b) in Fig.~\ref{fig:shift}, respectively. See the caption of Fig.~\ref{fig:anomaly_inflow_n2} for a detail explanation.}
\label{fig:anomaly_inflow_susy_nf3}
\end{figure}

Finally, let us take a look on more complicated string-wall composites in SUSY inspired model. All the features of the axion(-like) model found above are also common to the SUSY-like cases. In order to quickly see this, we take the previous examples for the $N_{\rm F}=3$ case. The background solution is the three-pronged domain wall junction shown in Fig.~\ref{fig:susy_dwj_n3_1}, and the fermion zero modes are shown in Fig.~\ref{fig:shift} with different overall mass parameter $\vec m_0$.
Here we focus on panels (f), (d), and (b) of Fig.~\ref{fig:shift}.
Corresponding supercurrents and anomalous currents are shown in Fig.~\ref{fig:anomaly_inflow_susy_nf3}.
In the top row the supercurrent flows within one of the three vacua, in the middle row it flows along the string, and in the bottom row it flows along one of the three walls, respectively. Again we observe interesting situations that the supercurrents flow not along the stings but on one of the domain walls or in one of the noncompact vacua. The anomalous electric currents radially flow in from the domain walls which do not overlap with the supercurrents.
\begin{figure}[htbp]
\centering
\begin{minipage}[b]{0.32\linewidth}
\centering
\includegraphics[keepaspectratio,scale=0.35]{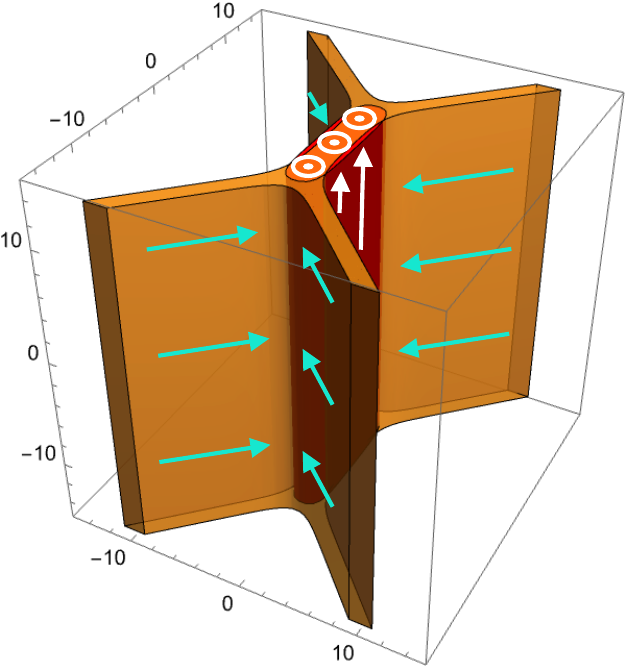}
\vspace*{.5cm}
\end{minipage}
\begin{minipage}[b]{0.32\linewidth}
\centering
\includegraphics[keepaspectratio,scale=0.35]{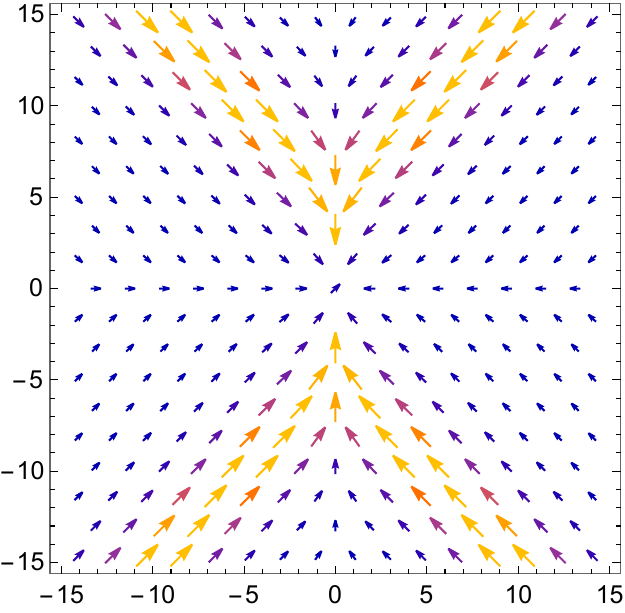}
\vspace*{.5cm}
\end{minipage}
\begin{minipage}[b]{0.32\linewidth}
\centering
\includegraphics[keepaspectratio,scale=0.35]{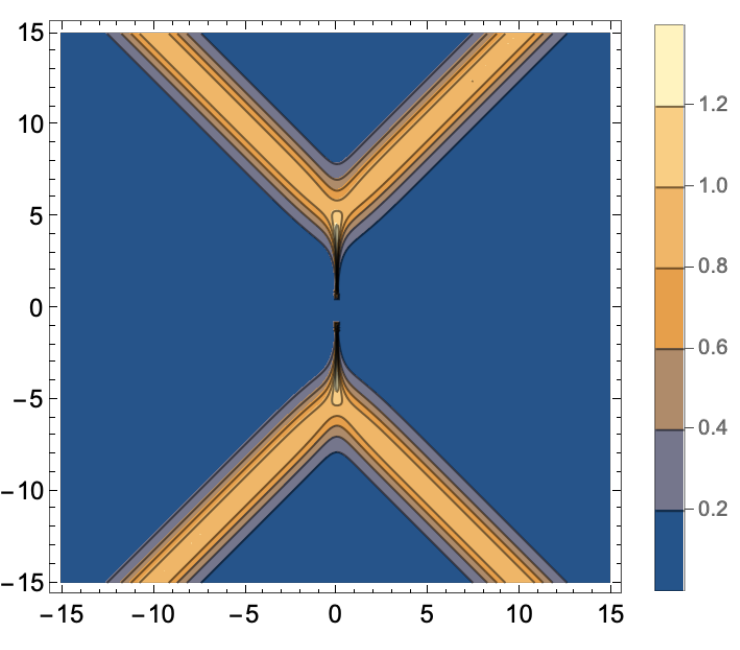}
\vspace*{.5cm}
\end{minipage}
\begin{minipage}[b]{0.32\linewidth}
\centering
\includegraphics[keepaspectratio,scale=0.35]{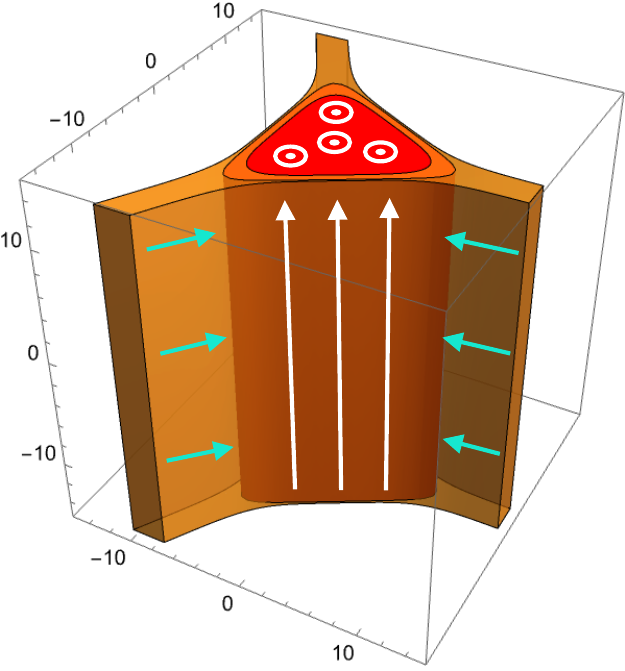}
\vspace*{.5cm}
\end{minipage}
\begin{minipage}[b]{0.32\linewidth}
\centering
\includegraphics[keepaspectratio,scale=0.35]{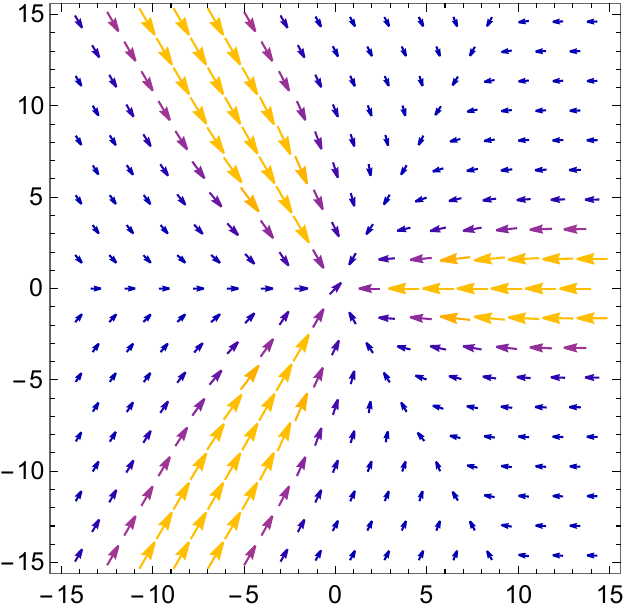}
\vspace*{.5cm}
\end{minipage}
\begin{minipage}[b]{0.32\linewidth}
\centering
\includegraphics[keepaspectratio,scale=0.35]{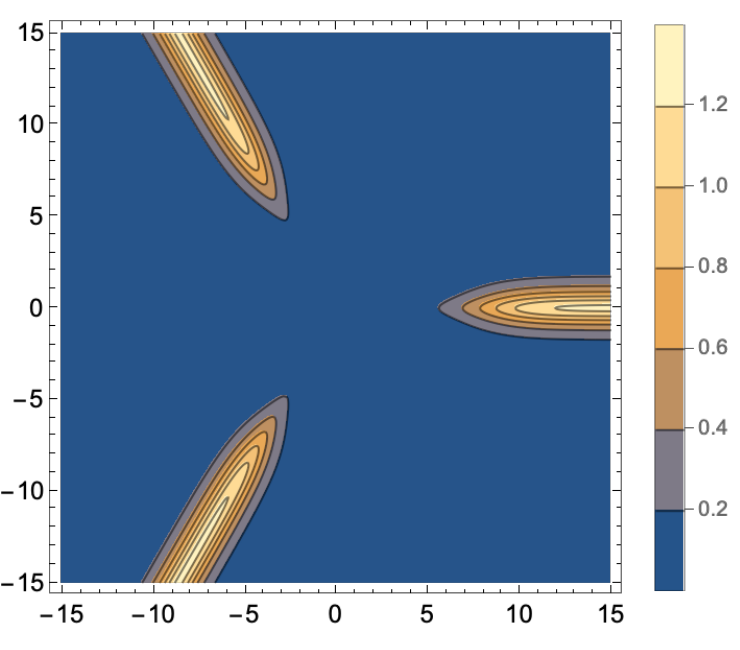}
\vspace*{.5cm}
\end{minipage}
\caption{The supercurrents and anomalous electric currents on the BPS string-wall composites in the SUSY-like models with $N_F=4$. 
The top (bottom) row corresponds to Fig.~\ref{fig:fermion_susy_nf4}(a) [(b)], respectively. See the caption of Fig.~\ref{fig:anomaly_inflow_n2} for a detail explanation.}
\label{fig:anomaly_inflow_susy_nf4}
\end{figure}
An advantage of the SUSY-like models is, however, the presence of more complicated composites of strings and walls. 
Here we focus on two special cases that can appear in the SUSY-like model with $N_{\rm F} \ge 4$ and have no similarities to the axion models. The first example is a supercurrent flowing on a domain wall of finite width given in the top row of Fig.~\ref{fig:susy_dwj_Nf4}.
The corresponding fermion zero mode is shown in Fig.~\ref{fig:fermion_susy_nf4}(a).
The second example is a supercurrent flowing inside a finite vacuum surrounded by the domain walls  given in the bottom row of Fig.~\ref{fig:susy_dwj_Nf4}. The corresponding fermion zero mode is shown in Fig.~\ref{fig:fermion_susy_nf4}(b).
These supercurrents flow on the finite domain wall and the compact $N$-gonal vacuum are shown in Fig.~\ref{fig:anomaly_inflow_susy_nf4}.
The anomalous electric currents flow in through the walls which do not have overlap with the supercurrents.

\section{Summaries and discussions}
\label{sec:summary}

In this work we have studied massless fermions couple to the string-wall composites parallel to the $z$-axis both in the axion(-like) models and in the bosonic part of the ${\cal N}=2$ supersymmetric Abelian-Higgs models.
We obtained analytic solutions to the Dirac equations couples to the non-trivial scalar field backgrounds through the Yukawa term.
Our solution is a natural extension of the domain wall fermion \cite{Jackiw:1975fn} and the string fermion \cite{Jackiw:1981ee,CALLAN1985427} established decades ago. With the solutions at hand, we succeeded in answering the elementary question of whether massless fermions exist on the string-wall composites and, if so, where they are localized. By giving the analytic solutions, we proved that there is a fermionic zero mode which is localized on a portion of the composite solitons. It is normalizable if the fermion mass $m_{\rm f} = h\varphi - \mu$ which depends on $x$ and $y$ through the scalar field $\varphi$ vanishes at a point in the $xy$ plane. The zero mode function peaks at the vanishing point.

In the axion models where the bulk fermion mass is absent $\mu=0$, we showed the massless fermions are localized on the axion string even after the QCD phase transition where the $N$ domain walls are attached to the string. The zero mode function is no longer axisymmetric but is $N$-gonal shape as shown in Figs.~\ref{fig:axion_zeromode_N=2}, \ref{fig:axion_zeromode_N=3and4}, \ref{fig:axion_zeromode_N=3_2}, and \ref{fig:axion_zeromode_N=1}.
When $\mu\neq0$ it is difficult to interpret $\arg \varphi$ as the QCD axion  since $U(1)_{\rm PQ}$ is badly broken. But there is still room for $\varphi$ to play a role as an axion-like particle. We found that the effect of nonzero $\mu$ is drastic. It sensitively controls localization of massless fermions. The fermion zero mode functions for the $N=0$ axisymmetric string are obtained analytically which are shown in Fig.~\ref{fig:axion_string_shift} (These agree well with the numerical solutions recently obtained in Ref.~\cite{Bagherian:2023jxy}).
We also numerically obtained the fermion zero mode functions for $N > 0$, see Fig.~\ref{fig:shift_n2} for $N=2$ as an example.
We found that in general the peak of the zero mode function is off the center of the string.
There are also extreme cases in which the fermion mass vanishes at one of the vacua or one of the domain walls.
In the former case, the massless fermions expand two-dimensionally over the corresponding vacuum domain as shown in Fig.~\ref{fig:shift_n2}(a).
In contrast, in the latter case it is localized one-dimensionally on the corresponding domain wall  as shown in Fig.~\ref{fig:shift_n2}(g).

We have emphasized importance of finding the analytic solutions throughout the paper. 
However, for the axion(-like) models, we could find the analytic solutions only for the $N=0$ case. On the other hand, it was only possible to solve Eq.~(\ref{eq:f}) numerically for $N>0$. We can be satisfied by the numerical solutions on the one hand, but on the other hand it is certainly insufficient as a mathematical proof for existence of a massless mode. We overcame this point by considering the BPS string-wall composites in the SUSY inspired Abelian-Higgs models which is one of the ideal platforms for studying the composite solitons.
There are several advantages: 1) the BPS solutions, no matter how complex they are, are given by the single gauge invariant scalar function $\Omega$ satisfying the master equation (\ref{eq:master}), 2) all the moduli are simply controlled by the moduli matrix $H_0$. These have been already established in the literatures~\cite{Isozumi:2004jc,Isozumi:2004va,Isozumi:2004vg,Eto:2004rz,Eto:2005cp,Eto:2006uw,Eto:2005yh,Eto:2005fm,Eto:2006pg,Eto:2007uc,Eto:2006bb,Eto:2006dx,Eto:2007yv,Eto:2006db,Eto:2008yi,Eto:2009zz,Eto:2020vjm,Eto:2020cys,Fujimori:2023wkd}. In addition to these virtues, 3) we found in this paper that for any BPS background configurations the fermion zero mode function is analytically given by $\Omega^{-h/2}$, see Eq.~(\ref{eq:susy_fermion_zero_mode}). Thus, we gave a proof of existence of massless fermion for all the BPS string-wall composites.
Our analytic solutions are also useful in practice. We gave concrete solutions in Figs.~\ref{fig:fermion_susy_nf3} and \ref{fig:shift}. We found the massless fermions localized on the string, one of the domain walls, or one of the noncompact vacuum domains. In addition, we also found the new zero modes peculiar to the BPS string-wall composites: The one is the zero mode localized on the finite domain wall and the other is the zero mode localized in the compact vacuum domain as shown in Fig.~\ref{fig:fermion_susy_nf4}.
We also obtained the fully analytic solutions in the strong gauge coupling limit $g^2\to\infty$, in which both the background BPS solutions and the fermionic zero mode functions are available in the analytic formula given in Eqs.~(\ref{eq:Omega_infty}) and (\ref{eq:fermion_strong}).

Finally, we showed the chiral superconducting currents and anomalous electric currents flowing on the string-wall composites.
In contrast to the axially symmetric ($N=0$) axion string, distributions of the both currents are very rich as shown in Figs.~\ref{fig:anomaly_inflow_n2}, \ref{fig:anomaly_inflow_n0_asym}, \ref{fig:anomaly_inflow_n2_asym}, \ref{fig:anomaly_inflow_susy_nf3}, and \ref{fig:anomaly_inflow_susy_nf4}.

There seem to be several possible extensions of this work. We briefly mention them in order below.

We proved the existence of the massless fermion on the string-wall composites by giving the analytic solution to the Dirac equation. However, this is only to show the presence of one massless state and we still do not know how many massless modes exist. To reveal this point, we can calculate index of the Dirac equation by generalizing the index theorem for the vortex-fermion system \cite{Weinberg:1981eu}.
On the other hand, the string-wall composites can be thought of as the domain walls with boundaries. As for the index theorem of the domain wall, recently, the Atiyah-Patodi-Singer (APS) index theorem \cite{atiyah_patodi_singer_1975} has been renewed in connection with the bulk-edge correspondence of symmetry protected topological insulators in \cite{Fukaya:2017tsq,Fukaya:2019qlf,Fukaya:2019myi}.
It would be interesting to study whether the APS index theorem can be applied to the domain wall fermions with the boundaries found in this paper.

In this paper we have focused on the string-wall-fermion systems. There are many other composite solitons, like monopole-string composites, monopole-wall composites, and monopole-string-wall composites etc are known in the literature, see for example \cite{Eto:2023gfn} and a comprehensive reference list there. However, it seems that there are almost no studies about the fermion-composite soliton systems except for the fermion-string-wall system studied in this paper. We will generalize the results in this paper to these generic fermion-composite soliton systems elsewhere. Even if it is limited to the string-wall composites, there are still different sorts of solitons. One of the direct generalization of this paper is to consider the BPS non-Abelian domain wall junctions in the ${\cal N}=2$ supersymmetric Yang-Mills-Higgs model \cite{Eto:2005fm}. We will also study the massless fermion on the non-Abelian domain wall junctions elsewhere. 

Domain walls are in general cosmological villains because they can soon dominate energy of the Universe. This is called the domain wall problem. One solution to the problem in axion cosmology is introducing a $Z_N$ breaking term called the bias in the potential \cite{Sikivie:1982qv,Chang:1998tb,Vilenkin:1981zs,Hiramatsu:2010yn}. 
It lifts the vacuum degeneracy and the domain walls collapse due to the  pressure force acting between different vacua. In axion(-like) models couple to the fermions, we found the equality among the degenerate vacua or the domain walls is {\it spontaneously} broken by the localization of massless fermions. 
Therefore, even if the bias in the potential is absent, the domain wall problem might be resolved by the presence of massless fermion.
It would be nice if we could elucidate this through large-scale numerical calculations with the massless fermions being dealt as a dynamical field.

Discovering the classification of materials by topological invariants is one of the recent highlights in physics \cite{Hasan:2010xy,Qi:2010qag}. It is a powerful tool for understanding various behaviors of matter.  A fundamental consequence of the topological classification is the bulk-edge correspondence, which relates the topological class of the bulk system to the number of gapless fermion edge states. The topological classification is generalized to topological defects where a generalization of the bulk-edge correspondence that relates  the topological classes to the defects to the presence of gapless modes on the defects is found \cite{Teo:2010zb}. At this point, it is not clear for us if the classification and the generalized bulk-edge correspondence can be applied to the composite solitons considered in this paper without generalizing.
However, such topological classification is certainly important and worth studying. It would be also interesting to figure out some relation between the massless fermions on the composite solitons and the so-called edge of edge state (or corner state) which appear at an intersection of edges of a topological material \cite{Hashimoto:2017tuh}.

Let us point out an impact of our results to the brane-world scenarios \cite{Arkani-Hamed:1998jmv,Antoniadis:1998ig,Randall:1999ee,Randall:1999vf}. In so-called fat brane-scenario our world is assumed to be realized on extended topological defects such as domain walls. There are several attempts that our world is realized on a domain wall junction in $5+1$ dimensions. For this scenario to work, at least one has to ensure all the physical particles are confined on the domain wall junction.  
The massless fermions are found but they are localized not on the domain wall junction but spread over the domain walls, so that they are non-normalizable \cite{Oda:1999az,Kakimoto:2003zu}. See the footnote \ref{foot:susy} for detail explanations.
In contrast, we found the massless fermions localized on the string (domain wall junction) which is normalizable and can play as physical particles on a low energy effective theory in $3+1$ dimensions. Since no realistic brane world models on the domain wall junction have been constructed so far, it is worth building realistic models using our solution.

Finally, in this paper we find the massless fermion can be localized on a half plane as Fig.~\ref{fig:shift}(b) and one-$N$th of the bulk as Fig.~\ref{fig:shift}(f). It is also possible to confine the massless fermion on a band with a finite size or inside a finite $N$-gonal prism as Fig.~\ref{fig:fermion_susy_nf4}.  These solutions would be useful for clarifying massless fermions in compact/non-compact systems with boundaries \cite{Biswas:2022dkg}.

\begin{acknowledgments}
This work is supported in part by JSPS KAKENHI No. JP22H01221 (ME).
\end{acknowledgments}

\bibliographystyle{jhep}
\normalem
\bibliography{references}

\providecommand{\href}[2]{#2}\begingroup\raggedright\begin{thebibliography}{10}

\bibitem{Vilenkin:2000jqa}
A.~Vilenkin and E.~S. Shellard, \emph{{Cosmic Strings and Other Topological
  Defects}}. Cambridge University Press, 7, 2000.

\bibitem{Peccei:1977hh}
R.~D. Peccei and H.~R. Quinn, \emph{{CP Conservation in the Presence of
  Instantons}}, \href{https://doi.org/10.1103/PhysRevLett.38.1440}{\emph{Phys.
  Rev. Lett.} {\bfseries 38} (1977) 1440}.

\bibitem{Weinberg:1977ma}
S.~Weinberg, \emph{{A New Light Boson?}},
  \href{https://doi.org/10.1103/PhysRevLett.40.223}{\emph{Phys. Rev. Lett.}
  {\bfseries 40} (1978) 223}.

\bibitem{Wilczek:1977pj}
F.~Wilczek, \emph{{Problem of Strong $P$ and $T$ Invariance in the Presence of
  Instantons}}, \href{https://doi.org/10.1103/PhysRevLett.40.279}{\emph{Phys.
  Rev. Lett.} {\bfseries 40} (1978) 279}.

\bibitem{Kibble:1976sj}
T.~W.~B. Kibble, \emph{{Topology of Cosmic Domains and Strings}},
  \href{https://doi.org/10.1088/0305-4470/9/8/029}{\emph{J. Phys. A} {\bfseries
  9} (1976) 1387}.

\bibitem{Zurek:1985qw}
W.~H. Zurek, \emph{{Cosmological Experiments in Superfluid Helium?}},
  \href{https://doi.org/10.1038/317505a0}{\emph{Nature} {\bfseries 317} (1985)
  505}.

\bibitem{WITTEN1985557}
E.~Witten, \emph{Superconducting strings},
  \href{https://doi.org/https://doi.org/10.1016/0550-3213(85)90022-7}{\emph{Nuclear
  Physics B} {\bfseries 249} (1985) 557}.

\bibitem{Jackiw:1981ee}
R.~Jackiw and P.~Rossi, \emph{{Zero Modes of the Vortex - Fermion System}},
  \href{https://doi.org/10.1016/0550-3213(81)90044-4}{\emph{Nucl. Phys. B}
  {\bfseries 190} (1981) 681}.

\bibitem{CALLAN1985427}
C.~Callan and J.~Harvey, \emph{Anomalies and fermion zero modes on strings and
  domain walls},
  \href{https://doi.org/https://doi.org/10.1016/0550-3213(85)90489-4}{\emph{Nuclear
  Physics B} {\bfseries 250} (1985) 427}.

\bibitem{Lazarides:1984zq}
G.~Lazarides and Q.~Shafi, \emph{{Superconducting Strings in Axion Models}},
  \href{https://doi.org/10.1016/0370-2693(85)91398-X}{\emph{Phys. Lett. B}
  {\bfseries 151} (1985) 123}.

\bibitem{Lazarides:1987rq}
G.~Lazarides, C.~Panagiotakopoulos and Q.~Shafi, \emph{{COSMIC SUPERCONDUCTING
  STRINGS AND COLLIDERS}},
  \href{https://doi.org/10.1016/0550-3213(88)90037-5}{\emph{Nucl. Phys. B}
  {\bfseries 296} (1988) 657}.

\bibitem{Carter:1993wu}
B.~Carter and X.~Martin, \emph{{Dynamic instability criterion for circular
  (Vorton) string loops}},
  \href{https://doi.org/10.1006/aphy.1993.1078}{\emph{Annals Phys.} {\bfseries
  227} (1993) 151} [\href{https://arxiv.org/abs/hep-th/0306111}{{\ttfamily
  hep-th/0306111}}].

\bibitem{Brandenberger:1996zp}
R.~H. Brandenberger, B.~Carter, A.-C. Davis and M.~Trodden, \emph{{Cosmic
  vortons and particle physics constraints}},
  \href{https://doi.org/10.1103/PhysRevD.54.6059}{\emph{Phys. Rev. D}
  {\bfseries 54} (1996) 6059}
  [\href{https://arxiv.org/abs/hep-ph/9605382}{{\ttfamily hep-ph/9605382}}].

\bibitem{Martins:1998gb}
C.~J. A.~P. Martins and E.~P.~S. Shellard, \emph{{Vorton formation}},
  \href{https://doi.org/10.1103/PhysRevD.57.7155}{\emph{Phys. Rev. D}
  {\bfseries 57} (1998) 7155}
  [\href{https://arxiv.org/abs/hep-ph/9804378}{{\ttfamily hep-ph/9804378}}].

\bibitem{Martins:1998th}
C.~J. A.~P. Martins and E.~P.~S. Shellard, \emph{{Limits on cosmic chiral
  vortons}}, \href{https://doi.org/10.1016/S0370-2693(98)01466-X}{\emph{Phys.
  Lett. B} {\bfseries 445} (1998) 43}
  [\href{https://arxiv.org/abs/hep-ph/9806480}{{\ttfamily hep-ph/9806480}}].

\bibitem{Carter:1999an}
B.~Carter and A.-C. Davis, \emph{{Chiral vortons and cosmological constraints
  on particle physics}},
  \href{https://doi.org/10.1103/PhysRevD.61.123501}{\emph{Phys. Rev. D}
  {\bfseries 61} (2000) 123501}
  [\href{https://arxiv.org/abs/hep-ph/9910560}{{\ttfamily hep-ph/9910560}}].

\bibitem{Fukuda:2020kym}
H.~Fukuda, A.~V. Manohar, H.~Murayama and O.~Telem, \emph{{Axion strings are
  superconducting}}, \href{https://doi.org/10.1007/JHEP06(2021)052}{\emph{JHEP}
  {\bfseries 06} (2021) 052}
  [\href{https://arxiv.org/abs/2010.02763}{{\ttfamily 2010.02763}}].

\bibitem{Abe:2020ure}
Y.~Abe, Y.~Hamada and K.~Yoshioka, \emph{{Electroweak axion string and
  superconductivity}},
  \href{https://doi.org/10.1007/JHEP06(2021)172}{\emph{JHEP} {\bfseries 06}
  (2021) 172} [\href{https://arxiv.org/abs/2010.02834}{{\ttfamily
  2010.02834}}].

\bibitem{Agrawal:2020euj}
P.~Agrawal, A.~Hook, J.~Huang and G.~Marques-Tavares, \emph{{Axion string
  signatures: a cosmological plasma collider}},
  \href{https://doi.org/10.1007/JHEP01(2022)103}{\emph{JHEP} {\bfseries 01}
  (2022) 103} [\href{https://arxiv.org/abs/2010.15848}{{\ttfamily
  2010.15848}}].

\bibitem{Ibe:2021ctf}
M.~Ibe, S.~Kobayashi, Y.~Nakayama and S.~Shirai, \emph{{On Stability of
  Fermionic Superconducting Current in Cosmic String}},
  \href{https://doi.org/10.1007/JHEP05(2021)217}{\emph{JHEP} {\bfseries 05}
  (2021) 217} [\href{https://arxiv.org/abs/2102.05412}{{\ttfamily
  2102.05412}}].

\bibitem{Sikivie:1982qv}
P.~Sikivie, \emph{{Of Axions, Domain Walls and the Early Universe}},
  \href{https://doi.org/10.1103/PhysRevLett.48.1156}{\emph{Phys. Rev. Lett.}
  {\bfseries 48} (1982) 1156}.

\bibitem{Chang:1998tb}
S.~Chang, C.~Hagmann and P.~Sikivie, \emph{{Studies of the motion and decay of
  axion walls bounded by strings}},
  \href{https://doi.org/10.1103/PhysRevD.59.023505}{\emph{Phys. Rev. D}
  {\bfseries 59} (1999) 023505}
  [\href{https://arxiv.org/abs/hep-ph/9807374}{{\ttfamily hep-ph/9807374}}].

\bibitem{Vilenkin:1981zs}
A.~Vilenkin, \emph{{Gravitational Field of Vacuum Domain Walls and Strings}},
  \href{https://doi.org/10.1103/PhysRevD.23.852}{\emph{Phys. Rev. D} {\bfseries
  23} (1981) 852}.

\bibitem{Eto:2005cp}
M.~Eto, Y.~Isozumi, M.~Nitta, K.~Ohashi and N.~Sakai, \emph{{Webs of walls}},
  \href{https://doi.org/10.1103/PhysRevD.72.085004}{\emph{Phys. Rev. D}
  {\bfseries 72} (2005) 085004}
  [\href{https://arxiv.org/abs/hep-th/0506135}{{\ttfamily hep-th/0506135}}].

\bibitem{Eto:2005fm}
M.~Eto, Y.~Isozumi, M.~Nitta, K.~Ohashi and N.~Sakai, \emph{{Non-Abelian webs
  of walls}}, \href{https://doi.org/10.1016/j.physletb.2005.10.017}{\emph{Phys.
  Lett. B} {\bfseries 632} (2006) 384}
  [\href{https://arxiv.org/abs/hep-th/0508241}{{\ttfamily hep-th/0508241}}].

\bibitem{Gibbons:1999np}
G.~W. Gibbons and P.~K. Townsend, \emph{{A Bogomolny equation for intersecting
  domain walls}},
  \href{https://doi.org/10.1103/PhysRevLett.83.1727}{\emph{Phys. Rev. Lett.}
  {\bfseries 83} (1999) 1727}
  [\href{https://arxiv.org/abs/hep-th/9905196}{{\ttfamily hep-th/9905196}}].

\bibitem{Carroll:1999wr}
S.~M. Carroll, S.~Hellerman and M.~Trodden, \emph{{Domain wall junctions are
  1/4 - BPS states}},
  \href{https://doi.org/10.1103/PhysRevD.61.065001}{\emph{Phys. Rev. D}
  {\bfseries 61} (2000) 065001}
  [\href{https://arxiv.org/abs/hep-th/9905217}{{\ttfamily hep-th/9905217}}].

\bibitem{Jackiw:1975fn}
R.~Jackiw and C.~Rebbi, \emph{{Solitons with Fermion Number 1/2}},
  \href{https://doi.org/10.1103/PhysRevD.13.3398}{\emph{Phys. Rev. D}
  {\bfseries 13} (1976) 3398}.

\bibitem{Schwarz:1979ur}
A.~S. Schwarz, \emph{Instantons and fermions in the field of instanton},
  \href{https://doi.org/10.1007/BF01221733}{\emph{Communications in
  Mathematical Physics} {\bfseries 64} (1979) 233}.

\bibitem{Montonen:1976yk}
C.~Montonen, \emph{{On Solitons with an Abelian Charge in Scalar Field
  Theories. 1. Classical Theory and Bohr-Sommerfeld Quantization}},
  \href{https://doi.org/10.1016/0550-3213(76)90537-X}{\emph{Nucl. Phys. B}
  {\bfseries 112} (1976) 349}.

\bibitem{Sarkar:1976vr}
S.~Sarkar, S.~E. Trullinger and A.~R. Bishop, \emph{{Solitary Wave Solution for
  a Complex One-Dimensional Field}},
  \href{https://doi.org/10.1016/0375-9601(76)90784-2}{\emph{Phys. Lett. A}
  {\bfseries 59} (1976) 255}.

\bibitem{Eto:2023gfn}
M.~Eto, Y.~Hamada and M.~Nitta, \emph{{Composite topological solitons
  consisting of domain walls, strings, and monopoles in O(N) models}},
  \href{https://doi.org/10.1007/JHEP08(2023)150}{\emph{JHEP} {\bfseries 08}
  (2023) 150} [\href{https://arxiv.org/abs/2304.14143}{{\ttfamily
  2304.14143}}].

\bibitem{Lambert:1999ix}
N.~D. Lambert and D.~Tong, \emph{{Kinky D strings}},
  \href{https://doi.org/10.1016/S0550-3213(99)00610-0}{\emph{Nucl. Phys. B}
  {\bfseries 569} (2000) 606}
  [\href{https://arxiv.org/abs/hep-th/9907098}{{\ttfamily hep-th/9907098}}].

\bibitem{Tong:2002hi}
D.~Tong, \emph{{The Moduli space of BPS domain walls}},
  \href{https://doi.org/10.1103/PhysRevD.66.025013}{\emph{Phys. Rev. D}
  {\bfseries 66} (2002) 025013}
  [\href{https://arxiv.org/abs/hep-th/0202012}{{\ttfamily hep-th/0202012}}].

\bibitem{Shifman:2002jm}
M.~Shifman and A.~Yung, \emph{{Domain walls and flux tubes in N=2 SQCD: D-brane
  prototypes}}, \href{https://doi.org/10.1103/PhysRevD.67.125007}{\emph{Phys.
  Rev. D} {\bfseries 67} (2003) 125007}
  [\href{https://arxiv.org/abs/hep-th/0212293}{{\ttfamily hep-th/0212293}}].

\bibitem{Shifman:2003uh}
M.~Shifman and A.~Yung, \emph{{Localization of nonAbelian gauge fields on
  domain walls at weak coupling (D-brane prototypes II)}},
  \href{https://doi.org/10.1103/PhysRevD.70.025013}{\emph{Phys. Rev. D}
  {\bfseries 70} (2004) 025013}
  [\href{https://arxiv.org/abs/hep-th/0312257}{{\ttfamily hep-th/0312257}}].

\bibitem{Lee:2002gv}
K.~S.~M. Lee, \emph{{An Index theorem for domain walls in supersymmetric gauge
  theories}}, \href{https://doi.org/10.1103/PhysRevD.67.045009}{\emph{Phys.
  Rev. D} {\bfseries 67} (2003) 045009}
  [\href{https://arxiv.org/abs/hep-th/0211058}{{\ttfamily hep-th/0211058}}].

\bibitem{Isozumi:2003rp}
Y.~Isozumi, K.~Ohashi and N.~Sakai, \emph{{Exact wall solutions in
  five-dimensional SUSY QED at finite coupling}},
  \href{https://doi.org/10.1088/1126-6708/2003/11/060}{\emph{JHEP} {\bfseries
  11} (2003) 060} [\href{https://arxiv.org/abs/hep-th/0310189}{{\ttfamily
  hep-th/0310189}}].

\bibitem{Isozumi:2003uh}
Y.~Isozumi, K.~Ohashi and N.~Sakai, \emph{{Massless localized vector field on a
  wall in D = 5 SQED with tensor multiplets}},
  \href{https://doi.org/10.1088/1126-6708/2003/11/061}{\emph{JHEP} {\bfseries
  11} (2003) 061} [\href{https://arxiv.org/abs/hep-th/0310130}{{\ttfamily
  hep-th/0310130}}].

\bibitem{Isozumi:2004jc}
Y.~Isozumi, M.~Nitta, K.~Ohashi and N.~Sakai, \emph{{Construction of
  non-Abelian walls and their complete moduli space}},
  \href{https://doi.org/10.1103/PhysRevLett.93.161601}{\emph{Phys. Rev. Lett.}
  {\bfseries 93} (2004) 161601}
  [\href{https://arxiv.org/abs/hep-th/0404198}{{\ttfamily hep-th/0404198}}].

\bibitem{Isozumi:2004va}
Y.~Isozumi, M.~Nitta, K.~Ohashi and N.~Sakai, \emph{{Non-Abelian walls in
  supersymmetric gauge theories}},
  \href{https://doi.org/10.1103/PhysRevD.70.125014}{\emph{Phys. Rev. D}
  {\bfseries 70} (2004) 125014}
  [\href{https://arxiv.org/abs/hep-th/0405194}{{\ttfamily hep-th/0405194}}].

\bibitem{Eto:2004vy}
M.~Eto, Y.~Isozumi, M.~Nitta, K.~Ohashi, K.~Ohta and N.~Sakai, \emph{{D-brane
  construction for non-Abelian walls}},
  \href{https://doi.org/10.1103/PhysRevD.71.125006}{\emph{Phys. Rev. D}
  {\bfseries 71} (2005) 125006}
  [\href{https://arxiv.org/abs/hep-th/0412024}{{\ttfamily hep-th/0412024}}].

\bibitem{Kakimoto:2003zu}
K.~Kakimoto and N.~Sakai, \emph{{Domain wall junction in N=2 supersymmetric QED
  in four-dimensions}},
  \href{https://doi.org/10.1103/PhysRevD.68.065005}{\emph{Phys. Rev. D}
  {\bfseries 68} (2003) 065005}
  [\href{https://arxiv.org/abs/hep-th/0306077}{{\ttfamily hep-th/0306077}}].

\bibitem{Isozumi:2004vg}
Y.~Isozumi, M.~Nitta, K.~Ohashi and N.~Sakai, \emph{{All exact solutions of a
  1/4 Bogomol'nyi-Prasad-Sommerfield equation}},
  \href{https://doi.org/10.1103/PhysRevD.71.065018}{\emph{Phys. Rev. D}
  {\bfseries 71} (2005) 065018}
  [\href{https://arxiv.org/abs/hep-th/0405129}{{\ttfamily hep-th/0405129}}].

\bibitem{Eto:2006pg}
M.~Eto, Y.~Isozumi, M.~Nitta, K.~Ohashi and N.~Sakai, \emph{{Solitons in the
  Higgs phase: The Moduli matrix approach}},
  \href{https://doi.org/10.1088/0305-4470/39/26/R01}{\emph{J. Phys. A}
  {\bfseries 39} (2006) R315}
  [\href{https://arxiv.org/abs/hep-th/0602170}{{\ttfamily hep-th/0602170}}].

\bibitem{Blaschke:2016lyj}
F.~Blaschke, \emph{{Exact BPS domain walls at finite gauge coupling}},
  \href{https://doi.org/10.1093/ptep/ptw168}{\emph{PTEP} {\bfseries 2017}
  (2017) 013B01} [\href{https://arxiv.org/abs/1603.00177}{{\ttfamily
  1603.00177}}].

\bibitem{Gauntlett:2000ib}
J.~P. Gauntlett, D.~Tong and P.~K. Townsend, \emph{{Multidomain walls in
  massive supersymmetric sigma models}},
  \href{https://doi.org/10.1103/PhysRevD.64.025010}{\emph{Phys. Rev. D}
  {\bfseries 64} (2001) 025010}
  [\href{https://arxiv.org/abs/hep-th/0012178}{{\ttfamily hep-th/0012178}}].

\bibitem{Eto:2020vjm}
M.~Eto, M.~Kawaguchi, M.~Nitta and R.~Sasaki, \emph{{Exact solutions of domain
  wall junctions in arbitrary dimensions}},
  \href{https://doi.org/10.1103/PhysRevD.102.065006}{\emph{Phys. Rev. D}
  {\bfseries 102} (2020) 065006}
  [\href{https://arxiv.org/abs/2001.07552}{{\ttfamily 2001.07552}}].

\bibitem{Eto:2020cys}
M.~Eto, M.~Kawaguchi, M.~Nitta and R.~Sasaki, \emph{{Exhausting all exact
  solutions of BPS domain wall networks in arbitrary dimensions}},
  \href{https://doi.org/10.1103/PhysRevD.101.105020}{\emph{Phys. Rev. D}
  {\bfseries 101} (2020) 105020}
  [\href{https://arxiv.org/abs/2003.13520}{{\ttfamily 2003.13520}}].

\bibitem{Bagherian:2023jxy}
H.~Bagherian, K.~Fraser, S.~Homiller and J.~Stout, \emph{{Zero Modes of Massive
  Fermions Delocalize from Axion Strings}},
  \href{https://arxiv.org/abs/2310.01476}{{\ttfamily 2310.01476}}.

\bibitem{Eto:2004rz}
M.~Eto, Y.~Isozumi, M.~Nitta, K.~Ohashi and N.~Sakai, \emph{{Instantons in the
  Higgs phase}}, \href{https://doi.org/10.1103/PhysRevD.72.025011}{\emph{Phys.
  Rev. D} {\bfseries 72} (2005) 025011}
  [\href{https://arxiv.org/abs/hep-th/0412048}{{\ttfamily hep-th/0412048}}].

\bibitem{Eto:2006uw}
M.~Eto, Y.~Isozumi, M.~Nitta, K.~Ohashi and N.~Sakai, \emph{{Manifestly
  supersymmetric effective Lagrangians on BPS solitons}},
  \href{https://doi.org/10.1103/PhysRevD.73.125008}{\emph{Phys. Rev. D}
  {\bfseries 73} (2006) 125008}
  [\href{https://arxiv.org/abs/hep-th/0602289}{{\ttfamily hep-th/0602289}}].

\bibitem{Eto:2005yh}
M.~Eto, Y.~Isozumi, M.~Nitta, K.~Ohashi and N.~Sakai, \emph{{Moduli space of
  non-Abelian vortices}},
  \href{https://doi.org/10.1103/PhysRevLett.96.161601}{\emph{Phys. Rev. Lett.}
  {\bfseries 96} (2006) 161601}
  [\href{https://arxiv.org/abs/hep-th/0511088}{{\ttfamily hep-th/0511088}}].

\bibitem{Eto:2007uc}
M.~Eto, T.~Fujimori, T.~Nagashima, M.~Nitta, K.~Ohashi and N.~Sakai,
  \emph{{Dynamics of Domain Wall Networks}},
  \href{https://doi.org/10.1103/PhysRevD.76.125025}{\emph{Phys. Rev. D}
  {\bfseries 76} (2007) 125025}
  [\href{https://arxiv.org/abs/0707.3267}{{\ttfamily 0707.3267}}].

\bibitem{Eto:2006bb}
M.~Eto, T.~Fujimori, T.~Nagashima, M.~Nitta, K.~Ohashi and N.~Sakai,
  \emph{{Effective Action of Domain Wall Networks}},
  \href{https://doi.org/10.1103/PhysRevD.75.045010}{\emph{Phys. Rev. D}
  {\bfseries 75} (2007) 045010}
  [\href{https://arxiv.org/abs/hep-th/0612003}{{\ttfamily hep-th/0612003}}].

\bibitem{Eto:2006dx}
M.~Eto, L.~Ferretti, K.~Konishi, G.~Marmorini, M.~Nitta, K.~Ohashi et~al.,
  \emph{{Non-Abelian duality from vortex moduli: A Dual model of
  color-confinement}},
  \href{https://doi.org/10.1016/j.nuclphysb.2007.03.040}{\emph{Nucl. Phys. B}
  {\bfseries 780} (2007) 161}
  [\href{https://arxiv.org/abs/hep-th/0611313}{{\ttfamily hep-th/0611313}}].

\bibitem{Eto:2007yv}
M.~Eto, J.~Evslin, K.~Konishi, G.~Marmorini, M.~Nitta, K.~Ohashi et~al.,
  \emph{{On the moduli space of semilocal strings and lumps}},
  \href{https://doi.org/10.1103/PhysRevD.76.105002}{\emph{Phys. Rev. D}
  {\bfseries 76} (2007) 105002}
  [\href{https://arxiv.org/abs/0704.2218}{{\ttfamily 0704.2218}}].

\bibitem{Eto:2006db}
M.~Eto, K.~Hashimoto, G.~Marmorini, M.~Nitta, K.~Ohashi and W.~Vinci,
  \emph{{Universal Reconnection of Non-Abelian Cosmic Strings}},
  \href{https://doi.org/10.1103/PhysRevLett.98.091602}{\emph{Phys. Rev. Lett.}
  {\bfseries 98} (2007) 091602}
  [\href{https://arxiv.org/abs/hep-th/0609214}{{\ttfamily hep-th/0609214}}].

\bibitem{Eto:2008yi}
M.~Eto, T.~Fujimori, S.~B. Gudnason, K.~Konishi, M.~Nitta, K.~Ohashi et~al.,
  \emph{{Constructing Non-Abelian Vortices with Arbitrary Gauge Groups}},
  \href{https://doi.org/10.1016/j.physletb.2008.09.007}{\emph{Phys. Lett. B}
  {\bfseries 669} (2008) 98} [\href{https://arxiv.org/abs/0802.1020}{{\ttfamily
  0802.1020}}].

\bibitem{Eto:2009zz}
M.~Eto, T.~Fujimori, S.~B. Gudnason, K.~Konishi, M.~Nitta, K.~Ohashi et~al.,
  \emph{{Constructing non-Abelian vortices with arbitrary gauge groups}},
  \href{https://doi.org/10.1063/1.3052002}{\emph{AIP Conf. Proc.} {\bfseries
  1078} (2009) 483}.

\bibitem{Fujimori:2023wkd}
T.~Fujimori, M.~Nitta and K.~Ohashi, \emph{{Moduli Spaces of Instantons in Flag
  Manifold Sigma Models -- Vortices in Quiver Gauge Theories}},
  \href{https://arxiv.org/abs/2311.04508}{{\ttfamily 2311.04508}}.

\bibitem{Weinberg:1981eu}
E.~J. Weinberg, \emph{{Index Calculations for the Fermion-Vortex System}},
  \href{https://doi.org/10.1103/PhysRevD.24.2669}{\emph{Phys. Rev. D}
  {\bfseries 24} (1981) 2669}.

\bibitem{atiyah_patodi_singer_1975}
M.~F. Atiyah, V.~K. Patodi and I.~M. Singer, \emph{Spectral asymmetry and
  riemannian geometry. i},
  \href{https://doi.org/10.1017/S0305004100049410}{\emph{Mathematical
  Proceedings of the Cambridge Philosophical Society} {\bfseries 77} (1975)
  43}.

\bibitem{Fukaya:2017tsq}
H.~Fukaya, T.~Onogi and S.~Yamaguchi, \emph{{Atiyah-Patodi-Singer index from
  the domain-wall fermion Dirac operator}},
  \href{https://doi.org/10.1103/PhysRevD.96.125004}{\emph{Phys. Rev. D}
  {\bfseries 96} (2017) 125004}
  [\href{https://arxiv.org/abs/1710.03379}{{\ttfamily 1710.03379}}].

\bibitem{Fukaya:2019qlf}
H.~Fukaya, M.~Furuta, S.~Matsuo, T.~Onogi, S.~Yamaguchi and M.~Yamashita,
  \emph{{The Atiyah\textendash{}Patodi\textendash{}Singer Index and Domain-Wall
  Fermion Dirac Operators}},
  \href{https://doi.org/10.1007/s00220-020-03806-0}{\emph{Commun. Math. Phys.}
  {\bfseries 380} (2020) 1295}
  [\href{https://arxiv.org/abs/1910.01987}{{\ttfamily 1910.01987}}].

\bibitem{Fukaya:2019myi}
H.~Fukaya, N.~Kawai, Y.~Matsuki, M.~Mori, K.~Nakayama, T.~Onogi et~al.,
  \emph{{The Atiyah\textendash{}Patodi\textendash{}Singer index on a lattice}},
  \href{https://doi.org/10.1093/ptep/ptaa031}{\emph{PTEP} {\bfseries 2020}
  (2020) 043B04} [\href{https://arxiv.org/abs/1910.09675}{{\ttfamily
  1910.09675}}].

\bibitem{Hiramatsu:2010yn}
T.~Hiramatsu, M.~Kawasaki and K.~Saikawa, \emph{{Evolution of String-Wall
  Networks and Axionic Domain Wall Problem}},
  \href{https://doi.org/10.1088/1475-7516/2011/08/030}{\emph{JCAP} {\bfseries
  08} (2011) 030} [\href{https://arxiv.org/abs/1012.4558}{{\ttfamily
  1012.4558}}].

\bibitem{Hasan:2010xy}
M.~Z. Hasan and C.~L. Kane, \emph{{Topological Insulators}},
  \href{https://doi.org/10.1103/RevModPhys.82.3045}{\emph{Rev. Mod. Phys.}
  {\bfseries 82} (2010) 3045}
  [\href{https://arxiv.org/abs/1002.3895}{{\ttfamily 1002.3895}}].

\bibitem{Qi:2010qag}
X.~L. Qi and S.~C. Zhang, \emph{{Topological insulators and superconductors}},
  \href{https://doi.org/10.1103/RevModPhys.83.1057}{\emph{Rev. Mod. Phys.}
  {\bfseries 83} (2011) 1057}
  [\href{https://arxiv.org/abs/1008.2026}{{\ttfamily 1008.2026}}].

\bibitem{Teo:2010zb}
J.~C.~Y. Teo and C.~L. Kane, \emph{{Topological Defects and Gapless Modes in
  Insulators and Superconductors}},
  \href{https://doi.org/10.1103/PhysRevB.82.115120}{\emph{Phys. Rev. B}
  {\bfseries 82} (2010) 115120}
  [\href{https://arxiv.org/abs/1006.0690}{{\ttfamily 1006.0690}}].

\bibitem{Hashimoto:2017tuh}
K.~Hashimoto, X.~Wu and T.~Kimura, \emph{{Edge states at an intersection of
  edges of a topological material}},
  \href{https://doi.org/10.1103/PhysRevB.95.165443}{\emph{Phys. Rev. B}
  {\bfseries 95} (2017) 165443}
  [\href{https://arxiv.org/abs/1702.00624}{{\ttfamily 1702.00624}}].

\bibitem{Arkani-Hamed:1998jmv}
N.~Arkani-Hamed, S.~Dimopoulos and G.~R. Dvali, \emph{{The Hierarchy problem
  and new dimensions at a millimeter}},
  \href{https://doi.org/10.1016/S0370-2693(98)00466-3}{\emph{Phys. Lett. B}
  {\bfseries 429} (1998) 263}
  [\href{https://arxiv.org/abs/hep-ph/9803315}{{\ttfamily hep-ph/9803315}}].

\bibitem{Antoniadis:1998ig}
I.~Antoniadis, N.~Arkani-Hamed, S.~Dimopoulos and G.~R. Dvali, \emph{{New
  dimensions at a millimeter to a Fermi and superstrings at a TeV}},
  \href{https://doi.org/10.1016/S0370-2693(98)00860-0}{\emph{Phys. Lett. B}
  {\bfseries 436} (1998) 257}
  [\href{https://arxiv.org/abs/hep-ph/9804398}{{\ttfamily hep-ph/9804398}}].

\bibitem{Randall:1999ee}
L.~Randall and R.~Sundrum, \emph{{A Large mass hierarchy from a small extra
  dimension}}, \href{https://doi.org/10.1103/PhysRevLett.83.3370}{\emph{Phys.
  Rev. Lett.} {\bfseries 83} (1999) 3370}
  [\href{https://arxiv.org/abs/hep-ph/9905221}{{\ttfamily hep-ph/9905221}}].

\bibitem{Randall:1999vf}
L.~Randall and R.~Sundrum, \emph{{An Alternative to compactification}},
  \href{https://doi.org/10.1103/PhysRevLett.83.4690}{\emph{Phys. Rev. Lett.}
  {\bfseries 83} (1999) 4690}
  [\href{https://arxiv.org/abs/hep-th/9906064}{{\ttfamily hep-th/9906064}}].

\bibitem{Oda:1999az}
H.~Oda, K.~Ito, M.~Naganuma and N.~Sakai, \emph{{An Exact solution of BPS
  domain wall junction}},
  \href{https://doi.org/10.1016/S0370-2693(99)01355-6}{\emph{Phys. Lett. B}
  {\bfseries 471} (1999) 140}
  [\href{https://arxiv.org/abs/hep-th/9910095}{{\ttfamily hep-th/9910095}}].

\bibitem{Biswas:2022dkg}
S.~Biswas and G.~W. Semenoff, \emph{{Massless fermions on a half-space: the
  curious case of 2+1-dimensions}},
  \href{https://doi.org/10.1007/JHEP10(2022)045}{\emph{JHEP} {\bfseries 10}
  (2022) 045} [\href{https://arxiv.org/abs/2208.06374}{{\ttfamily
  2208.06374}}].

\end{thebibliography}\endgroup

\end{document}